\documentclass[11pt, a4paper]{article}
\pdfoutput=1
\usepackage{jcappub}
\usepackage{aas_macros}
\usepackage{graphicx}
\usepackage{enumerate}
\usepackage{color}
\usepackage{amsmath}
\usepackage{amssymb}
\usepackage{hyperref}
\usepackage{xspace}
\usepackage{booktabs}

\newcommand{\GeV}{\ensuremath{{\rm\,GeV}}}
\newcommand{\run}[1]{\textsc{Run#1}}
\newcommand{\fex}{\textit{e.g.}}
\newcommand{\vect}[1]{\boldsymbol{#1}}
\newcommand{\Fermi}{\textit{Fermi}\xspace}

\newcommand{\sconf}[5]{
  $\left[\begin{smallmatrix}
      {#1} & {#2} & {#3} \\ {#4} & {#5} & \cdot
  \end{smallmatrix}\right]$
}

\begin{document}
\begin{flushright}
LAPTH-015/17
\end{flushright}

\title{SkyFACT: High-dimensional modeling of gamma-ray
emission with adaptive templates and penalized likelihoods}

\author[a]{Emma Storm}
\author[a]{Christoph Weniger}
\author[b]{Francesca Calore}
\emailAdd{e.m.storm@uva.nl}
\emailAdd{c.weniger@uva.nl}
\emailAdd{francesca.calore@lapth.cnrs.fr}
\affiliation[a]{GRAPPA, Institute of Physics, University of Amsterdam, Science Park 904, 1090 GL Amsterdam, Netherlands}
\affiliation[b]{LAPTh, CNRS, 9 Chemin de Bellevue, BP-110, Annecy-le-Vieux, 74941, Annecy Cedex, France}

\date{Compiled: \today}

\abstract{
  We present \textsc{SkyFACT} (\textbf{Sky} \textbf{F}actorization with \textbf{A}daptive \textbf{C}onstrained \textbf{T}emplates), a new approach for studying, modeling and decomposing diffuse gamma-ray emission.  Like most previous analyses, the approach relies on predictions from cosmic-ray propagation codes like GALPROP and DRAGON.  However, in contrast to previous approaches, we account for the fact that models are not perfect and allow for a very large number ($\gtrsim10^5$) of nuisance parameters to parameterize these imperfections. We combine methods of image reconstruction and adaptive spatio-spectral template regression in one coherent hybrid approach.  To this end, we use penalized Poisson likelihood regression, with regularization functions that are motivated by the maximum entropy method.  We introduce methods to efficiently handle the high dimensionality of the convex optimization problem as well as the associated semi-sparse covariance matrix, using the L-BFGS-B algorithm and Cholesky factorization.  We test the method both on synthetic data as well as on gamma-ray emission from the inner Galaxy, $|\ell|<90^\circ$ and $|b|<20^\circ$, as observed by the \Fermi Large Area Telescope.  We finally define a simple reference model that removes most of the residual emission from the inner Galaxy, based on conventional diffuse emission components as well as components for the \Fermi bubbles, the \Fermi Galactic center excess, and extended sources along the Galactic disk.  Variants of this reference model can serve as basis for future studies of diffuse emission in and outside the Galactic disk.
}


\maketitle
\tableofcontents


\section{Introduction}
\label{sec:intro}

Two important activities in the analysis of astronomical images are parametric modeling and image reconstruction.  Identifying adequate parametric models plays a central role in the astrophysical interpretation of observations, in particular for model discrimination and parameter regression.  In the case of gamma rays, modeling observational data is very often done using template regression techniques, where fits with linear combinations of physically or observationally motivated spatial templates are used to perform component separation, often as function of energy~\citep{Bennett:2003ca, Finkbeiner:2003im, Dobler:2007wv, Ade:2012nxf, Selig:2014qqa, Calore:2014xka}.  An orthogonal approach uses spectral templates to infer spatial characteristics of the various emission components~\citep{Selig:2014qqa, Huang:2015rlu, deBoer:2016esu}.  Template regression does in general allow a straightforward incorporation of observations at other frequencies or sky regions.

A more ambitious approach is the full spatio-spectral modeling of gamma-ray emission, using numerical codes that simulate the generation, propagation and interaction of a multitude of cosmic-ray species in the Galaxy.  This approach was pioneered with GALPROP~\citep[see, \fex,][]{Strong:1998fr, Ackermann:2012pya, TheFermi-LAT:2015kwa}; modern independent codes like DRAGON~\citep{Gaggero:2014xla, Gaggero:2015nsa, Evoli:2016xgn} and PICARD~\cite{Kissmann:2014sia,Werner:2014sya} build on that success (for a description of similarities and differences between the codes see ref.~\cite{Evoli:2016xgn}).  Both template regression and the full modeling of gamma-ray emission are used for parameter regression, usually of up to a few dozen parameters~\cite{Ackermann:2012pya}.  The overall performance of the models can be considered as reasonably good, with residuals typically at the level of $\sim30\%$~\cite{Ackermann:2012pya}.  However, given the uncertainties in various emission components, which are, \fex, related to errors in estimated gas densities~\cite{1976ApJ...208..346G,1988ApJ...324..248B,Pohl:2007dz,2016PASJ...68....5N}, poorly constrained interstellar radiation field~\cite{Porter:2008ve,Moskalenko:2005ng,2010A&A...524A..51D}, simplifying assumptions in the physical cosmic-ray propagation models~\citep[\fex,][]{Gaggero:2014xla,Evoli:2012ha,Acero:2016qlg}, it is not a surprise that the models are far away from providing formally acceptable fits to the data~\citep{Calore:2014xka}.

One possible consequence of imperfect modeling is that estimators can be severely biased.  At high Galactic latitudes, where the spatial characteristics of the background are reasonably well constrained and the relevant emission components (with the exception of, \fex, Loop I~\cite{1962MNRAS.124..405L,2009arXiv0912.3478C}) are reasonably well known~\cite[see, \fex][]{Ackermann:2014usa}, one could argue that biases are under control and comparable to statistical uncertainties.  However, close to and in the Galactic disk, where modeling the Galactic emission becomes increasingly important and statistical errors are small, bias can be rather significant.  Strong residuals along the Galactic disk~\citep[see, \fex][]{Calore:2014xka, Acero:2016qlg} can potentially lead to wrong conclusions in particular about extended emission features.  One possible approach to incorporate the characteristic magnitude of residuals as systematic error in statements about extended emission components was presented in~\citep{Calore:2014xka}, and based on a principal component analysis.

\medskip

Model-agnostic image reconstruction is a central tool to aid the visual interpretation of complex data.  It is hence no surprise that the use of image reconstruction techniques spans disciplines.  Their relevance extends across many research fields, including optical interferometry~\citep[\fex,][]{Schutz:2014a}, the analysis of gamma rays~\citep[\fex,][]{1979MNRAS.187..145S, Selig:2014qqa}, and image reconstruction in medical positron emission tomography (PET)~\cite[\fex,][]{Shepp:1982a}.  Interestingly, the underlying formal problems are often very similar, which allows the transfer of algorithms and techniques.  Image reconstruction is usually based on penalized likelihood maximization (or, in a Bayesian interpretation, maximum-a-posteriori reconstruction), often using Poisson or Gaussian likelihoods or one of their bias-reducing variants~\cite{Slambrouck:2015a}.

The number of parameters is usually very high ($\geq10^4$, often $\gg10^5$), as parameters correspond to individual pixels in the image reconstruction domain.  This requires additional regularization conditions to (a) reduce shot noise, (b) prevent over-fitting of the data, and (c) define unique solutions in potentially under-constrained problems~\cite[\fex,][]{Sra:2008a, Afonso:2011a, 2013JOSAA..30..160T}.  The regularization has the form of penalty terms in the likelihood, and its impact on the fit is controlled by various hyper-parameters.  In a Bayesian interpretation, they correspond to priors on the regression parameters.  Depending on the application, on the desired image properties, and on the available optimization algorithms, the regularization terms can take very different forms ~\cite{Afonso:2011a, 2014MNRAS.439.3591C}.  The arguably simplest regularizer is the $\ell_2$ norm, sometimes called `least square error method' or `energy penalty', which directly penalizes the variance of the model parameters~\cite{Afonso:2011a, Sra:2008a}.  A traditional regularization technique popular in astronomy is the maximum entropy method (MEM)~\cite{1979MNRAS.187..145S, Cornwell:1985a, Strong:1992a, Buscher:1994a, Hobson:1998a, Lalush:2000a, Bennett:2003ca, Sra:2008a}.  It aims to minimize the configurational entropy, providing a reconstructed image that is as `uninformative' and featureless as possible, while still being compatible with the data.  However, non-smooth and/or sparsity-enhancing regularizers have become increasingly important, since they aid in feature selection.  In particular the simple $\ell_1$ regularization enhances the sparsity of features in the image, and is connected to compressed sensing~\cite{Zhang:2013a}.  The similar `total variation regularization' enhances the sparsity of the image in the gradient domain~\cite{Afonso:2011a, 2014MNRAS.439.3591C}.  More complex variants of this approach enforce sparsity in the wavelet space~\cite{Schutz:2015a}.  However, explicit smoothing regularizers, based on finite differences and the $\ell_2$ norm, are also used.  The induced effective smoothing has an exponential profile and is in general non-uniform, since it depends on the varying signal-to-noise ratio of the image~\cite{Fessler:1996a, Nuyts:2002a, Asma:2004a, Schutz:2014a}.

\medskip

In gamma-ray data analysis, a combination of various aspects of image reconstruction and parametric modeling is desirable for a number of reasons.  In the case of existing spectral templates or models, fits in the spatial domain are practically an image reconstruction problem; a typical example is the \Fermi bubbles~\cite{2010ApJ...724.1044S,Fermi-LAT:2014sfa}.  But even in situations where spatial information about an emission component is available (from, \fex, observations at other frequencies), the associated predictions have in general correlated uncertainties that can be fully modeled only with a very large number of nuisance parameters.  An analysis of these nuisance parameters resembles again an image reconstruction problem.  Indeed, various approaches that effectively combine aspects of image reconstruction and modeling exist in the literature.  One example is the D$^3$PO algorithm~\citep{Selig:2014qqa, Huang:2015rlu}, which is based on a Bayesian analysis using information field theory.  The authors of this technique perform component separation, based on spectral information as well as priors on the smoothness of diffuse emission, to separate point source and diffuse components.  Another example is ref.~\cite{Bennett:2003ca}, where the authors perform an analysis of multi-frequency WMAP data, using spectral models and the MEM for spatial regularization.

In most gamma-ray analyses to date, nuisance parameters for spatial and/or spectral components of the models are not accounted for.  Instead, a common approach is to `bracket uncertainties' by focusing on a discrete set of plausible astrophysical configurations and analysis decisions.  Each of these scenarios leads then to a different result for the observable of interest, and the envelope of all results defines the `systematic uncertainty band'~\citep[\fex,][]{Ackermann:2012pya,Ackermann:2014usa,Calore:2014xka,TheFermi-LAT:2017vmf}.  
This approach is problematic in a number of ways.  Usually, none of the models provides a statistically acceptable fit to the data.  This raises the question why the consideration of various deficient models should lead to correct conclusions.  Furthermore, it greatly hinders any rigorous model comparison, because the typically adopted statistical approaches (comparing the goodness-of-fit, considering likelihood ratios, or even Bayesian model comparison) exhibit undefined behaviour.  In many cases, heuristic arguments can be made to support conclusions, but this does not replace a robust statistical approach.  Lastly, the nuisance parameters are not only accounting for uncertainties in the model predications, but usually provide valuable information about the underlying astrophysical cause for the model deviations.  This information is otherwise lost, or has to be guessed from residual maps.  With the present work, we aim at providing a \emph{first step} to overcome these limitations.

\medskip

In this article, we propose a new hybrid approach, which attempts to generalize and combine the benefits of many of the above techniques.  Our approach incorporates intrinsic uncertainties of various model components as penalty terms on spatial and spectral template modulation parameters.  It effectively allows the application of image reconstruction techniques to excesses above partially constrained backgrounds.  In particular, we aim to (a) facilitate component separation in scenarios where only partial knowledge about the spatio-spectral characteristics of the components is available, and (b) introduce a sufficient number of nuisance parameters in the analysis such that we can obtain formally good fits. A good fit to the data is a necessary criterion for reliable model comparison and bias-free parameter estimation.  Our approach is based on a penalized Poisson likelihood~\cite[\fex,][]{Asma:2004a, He:2016a} optimization with adaptive spatio-spectral (2-dim + 1-dim) templates.  The spatial and spectral modulation parameters for the adaptive templates can be in general regularized with MEM, $\ell_2$ or explicit first- or second-order smoothing terms.  Baring avoidable component degeneracies, we show that the underlying optimization problem is convex and hence has a unique (local and global) solution, despite the potentially millions of parameters.  We perform analytic calculations of the gradient, which allows the use of the efficient optimization algorithm L-BFGS-B~\cite{Byrd:1995a, Zhu:1997a, Morales:2011a}.  Uncertainties and covariances of the fit parameters are sampled from the inverse of the Fisher information matrix, using sparse Cholesky decomposition~\cite{Davis:2009a}.  We study the properties of our approach with synthetic data.

As the first application of our method, we analyze gamma-ray data from the \Fermi Large Area Telescope (LAT), focusing on a Region of Interest (RoI) along the Galactic plane including the Galactic center.  Our analysis is based on predictions from the cosmic-ray propagation code DRAGON~\cite{Evoli:2016xgn}, as well as a number of \textit{ad hoc} templates that absorb residual emission in the inner Galaxy.  We focus here on the technical aspects of obtaining a good fit to gamma-ray data, and will present a detailed physical interpretation of the results elsewhere.  One of the purposes of this work is hence to define a simple reference Galactic diffuse emission model that we can refer to (and further improve) in future work.

\medskip

This paper is organized as follows:  In section~\ref{sec:skyfact}, we describe the adaptive template model, regularization terms in the likelihood function, the fitting strategy as well as our treatment of the covariance matrix of the problem. In section~\ref{sec:gde}, we describe some details of the modeling of Galactic diffuse emission that we use in this work.  In section~\ref{sec:tomin} we provide a step-by-step example for how to construct a model for gamma-ray emission in our RoI.  In section~\ref{sec:min} we will discuss selected aspects of this model. In section~\ref{sec:discussion}, we discuss the fit quality of the reference model. In section~\ref{sec:conclusions}, we present our conclusions. In appendix~\ref{apx:synthetic} we present several studies using synthetic data to estimate possible biases of the method.

\section{SkyFACT -- Concepts and definitions}
\label{sec:skyfact}

SkyFACT (Sky Factorization with Adaptive Constrained Templates) is based on penalized maximum likelihood regression, using spatial and spectral templates with additional nuisance degrees of freedom that account for uncertainties of these templates.  Penalization terms act as priors on the nuisance parameters.  Errors are estimated by sampling from the inverse Fisher information matrix of the best fit point.  We discuss model definition, penalization terms, the optimization procedure and error calculations in detail in this section.

\subsection{Multi-linear modeling with adaptive templates}

The diffuse component of the photon flux (per time and area) in energy bin $b$ and spatial pixel $p$ is given by a tri-linear model,
\begin{equation}
    \phi_{pb} = \sum_k
    T_{p}^{(k)} \tau^{(k)}_{p} \cdot
    S_{b}^{(k)} \sigma^{(k)}_{b} \cdot
    \nu^{(k)}\;,
    \label{eqn:diffmodel}
\end{equation}
where $T_p^{(k)}$ and $S_b^{(k)}$ refer, respectively, to the spatial and spectral template for emission component $k$.  Furthermore, $\tau_p^{(k)}$ and $\sigma_b^{(k)}$ are spatial and spectral modulation parameters, respectively, and $\nu^{(k)}$ is an overall component normalization factor.   Note that, in general, the overall normalization factor cannot be absorbed into the spatial and/or spectral modulation parameters: In some cases the spatial and spectral uncertainties might be very small, while only the overall normalization remains as unconstrained fitting parameter.

The only \emph{strict} physical constraint on the modulation parameters and the normalization is non-negativity,
\begin{equation}
  \sigma_b^{(k)}, \tau_p^{(k)}, \nu^{(k)} \geq 0\;.
\end{equation}
Further constraints are introduced by regularization terms in the likelihood, and will be discussed below.  For diffuse components, 
the expected photon count in pixel $p$ and energy bin $b$ is given by
\begin{equation}
    \mu_{pb}^{\rm D} = \sum_{p'} \mathcal{P}_{bpp'}
    \mathcal{E}_{bp'}
    \phi_{p'b}\;,
\end{equation}
where $\mathcal{E}_{bp}$ denotes the exposure in the energy bin $b$ and $\mathcal{P}_{bpp'}$ the effect of the instrument Point Spread Function, PSF (the probability that a photon from pixel $p'$ is measured in pixel $p$).  We neglect here the effects of energy dispersion, although there is no conceptual difficulty in adding this when required.

The modeling of point sources does not require spatial modulation parameters, so the model is only bi-linear.  
The expected number of signal photons in this case is given by
\begin{equation}
    \mu_{pb}^{\rm P} = \sum_{s} \mathcal{P}_{bp}(\vect{\Omega_s})
    \mathcal{E}_b(\vect{\Omega_s})
    \cdot
    S_b^{(s)} \sigma_b^{(s)} \cdot \nu^{(s)}\;,
\end{equation}
where the sum is over point sources $s$ at angular locations $\vect{\Omega_s}$, and the PSF and exposure are evaluated at the source position. Note that the PSF matrix depends on the exact position of the source within the analysis pixel (whether it is at the center or at the edges). The total expected photon count is given by $\mu_{pb} = \mu_{pb}^{\rm D} + \mu_{pb}^{\rm P}$. To keep the notation compact, we use the symbol of the upper index ($s$ vs $k$) to discriminate between parameters related to point sources and related to diffuse components.

\subsection{Poisson likelihood and regularization terms}

\begin{figure}[h]
  \centering
  \includegraphics[width=0.6\linewidth]{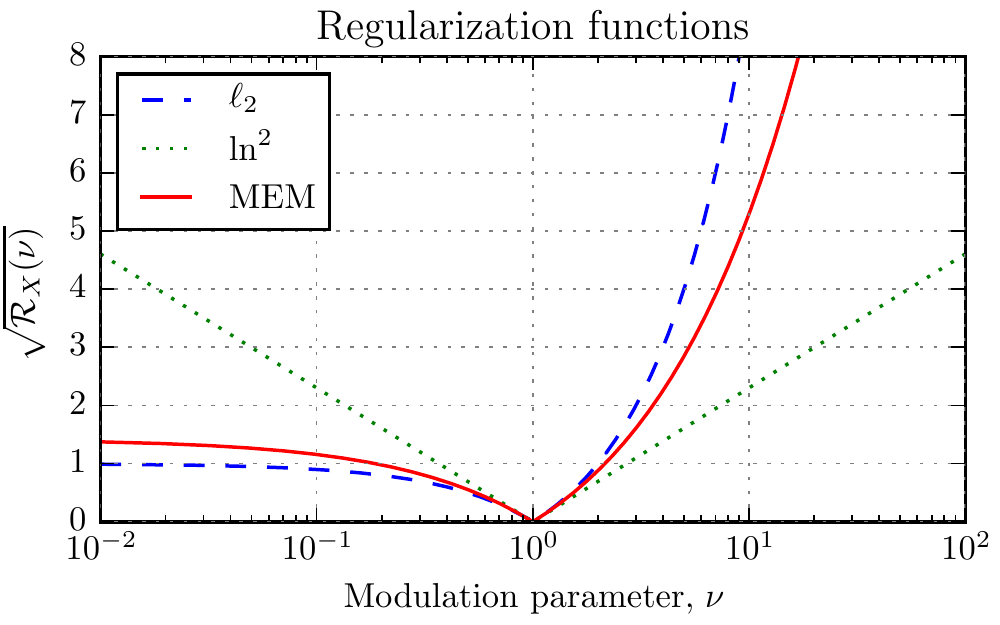}
  \caption{Various parameter regularization functions, as shown in Eqs.~\eqref{eqn:R_l2}, \eqref{eqn:R_ln2} and \eqref{eqn:R_MEM}, assuming $\lambda=1$. We plot $\sqrt{\mathcal{R}_X}$ rather than $\mathcal{R}_X$, since this allows a simple interpretation of the associated penalty in terms of standard deviations.  Throughout the paper, we adopt the convex MEM regularizer, which is motivated by the Maximum Entropy Method, as discussed in the text.  It is a compromise between the classical $\ell_2$ regularizer, and the non-convex $\ln^2$ regularizer.}
  \label{fig:MEM}
\end{figure}

The likelihood function used in this work has two major parts,
\begin{equation}
    \ln\mathcal{L} = \ln\mathcal{L}_{\rm P}+ \ln\mathcal{L}_{\rm R}\;,
\end{equation}
where $\mathcal{L}_{\rm P}$ corresponds to the Poisson likelihood that confronts model predictions with observations, and $\mathcal{L}_{\rm R}$ contains various regularization terms for the modulation parameters.

The Poisson likelihood reads
\begin{equation}
    \ln\mathcal{L}_{\rm P} = \sum_{pb} c_{pb} - \mu_{pb} + c_{pb}\ln
  \frac{\mu_{pb}}{c_{pb}}\;,
  \label{eqn:lnLP}
\end{equation}
where $\mu_{pb}$ and $c_{pb}$ refer, respectively, to the expectation value and the observed number of events.  We emphasize that we use the full Poisson likelihood in our fits, and do not assume Gaussian errors.  The definition is slightly different from the commonly adopted Cash-statistic~\cite{Cash:1979vz}, since we are actually considering the likelihood ratio conditioned on the observed data (in the sense that $\ln \mathcal{L}_P = 0$ for $\mu_{pb} = c_{pb}$). This version of the Poisson likelihood is commonly used in X-ray astronomy and is called the ``c-statistic'' in the widely-used spectral fitting tool XSPEC\footnote{https://heasarc.nasa.gov/docs/xanadu/xspec/} \cite{1996ASPC..101...17A}. It facilitates goodness-of-fit calculations, but has no effect on parameter regression.

\medskip

A central ingredient of our method are the large number of modulation parameters.
They can either act as nuisance parameters that effectively account for systematic uncertainties in the adopted templates, or they permit functionality that is akin to image reconstruction.  However, even in the case of image reconstruction, it remains important to regularize the modulation parameters in order to reduce the effects of Poisson noise and avoid over-fitting of the data.  In the present work, the regularization terms are defined as
\begin{multline}
  -2\ln\mathcal{L}_R = \sum_{k}
  \lambda_k \mathcal{R}_X(\vect\tau^{(k)})+
  \lambda'_k \mathcal{R}_X(\vect\sigma^{(k)})+
  \lambda''_k \mathcal{R}_X(\nu^{(k)})+
  \eta_k \mathcal{S}_1(\vect\tau^{(k)})+
  \eta'_k \mathcal{S}_2(\vect\sigma^{(k)})\\
  +\sum_{s}
  \lambda'_s \mathcal{R}_X(\vect\sigma^{(s)})+
  \lambda''_s \mathcal{R}_X(\nu^{(s)})+
  \eta'_s \mathcal{S}_2(\vect\sigma^{(s)})\;,
  \label{eqn:lnLR}
\end{multline}
where $X$ refers to the type of regularization, for each term separately. The first three regularization hyper-parameters, $\lambda_k$, $\lambda'_k$ and $\lambda''_k$, control respectively the constraints on the spatial, spectral and overall modulation parameters, for the diffuse components.  The regularization hyper-parameters $\eta_k$ and $\eta'_k$ control the strength of the spatial and spectral smoothing, respectively.  Explicit  smoothing is for instance useful to reduce the impact of Poisson noise on the spatial modulation parameters, or to enforce a physically motivated smoothness of a given emission component.  The point source regularization hyper-parameters, $\lambda'_s$, $\lambda''_s$ and $\eta'_s$, have an analogous functionality for point sources.  We use the vector notation, $\vect\tau^{(k)}$ etc, to indicate that the regularization and smoothing terms depend in general on all parameters simultaneously.  We briefly discuss three choices for the form of the regularization function $\mathcal{R}_X$, although only one of them (MEM) is used in the rest of this paper. The three choices are equivalent for large values of the regularization parameters, but differ in the tails, which is relevant if the regularization hyper-parameters are small and therefore the modulation parameters not tightly constrained.

\medskip

The $\ell_2$-norm regularization is quite commonly used in the literature~\cite{Afonso:2011a, Sra:2008a}. For our parameter choices, it has the simple form
\begin{equation}
  \lambda \mathcal{R}_{\ell_2}(\vect x) = \lambda \sum_i (x_i-1)^2\;,
  \label{eqn:R_l2}
\end{equation}
where $i$ runs over the elements of $\vect x$.  This regularization is a convex function, with the minimum at $x_i=1$; the associated variance is $1/\sqrt{\lambda}$. It can be used to constrain the element-wise variance of $\vect x$, and corresponds to Gaussian priors in a Bayesian interpretation.  However, it behaves poorly when parameters are supposed to lie within a logarithmic range, \fex, one or two dex. For that reason we do not use this regularization in the present work.

A naive modification of the above $\mathcal{R}_{\ell_2}(\cdot)$ regularization that constrains parameters on logarithmic scales is obtained by applying the $\ell_2$ in log-space of the modulation parameters.  This yields,
\begin{equation}
  \lambda\mathcal{R}_{\ln^2}(\vect x) = \lambda \sum_i \ln^2 x_i\;,
  \label{eqn:R_ln2}
\end{equation}
and has a global minimum at $x_i=1$.  The second derivative is given by $2\lambda(1-\ln x_i)/x_i^2$.  This means that again for large values of $\lambda$, this regularization is equivalent to the above linear $\ell_2$ one.  For small values of $\lambda$, it corresponds to a quadratic penalty term in the log-space of $x_i$, which is useful to constrain parameter ranges within a few dex, if necessary.  However, a disadvantage of the regularization term is that it is not convex (the second derivative becomes non-positive for $\ln x_i \geq 1$).  This spoils the convexity of the entire optimization problem, potentially introducing multiple local minima.  For this reason, we do not consider this regularizer further in the present work.

\medskip

When the magnitude of an emission component is largely unknown and should be inferred from the data, a well motivated regularization is based on the maximum entropy method~\cite{1979MNRAS.187..145S, Strong:1992a, Buscher:1994a}.  It reads
\begin{equation}
  \lambda\mathcal{R}_{MEM}(\vect x) = 2\lambda\sum_i 1- x_i + x_i \ln x_i\;.
  \label{eqn:R_MEM}
\end{equation}
The MEM regularizer is convex, and has a minimum at $x_i=1$.  The variance for large values of $\lambda$, as derived from the second derivative at $x_i=1$, is $1/\sqrt{\lambda}$.  It hence becomes equivalent to the $\ell_2$ and $\ln^2$ regularization in the large-$\lambda$ limit.  For smaller values of $\lambda$, the main effect of this regularization term is to smooth out regions with low flux levels, while still permitting pronounced bright regions if they are preferred by the data.  A good intuition can be obtained by noticing that its first and second derivatives are given by $2\lambda\ln x_i$ and $2\lambda/x_i$, respectively.  For very small values of $x_i$, $\mathcal{R}(\vect x)_{MEM}$ hence steepens infinitely and prevents $x_i$ from completely approaching zero.  For very large values of $x_i$, on the other hand, the Poisson part of the likelihood can take control, such that compact bright regions remain relatively unconstrained.  Due to its numerous practical properties, we use the MEM regularizer in the current work, not only for the spatial modulation parameters, but for all parameters. In figure~\ref{fig:MEM}, we show how the MEM regularizer behaves relative to the other regularizers discussed above.

\medskip

The smoothness of components can be imposed by constraining the \emph{gradient} of the modulation parameters instead of their range.  For the spatial modulation parameters, this is enforced by terms of the form
\begin{equation}
  \eta\mathcal{S}_1(\vect x) = \eta\sum_{(p,p')\in \mathcal{N}} (\ln x_p - \ln x_{p'})^2\;,
\end{equation}
where $\mathcal{N}$ refers to the set of nearest-neighbour pixel pairs.  Here, only pairs are included where both pixels are actually part the template (which can only cover a fraction of the sky).  Note that we constrain the gradient of the \emph{logarithm} of the modulation parameters instead of the modulation parameters directly.  This has the advantage that the overall normalization of the modulation parameters drops out.  The disadvantage is a loss of convexity, like for the above $\ln^2$ regularization.  This will be discussed below.

The spectral smoothing terms that we adopt in this work, on the other hand, constrain the \emph{second} derivative of the logarithm of the modulation parameter, and not the gradient.  The corresponding regularization term has the form
\begin{equation}
  \eta\mathcal{S}_2(\vect x) =
  \eta \sum_b (\ln x_{b-1} - 2 \ln x_b + \ln x_{b+1})^2\;.
\end{equation}
Assuming that the energy bins are logarithmically separated, this regularization favours power-law spectra. Since many astrophysical spectra actually have the form of a (potentially rolling) power-law, this is a reasonable behaviour that favours in many cases `more physical' spectra.  However, it has to be used with care since it will cause an automatic `power-law extrapolation' when the shot-noise level increases at high energies.

\medskip

The strength of all regularization terms are controlled with the regularization hyper- parameters $\lambda_k$, $\lambda'_k$, $\lambda''_k$, $\eta_k$ and $\eta'_k$ for diffuse components, and $\lambda'_s$, $\lambda''_s$, $\eta'_s$ for point sources.  We will discuss below, and in appendix~\ref{apx:synthetic}, a few examples of typical values and their effects on the resulting reconstruction.

\subsection{Parameter optimization with L-BFGS-B and convergence conditions}

Internally, the tri-linear model for the calculation of expected photon counts from diffuse emission components is represented by three matrices $A^{(1)}$, $A^{(2)}$ and $A^{(3)}$.  These map model parameters $\vect\theta \equiv (\vect\tau^{(k)}, \vect\sigma^{(k)}, \nu^{(k)}, \vect\sigma^{(s)}, \nu^{(s)})^T$ onto the expected photon fluxes $\vect \phi^D \equiv (\phi_{bp})$ via
\begin{equation}
  (\vect \phi^D)_i = (A^{(1)} \vect\theta)_i (A^{(2)} \vect\theta)_i (A^{(3)} \vect\theta)_i\;.
\end{equation}
The transformed vectors are multiplied element-wise, as indicated by the index $i$.  Although the matrices $A^{(1,2,3)}$ are very large (in our example of the order $10^5\times10^5$), they are rather sparse (in our example with a sparsity well above $99\%$) and can be efficiently stored in the compressed sparse column format.  The use of optimized sparse matrix routines leads then to very fast evaluation times.  Another advantage of the above representation is its simple analytical form, which facilitates the calculation of gradients and the Fisher information matrix.  The final expected number of photons, $\vect\mu^D = (\mu_{bp}^D)$ are obtained from the matrix product
\begin{equation}
  \vect\mu^D = \sum_j P_{ij} (\phi^D)_j (\vect E)_j\;,
\end{equation}
where $P$ and $\vect E$ represent the PSF and the exposure map convolution in data space respectively. For the PSF convolution, we assume periodic boundary conditions for simplicity, which leads to minor artefacts at the borders of the RoI, but is not relevant for the present discussion (although it will some impact on the formal goodness-of-fit).  Point sources are calculated in a similar way, but here two matrices $B^{(1)}$ and $B^{(2)}$ are sufficient, since the corresponding model is only bi-linear.  For the models below, we find that the evaluation of the likelihood function and the gradient takes less than a second on one core of a typical modern computer.  Sparse matrix multiplication is not straightforward to parallelize on shared or distributed memory systems, which makes it somewhat challenging (but certainly not impossible~\cite[\fex,][]{ballard2013communication}) to reduce the computation time further, even on multi-core systems.

\medskip

Parameter optimization in SkyFACT is performed with the L-BFGS-B method~\cite{Byrd:1995a, Zhu:1997a, Morales:2011a}.  The L-BFGS-B algorithm is based on the quasi-Newtonian method of Broyden, Fletcher, Goldfarb and Shanno (BFGS)~\cite{Wright:1999a}.  Its limited-memory version (L-BFGS) does not directly store numerical approximations to the Hessian matrix, but rather a number of previously calculated gradients to construct approximate Hessian matrices on the fly.  This makes it possible to use it for very large number of parameters ($10^6$ and more).  Furthermore, the version used here (L-BFGS-B) supports parameter boundaries, using the active-set method.  This is essential in order to be able to impose the non-negativity constraints.  We use here the implementation from Refs.~\cite{Zhu:1997a, Morales:2011a, Jones:2001--a}.  The algorithm takes both the objective function and its derivative as input, which are both straightforward to calculate based on the above analytical expressions.

There are two main convergence criteria that we consider in this work.  In the simplest case, one can just require that the vertical step-size of the last iteration is below a certain threshold value (this is one of the criteria of the adopted L-BFGS-B algorithm). A more informative quantity is however the estimated (vertical) distance to the minimum (EDM), which is known from the popular Minuit algorithm~\cite{James:1975dr}.  To estimate the EDM properly requires information about the full Hessian, which is however not easily available outside the L-BFGS-B implementation that we use.  We therefore adopt an extremely simple heuristic definition of an `effective EDM' that is motivated by the one-dimensional case.  It is defined as
\begin{equation}
  \text{EDM} \equiv
  \frac{||\vect\nabla f ||_2^2}{2}\frac{||\delta\vect x||_2}
  {||\delta \vect\nabla f||_2}\;,
\end{equation}
where $\delta \vect x$ refers to the parameter changes in the last step, $\delta \vect\nabla f$ to the change of the gradient, and $||\cdot||_2$ is the $\ell_2$ norm. We find that although it is only a crude (noisy, not monotonically decreasing) estimate of the vertical distance to the minimum, it is a useful dimensionless estimator to determine how close the fit is to the global minimum.

\subsection{Error estimation with Cholesky decomposition}

Estimating parameter uncertainties requires knowledge of the full covariance matrix of the model.  Here, we approximate the covariance matrix by the inverse of the Fisher information matrix. The Fisher matrix encodes the \emph{expected} error, and is independent of the data (it is numerically faster to calculate than the full Hessian matrix of the likelihood function, and numerically more stable even when the minimum of the system has not been exactly reached).  Formally, it is given by
\begin{equation}
    \mathcal{I}_{ij}
    = -\left\langle\frac{\partial^2}{\partial \theta_i \partial \theta_j} \ln\mathcal{L}\right\rangle_{\mathcal{D}(\vect\theta)}\;,
\end{equation}
where the average is taken over mock data from the best-fit model, $\mathcal{D}(\vect\theta)$.  The average can be taken analytically, and the resulting expressions are relatively simple to handle (see, \fex, ref.~\cite{Fisher} for a detailed discussion in context of Poisson likelihoods).  Since the Fisher matrix is composed of sparse matrices, it is sparse as well, at the 99\% level in the examples discussed below.

We estimate the errors of individual model parameters as well as predicted fluxes and other quantities using a sampling method.  This circumvents the need to explicitly calculate the covariance matrix (this would be computationally very costly, since the inverse of a sparse matrix is not sparse itself).  The procedure is based on sparse Cholesky decomposition~\cite{Davis:2009a}:   The Fisher information matrix is decomposed in the form $P\mathcal{I}P^T = L D L^T$, where $L$ is a lower unit triangular matrix, $D$ is a positive-definite diagonal matrix, and $P$ is a fill-reducing permutation matrix that maximizes the sparsity of $L$.   Based on the matrices $L$, $D$ and $P$, one can efficiently sample model parameter vectors $\delta\vect\theta$ such that their covariance is given by the inverse, $\mathcal{I}^{-1}$.  These model parameters describe then deviations from the best-fit model, similar to samples from Bayesian posterior distributions.

The sampling procedure works as follows.  We select a vector of standard normal distributed parameters, $\vect x \sim \mathcal{N}(0, 1)$, with the same length as the number of model parameters.  Next, we obtain $\vect y$ as solution of $L^T \vect y = D^{-1/2} \vect x$, which can be done by back-insertion.  Finally, we calculate $\delta \vect\theta = P^T \vect y$.  We repeat the process as many times as necessary to obtain a reasonably-sized sample.  As mentioned above, the full covariance matrix, $\Sigma_{ij}$, of the model parameters $\vect\theta$ can be estimated from $\langle \delta \theta_i \delta\theta_j \rangle\approx \Sigma_{ij}$, where the average is taken over many samples.  In practice, we are usually interested in mean values and variances of model fluxes, which are functions of $\vect\theta$.  Mean values for fluxes and other model predictions are derived from the best-fit value of $\vect\theta$.  Variances are derived by calculating model predictions for $\vect\theta+\delta\vect\theta$ and averaging over many samples.

\subsection{Convex optimization and non-degeneracy conditions}

Convex optimization is the minimization of convex functions, and has a number of useful properties.  In particular, any local minimum is also a global minimum, meaning that there is only one best-fit solution to the problem.  We will show here that -- as long as the model components remain sufficiently non-degenerate -- the objective function is indeed convex.  We will point out qualitatively under what conditions problems can occur (one can also derive quantitative conditions, but the resulting expressions are somewhat lengthy and we do not reproduce them here).

We first consider the Poisson likelihood $\mathcal{L}_P$.  To simplify the notation, it is useful to write the $\ln \mathcal{L}_P$ in the unbinned form (we refer to ref.~\cite{Fisher} for a discussion of unbinned Poisson likelihoods).  For simplicity, we neglect point source degrees of freedom (which are very sparse in the sky and not expected to be a problem in the present scenario), as well as the overall normalization factor.  We are in particular interested in the differential version of the Hessian.  It is given by
\begin{equation}
  -\sum_{ij} \delta \theta_i \delta \theta_j \frac{\partial^2}{\partial \theta_i\partial \theta_j} \ln\mathcal{L}_P
  = \int d\Omega\, dE\,
  \frac{\mathcal{C}}{\Phi^2}\delta \Phi^2 + \frac{\Phi-\mathcal{C}}{\Phi}
  \sum_k \frac{\delta\tau_k}{\tau_k}\frac{\delta\sigma_k}{\sigma_k}\Phi_k\;.
  \label{eqn:hessianP}
\end{equation}
Here, we defined the `unbinned counts map' $\mathcal{C}(\Omega,E) \equiv \sum_i \delta^{(2)}(\Omega-\Omega_i)\delta(E-E_i)$, where $E_i$ and $\Omega_i$ are the energy and position of photon $i$, and $\delta(\cdot)$ refers to Dirac delta distributions.
Furthermore, $\delta \Phi \equiv \delta\vect \theta \cdot \vect\nabla_\theta \Phi$, and $\delta \tau_k$ and $\delta\sigma_k$ are defined analogously.  Finally, $\Phi_k$ is the contribution to $\Phi$ from component $k$.

It is clear that the first term in Eq.~\eqref{eqn:hessianP} is always non-negative,  and it is strictly positive unless two or more model components are exactly degenerate (in that case some change in component $k$ can be exactly undone by the opposite change in component $k'$, leading to $\delta\Phi=0$ in that particular direction of the parameter vector $\delta\vect\theta$).  The second term can, depending on $\mathcal{C}$, however become negative and spoil the convexity of the objective function.  However, it depends on the simultaneous change of spectral and spatial modulation parameters of the \textit{same} component.  It is exactly zero for any component that has either spatial or spectral degrees of freedom, but not both.  As long as the model consists only of components that each have either a fixed distinct morphology or spectrum, the optimization problem is strictly convex.

However, if both spectral and spatial variations are allowed for at least two components in the fit, one can find in general degenerate transformations with $\delta\vect \Phi = 0$ while $\delta\tau_k\neq0$ and $\delta\sigma_k\neq0$.  In that case, the second term of Eq.~\eqref{eqn:hessianP} can potentially become negative, and the objective function could have multiple local minima.  This corresponds to the case where components have so much spectral and spatial freedom that they can effectively exchange their roles in the fit.  Situations like that are usually not desired, and should be prevented by the regularization of the modulation parameters.

As discussed above, both the MEM and $\ell_2$ regularizations are completely convex, and hence cannot spoil the convexity of the problem.  Since we use the MEM regularization, increasing the value of the corresponding hyper-parameters $\lambda$ can only increase the convexity of the problem, while breaking the degeneracy between model components.

Note, however, that the smoothing terms $\mathcal{S}$ are \emph{not} generally convex.  This is caused by the fact that the logarithm of the modulation parameters (\fex, $\ln \tau_p^{(k)}$) enters their definition, which is similar to the above $\ln^2$ regularization.  For this reason, it is advisable to check explicitly that the global minimum is actually reached.  This can be done by confirming that reducing the smoothing hyper-parameters $\eta$ and $\eta'$ does not qualitatively change the result (although it will reduce the smoothness of the components in the fit; see appendix~\ref{apx:run5}.)

\section{Modeling Galactic diffuse emission}
\label{sec:gde}

\subsection{Data selection}

\begin{figure}[t]
    \centering
    \includegraphics[width=0.99\linewidth]{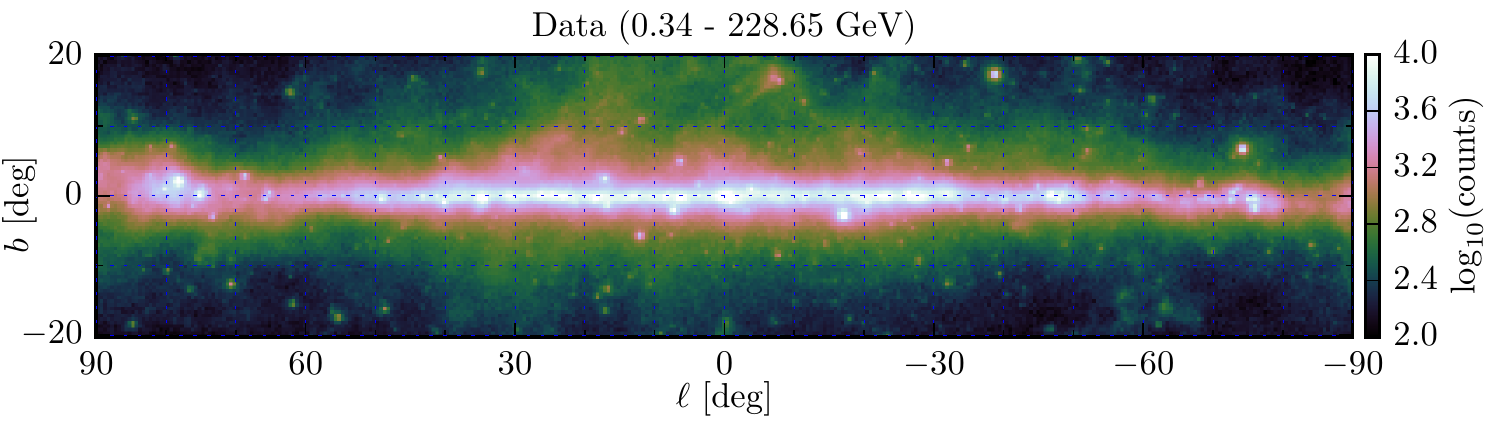}
    \caption{Counts map of data used in the analysis.}
    \label{fig:data}
\end{figure}

We use $7.6$ years of \Fermi-LAT Pass 8 data from 2008 August 4 and 2016 March 3.  We apply standard data cuts to events recommended by the \Fermi Science Support Center (``\verb+(DATA_QUAL>0) && (LAT_CONFIG==1)+'' and zenith angles $>90^{\circ}$).  We select only ULTRACLEAN events (evclass=512) and use both FRONT and BACK converted events (evtype=3).  We use the \Fermi Science Tools\footnote{http://fermi.gsfc.nasa.gov/ssc/data/analysis/} (v10r0p5) to build count and exposure maps binned into 25 log-spaced energy slices in the energy range $0.34-228.65$~GeV. Our RoI spans $|\ell| \leq 90^\circ$ and $|b|\leq 20.25^\circ$. The data are binned in Cartesian coordinates with a pixel size of $0.5^\circ$.  In figure~\ref{fig:data}, we show the data used in this analysis.  Note that the adopted pixelization ensures that the bright Galactic disk falls onto the central row of pixels.  Since we use a relatively small number of pixels in the analysis (the number of spatial nuisance parameters would otherwise explode in our current treatment), using a pixelization scheme that takes into account the characteristic emission gradient towards the disk is particularly important.

\subsection{Foreground modeling}

We model the foreground components following standard methods and codes.  In particular, we rely on the prescriptions of the GALPROP~\cite{Moskalenko:2001ya} and DRAGON~\cite{Evoli:2008dv} codes, and refer to the \Fermi-LAT Collaboration works, \fex~\cite{Ackermann:2012pya}.  Details of the modeling are explained below.

The gamma-ray emission from interactions of cosmic rays (CR) with the interstellar medium (ISM) and low-energy ambient photons is the so-called Galactic diffuse emission (GDE).  At GeV energies, the GDE originates mainly from three processes: (1) the decay of neutral pions produced by collisions of CR protons with ISM; (2) the bremsstrahlung radiation of electrons in the electric field of ISM charged particles; and (3) the up-scattering to GeV energies of the low-energy ambient photons of the interstellar radiation field (ISRF) due to Inverse Compton scattering (ICS) off CR electrons.

The GDE depends on the production of CR is the Galaxy, their distribution and spectra, as well as on the propagation mechanisms in the ISM and the interactions with the environment.  In the present work, we model the GDE in a relatively minimalistic way, and explore what ingredients we need to obtain a reasonably good fit to the data (after nuisance parameters are taken in account).

\paragraph{Hadronic $\pi^0$ emission.} The spatial distribution of the hadronic emission from $\pi^0$ decay traces closely the target material distribution, i.e.~the distribution of hydrogen in the Galaxy.  The hydrogen in the Galaxy is in the form of atomic (HI), molecular (H$_2$), and ionized (HII) gas, traced by different observables, namely the 21 cm line, the 2.6 mm CO emission line, and pulsar dispersion measurements respectively~\cite{Ferriere:2001rg}.  The most abundant component is represented by the molecular and atomic gas (almost 70\% by mass), while ionized H is subdominant in mass (only few percent on average).

The total H column density (in cm$^{-2}$) can be approximated by the sum of atomic (HI) and molecular hydrogen (H$_2$):
\begin{equation}
  N_{\rm H} \simeq N_{\rm HI} + 2 N_{\rm H_2} = N_{\rm HI} + 2 X_{\rm CO} W_{\rm CO} \, ,
\end{equation}
where the molecular H$_2$ column density is assumed to be proportional to the integrated
line intensity of CO, $W_{\rm CO}$ (in K km/s ), through the $X_{\rm CO}$ coefficient (in cm$^{-2}$/(K km/s)).
The $X_{\rm CO}$ is typically considered to be constant thorough the Galaxy. However, there are no \textit{a priori} reasons for it to be such and, indeed,
gamma-ray analyses have shown that, most likely, it is spatial dependent~\cite{Strong:2004td,2016PASJ...68....5N,Yang:2016jda,Acero:2016qlg}.

In the present work, we build the $\pi^0$ template as the sum of the gas column densities for atomic and molecular hydrogen, available within the GALPROP public release.\footnote{\url{http://galprop.stanford.edu}}  The HI column density and $W_{\rm CO}$ integrated line intensity maps are based on the corresponding 2D analytical models developed in ref.~\cite{Moskalenko:2001ya}, renormalised to the Leiden-Argentine-Bonn 21 cm line survey~\cite{Kalberla:2005ts} and to the 2.6 mm CO line survey~\cite{Dame:2000sp}, respectively.  For the HI, the spin temperature used to correct for opacity of the 21 cm line is $T_S = 125$ K.  No dust-reddening correction is applied to the column density.  Furthermore, \emph{we do not apply a dark gas correction}, since we are instead interested in seeing how dark gas affects the spatial modulation parameters associated with hadronic emission~\cite{1982A&A...105..159S,1982A&A...115..404S,2005Sci...307.1292G}.  The GALPROP gas maps are binned in Galacto-centric radial annuli, using the distance information extracted from the gas velocity and rotation curve (for details about the methodology see appendix B of ref.~\cite{Ackermann:2012pya}).
In practice, we obtain the gas template by adding the hydrogen column densities as: $N_{\rm HI} + 2 N_{\rm H_2}$, assuming a constant $X_{\rm CO} = 1.9 \times 10^{20}$ cm$^{-2}$/(K km/s). We build three gas templates using the following radial binning: the first gas ring corresponds to the emission within 0--3.5 kpc from the GC, the second ring to 3.5--6.5 kpc, and the third one to 6.5--19 kpc.  The advantage of using radial rings is that we can combine them arbitrarily to get maps for different spherical shells about the GC and study how this affects our conclusions.

\paragraph{Leptonic ICS emission.} To model the ICS emission we use the publicly available DRAGON code~\cite{Evoli:2008dv} and GammaSky~\cite{Evoli:2012ha,2013JCAP...03..036D}, a custom numerical package interfaced to DRAGON to compute the gamma-ray emission from the CR distribution output of DRAGON.  We use the ``Ferri\`{e}re'' model for the source distribution from the primary components~\cite{Ferriere:2001rg}, and propagation parameters corresponding approximately to the ``KRA4'' model in ref.~\cite{2013JCAP...03..036D}.\footnote{ We use the run\_2D\_KRA.xml model file available at~\url{https://github.com/cosmicrays/DRAGON/examples}.} The ISRF model adopted by DRAGON/GammaSky is documented in~\cite{Porter:2005qx} and available for download from the GALPROP public website (v54).

\paragraph{Leptonic bremsstrahlung emission.} Starting from the same DRAGON model file and using the GALPROP gas maps as target for interactions, we model the bremsstrahlung emission of electrons and positrons off interstellar ions and nuclei.  The gamma-ray emission from bremsstrahlung is typically suppressed with respect to that from $\pi^0$ and ICS. However, it might play a significant role in the very inner region of the Galaxy; see for example ref.~\cite{2013ApJ...762...33Y}. We tested the effect of including this additional template in our baseline model, without improvement of the fit that would be relevant for the current work (for our purposes, it is enough that bremsstrahlung can be absorbed by the $\pi_0$ templates). Therefore, we do not include the bremsstrahlung template in our baseline model.

\medskip

\begin{figure}[t]
    \centering
    \includegraphics[width=0.49\linewidth]{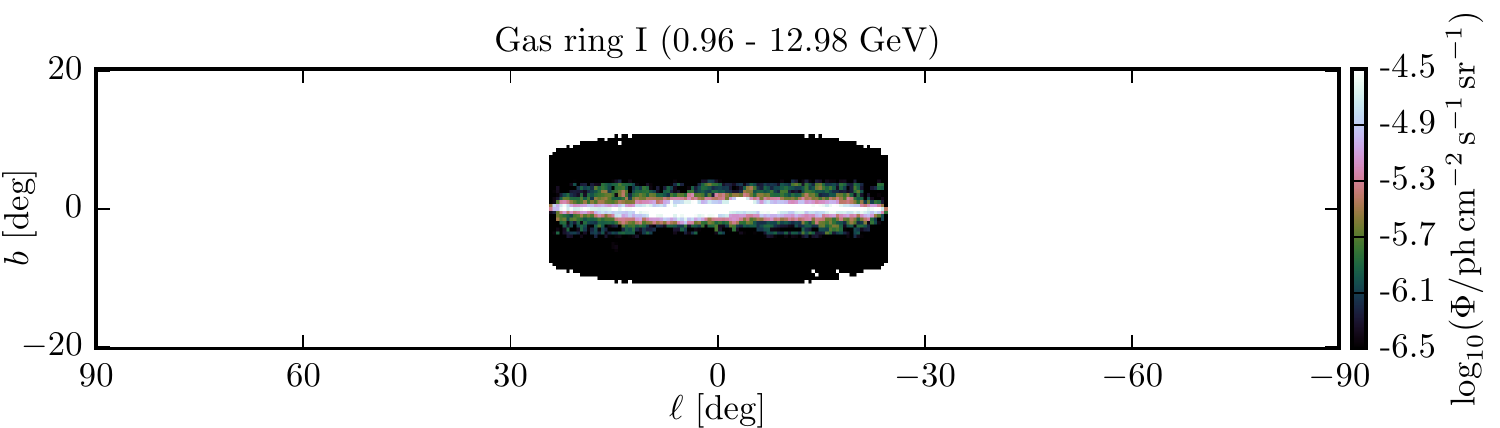}
    \includegraphics[width=0.49\linewidth]{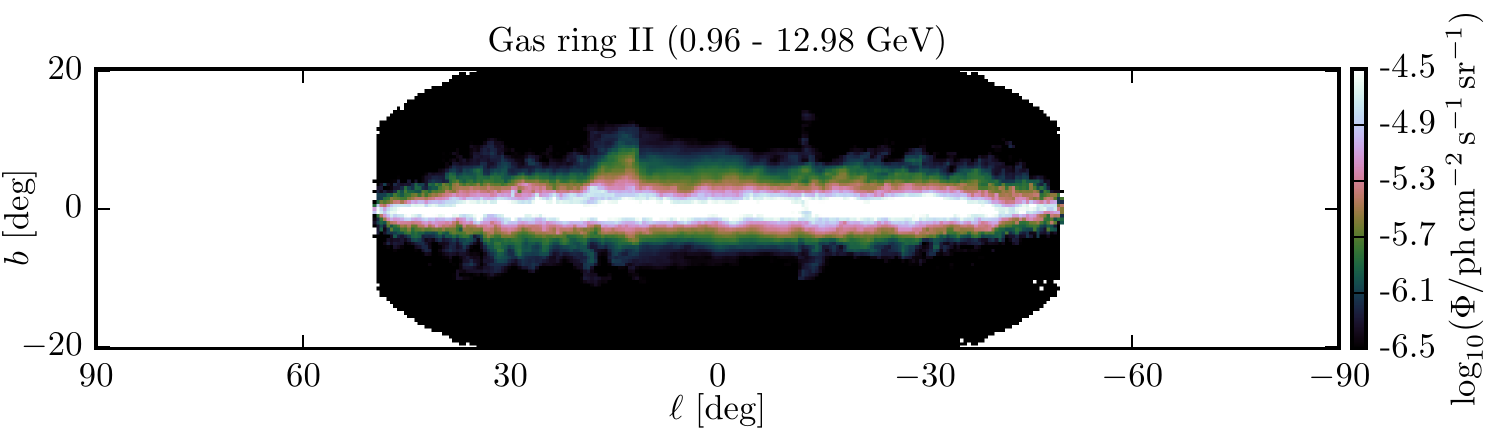}
    \includegraphics[width=0.49\linewidth]{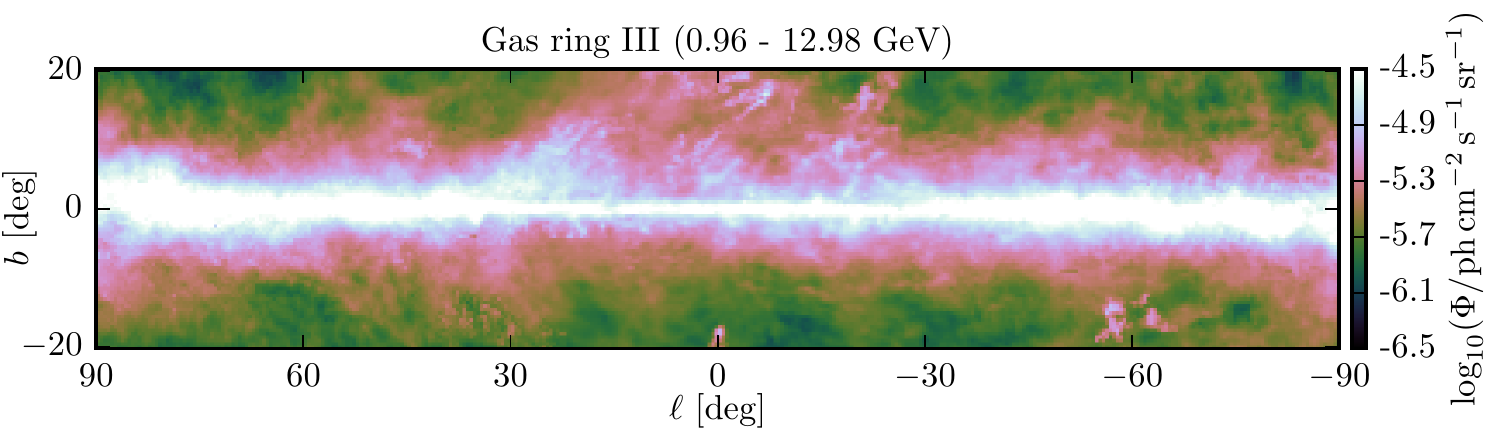}
    \includegraphics[width=0.49\linewidth]{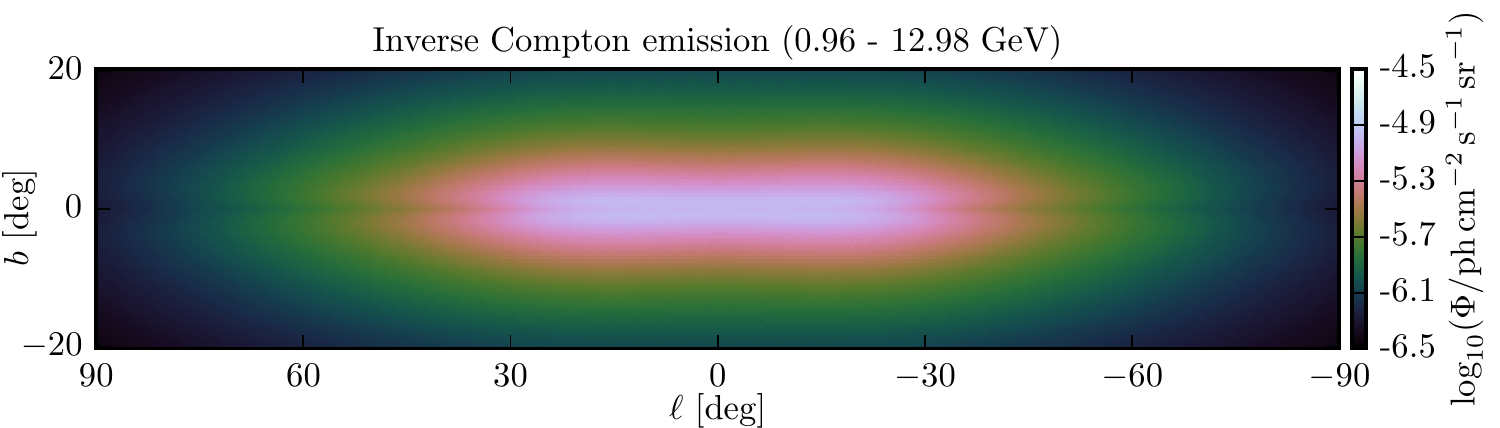}
    \caption{Initial model templates for fits of \run3--\run5. For \run1 and \run2, only one gas emission template is used, which is equal to the sum of the three gas templates shown here. Gas ring I corresponds to the emission from the innermost shell, 0--3.5 kpc from the Galactic center. Gas ring II correponds to emission from 3.5--6.5 kpc, and gas ring III to emission from 6.5--19 kpc.}
    \label{fig:templates}
\end{figure}

The spatial templates used as starting points in the fits are shown in figure~\ref{fig:templates}. The energy range shown is 0.96 -- 12.98 GeV, chosen to highlight both the $\pi^0$ bump as well as a significant fraction of the Galactic center excess emission (discussed below).

\medskip

Although DRAGON also produces the energy spectra of the various emission components, we do not use them in the current work.  The energy spectra depend critically on the energy spectra of hadronic and leptonic CRs, and hence on propagation and injection.  Instead, we use the $\pi^0$ and ICS component spectra from~\cite{Ackermann:2012pya} as references.  These spectra are obtained from the gamma-ray analysis of CR propagation models that are also in agreement with local CR measurements, and hence represent for us a good starting point. However, CR spectra in the Galaxy might be in general quite different from what we measure locally.  For this reason, we use the reference spectra only as starting points in the fit, and allow considerable freedom such that they can adapt to the gamma-ray data if needed.

\section{Towards a simple model for the Galactic disk emission}
\label{sec:tomin}

In this section, we derive a simple reference model for Galactic diffuse emission in our main RoI.  Since even this reference model will have $10^5$ parameters, the meaning of the term `simple' is not automatically evident.  What we actually aim to find is a minimal set of physical components, with distinct spatial or spectral characteristics (these contribute a very low number of parameters), augmented with realistically sized nuisance parameters (this number is necessarily very large).  The goal is to remove most of the residuals in our RoI in a physically motivated way.  Our reference model is not meant to be the final answer to modeling all of the gamma-ray emission in our RoI, but rather as a base for a number of follow-up studies.

The starting point for our reference model will be the numerical predictions for hadronic and leptonic emission in the Galaxy from DRAGON, based on locally measured CR fluxes and HI and CO gas maps, as discussed in section~\ref{sec:gde}.  We will study how much of the observed gamma-ray emission is already explained by these \textit{a priori} predictions, both directly and when augmented with additional nuisance parameters that account for expected spatial and spectral uncertainties in the predictions.  For a number of irreducible residuals, particularly in the bulge region, we will introduce additional templates.

\subsection{Conventional diffuse emission components with nuisance parameters}

\begin{table}[t]
    \centering
    \begin{tabular}{lcccccccc}
        \toprule
        Components && \run1 & \run2 & \run3 & \run4 & \run5 \\
                   && \multicolumn{5}{c}{Regularization hyper-parameters: \sconf{\lambda}{\lambda'}{\lambda''}{\eta}{\eta'} }\\
        \midrule
        IGRB &
             & \sconf\infty{16}\infty00
             & \sconf\infty{16}\infty00
             & \sconf\infty{16}\infty00
             & \sconf\infty{16}\infty00
             & \sconf\infty{16}\infty00
        \\[0.1cm]
        3FGL PSC &
                 & \sconf\cdot{25}0\cdot0
                 & \sconf\cdot{25}0\cdot0
                 & \sconf\cdot{25}0\cdot0
                 & \sconf\cdot{25}0\cdot0
                 & \sconf\cdot{25}0\cdot0
        \\[0.1cm]
        Gas (0--19 kpc) &
                        & \sconf\infty{16}000
                        & \sconf{10}{16}0{25}0
                        & ---
                        & ---
                        & ---
        \\[0.1cm]
        Gas ring I (0--3.5 kpc) &
                        & ---
                        & ---
                        & \sconf{10}{16}0{25}0
                        & \sconf{10}{16}0{25}0
                        & \sconf{10}{16}0{25}0
        \\[0.1cm]
        Gas ring II (3.5--6.5 kpc) &
                        & ---
                        & ---
                        & \sconf{10}{16}0{25}0
                        & \sconf{10}{16}0{25}0
                        & \sconf{10}{16}0{25}0
        \\[0.1cm]
        Gas ring III (6.5--19 kpc) &
                        & ---
                        & ---
                        & \sconf{4}{16}0{25}0
                        & \sconf{4}{16}0{25}0
                        & \sconf{4}{16}0{25}0
        \\[0.1cm]
        Extended sources &
                         & ---
                         & ---
                         & ---
                         & \sconf01\infty40
                         & \sconf01\infty40
        \\[0.1cm]
        Inverse Compton &
                        & \sconf\infty{16}000
                        & \sconf1{16}0{100}0
                        & \sconf1{16}0{100}0
                        & \sconf1{16}0{100}0
                        & \sconf1{16}0{100}0
        \\[0.1cm]
        \Fermi bubbles &
                       & ---
                       & ---
                       & ---
                       & \sconf0{400}\infty40
                       & \sconf0{400}\infty40
        \\[0.1cm]
        511 keV template &
                         & ---
                         & ---
                         & ---
                         & ---
                         & \sconf{25}0\infty00
        \\[0.1cm]
        \midrule
        Naive model parameters, $N_\text{param}$ && 20253 & 78573 & 97838 & 104596 & 107639 \\
        Naive DOF  && 708747 & 650427 & 631162 & 624404 & 621361 \\
        \midrule
        Eff.~model parameters, $N_\text{param}^\text{eff}$ && 1900 & 11800  & 10200 & 12700 & 12800 \\
        Eff.~data bins, $N_\text{data}^\text{eff}$ && 624700 & 622700 & 620200  & 618800 & 619000  \\
        Eff.~DOF, $k$ && 622800 & 610900 & 610000 & 606100 & 606200 \\
        \midrule
        $-2\ln\mathcal{L}_P$ && 1016041 & 637742  & 633334  & 627206 & 626998 \\
        $-2\ln\mathcal{L}_R$ && 14152 & 24652 & 23709 & 21640 & 20988  \\
        Model fidelity, $\mathcal{F}$  && 627 & 164 & 153 & 145 & 144 \\
        \bottomrule
    \end{tabular}
    \caption{Summary of components and associated regularization hyper-parameters used for the models \run1 to \run5 in this work.  For each component, we show in the compact matrix notation the adopted regularization hyper-parameters for the spatial ($\lambda$), spectral ($\lambda'$) and overall ($\lambda''$) modulation parameters, and for the spatial ($\eta$) and spectral ($\eta'$) smoothing parameters.  The definition of hyper-parameters is shown in Eq.~\eqref{eqn:lnLR}. For the spatial and spectral hyper-parameters, an `$\infty$' indicates that the corresponding modulation parameters are kept fixed to one, while for the smoothing parameters, a `0' indicates no smoothing on that component. 
      We also show the total number of fit parameters $N_\text{param}$, the naive number of DOF ($N_\text{ebin}\times N_\text{pix} - N_\text{par}$), the effective number of DOF as estimated from mock data along with the effective number of model parameters and data bins, the values of the Poisson and regularization parts of the likelihood function, and the model fidelity, $\mathcal{F}$ as an indicator for the goodness-of-fit (see section~\ref{sec:gof}). Effective fit parameters are rounded to the nearest hundred.
    }
    \label{tab:fits}
\end{table}

In table~\ref{tab:fits}, we list five increasingly complex models, their components and the associated regularization parameters (\run1 to \run5).  They provide a step-by-step illustration for how additional components and nuisance parameters account for residuals when modeling the gamma-ray emission observed by \Fermi-LAT in our RoI.  We will discuss each of the five models separately.

\begin{figure}[t]
  \centering
  \includegraphics[width=0.49\linewidth]{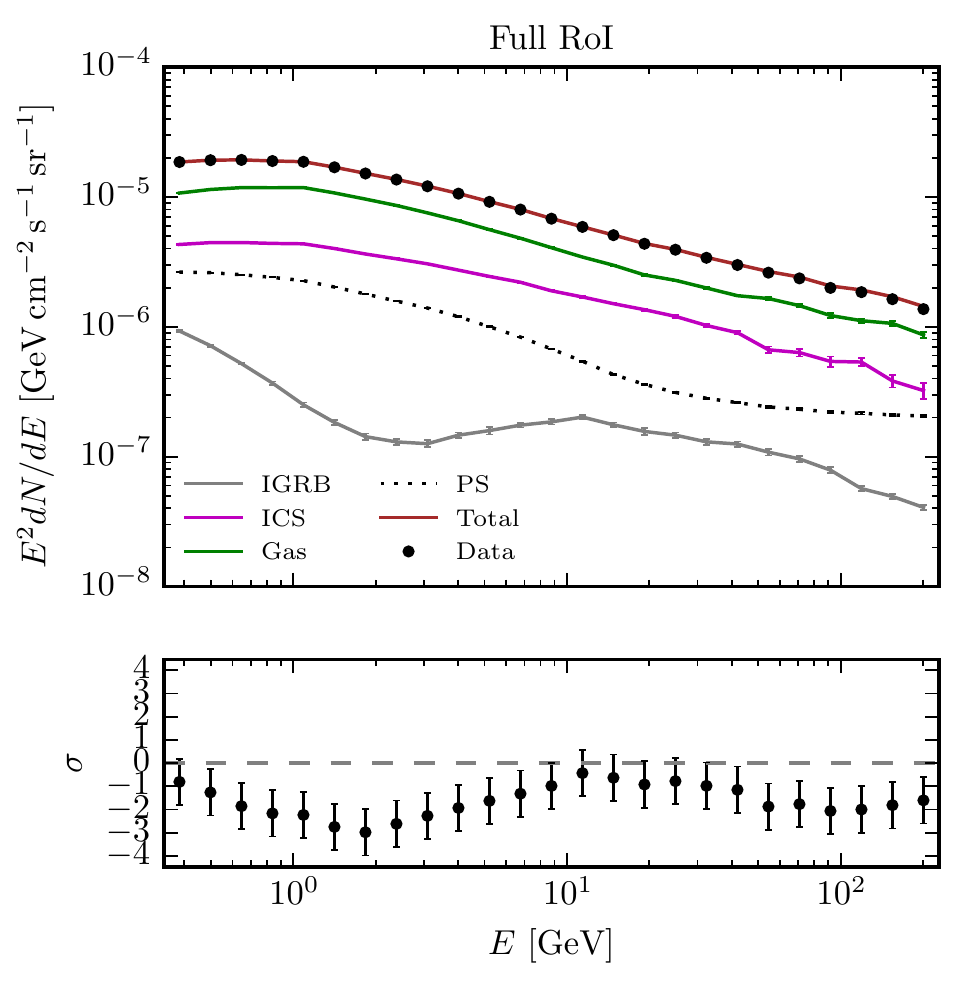}
  \includegraphics[width=0.49\linewidth]{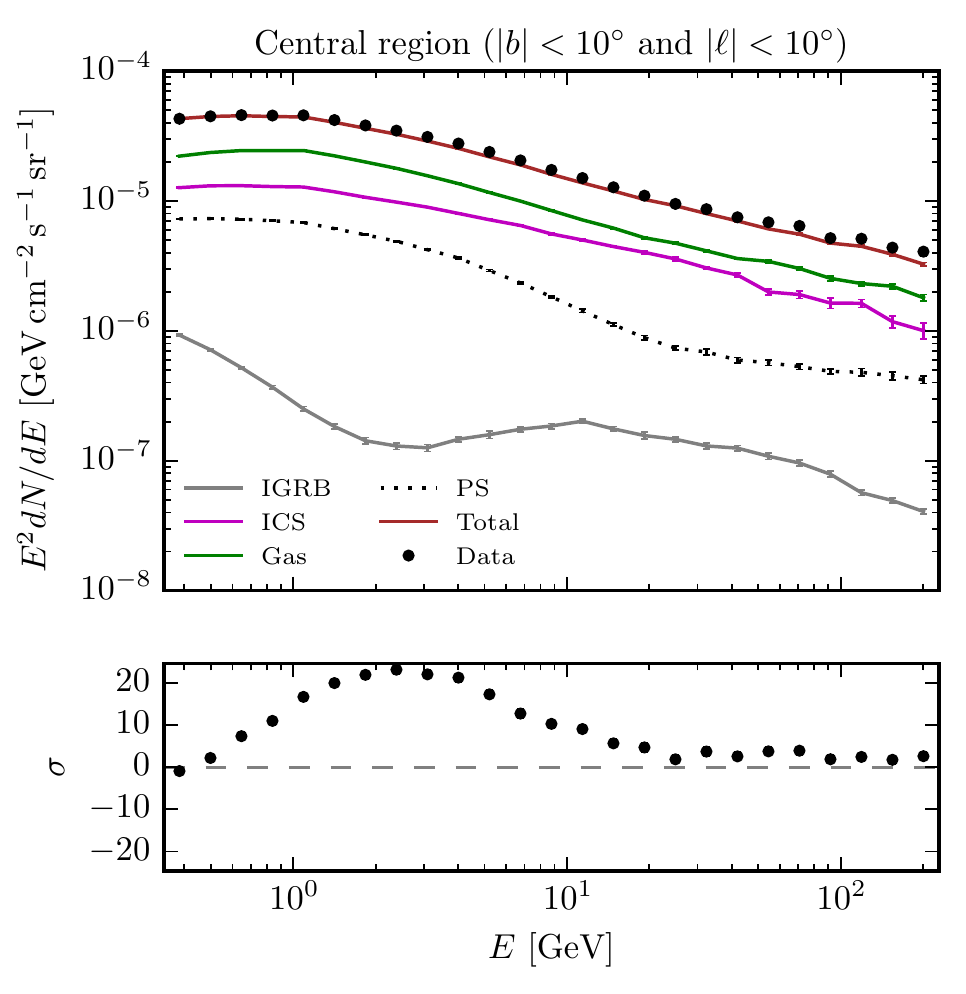}
  \caption{Baseline spectra in our initial \run1, in two different sky regions.  As show in table~\ref{tab:fits}, only the normalization of the various components is left free to vary. Residuals are in units of standard deviations, defined as $\sigma = (\text{data}-\text{model})/\sqrt{\text{model}}$.}
  \label{fig:run1_spectra}
\end{figure}

\medskip

\run1.  This is our initial model, mostly based on the \textit{a priori} information that enters the modeling with DRAGON.  It includes just one gas and one ICS component, along with a component for the isotropic gamma-ray background (IGRB) as measured in ref.~\cite{Ackermann:2014usa} and the 3FGL point sources~\cite{Acero:2015hja}.  In the case of the IGRB, we allow spectral uncertainties with a regularization of $\lambda = 16$ (this corresponds to $1/\sqrt{\lambda}=25\%$ uncertainty) that are somewhat smaller than the systematic errors quoted in ref.~\cite{Ackermann:2014usa}. This prevents large deviations from the measured spectrum. The spectra of the gas and ICS components are constrained to be close to the spectra corresponding to the locally measured CRs (again within $25\%$), and are taken from ref.~\cite{Ackermann:2012pya}. For all of the components of \run1, except the IGRB, we keep the overall normalization, $\nu^{(k)}$, free in the fit.  The same is the case for the point sources.  Furthermore, we use a weak regularization to keep the source spectrum close to its original spectrum in the 3FGL (to within $20\%$).  The total number of model parameters is 20253 (77 for the diffuse components, and the remaining 20176 for the 776 3FGL point sources).  We show the resulting spectra and the spectral residuals in figure~\ref{fig:run1_spectra}.  It can be clearly seen that very significant residuals remain in particular in the central region.  The overall spectrum integrated over the full RoI is better recovered, however only at the expense of a strongly distorted IGRB spectrum.

\begin{figure}[t]
    \centering
    \includegraphics[width=0.49\linewidth]{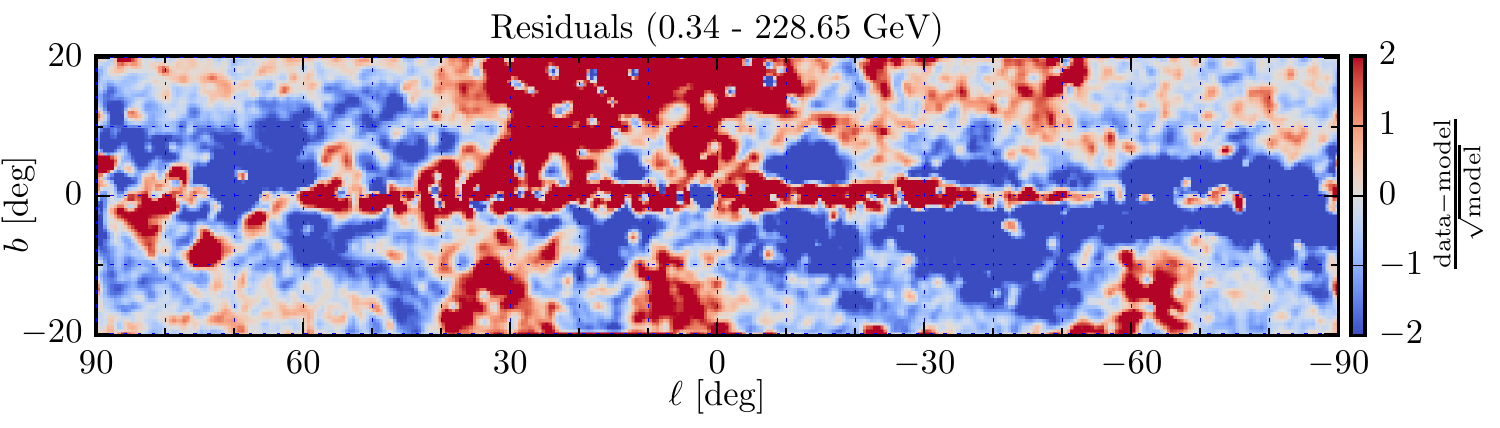}
    \includegraphics[width=0.49\linewidth]{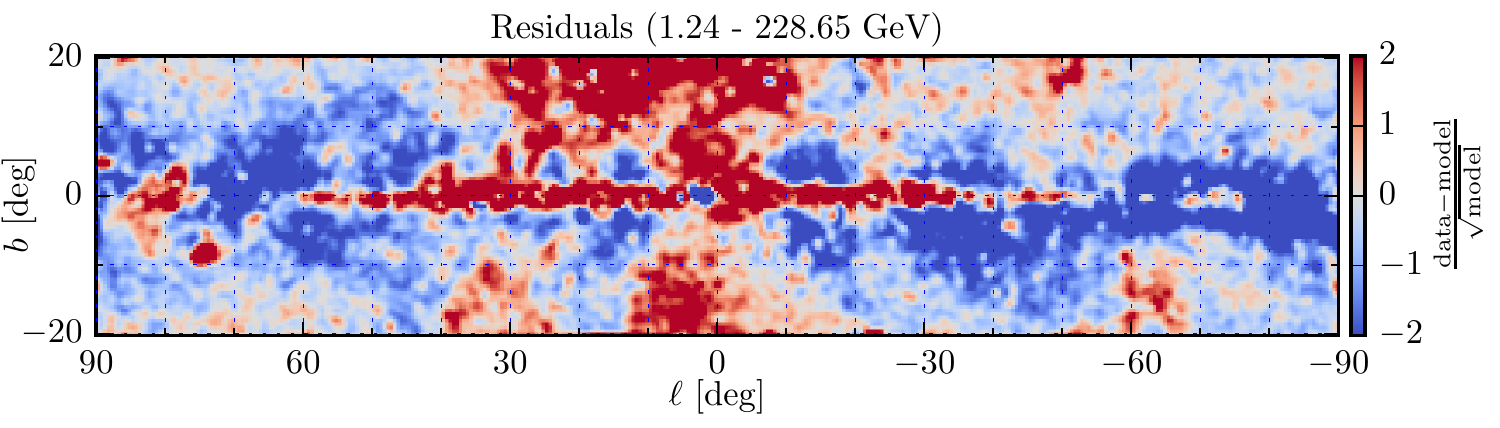}
    \includegraphics[width=0.49\linewidth]{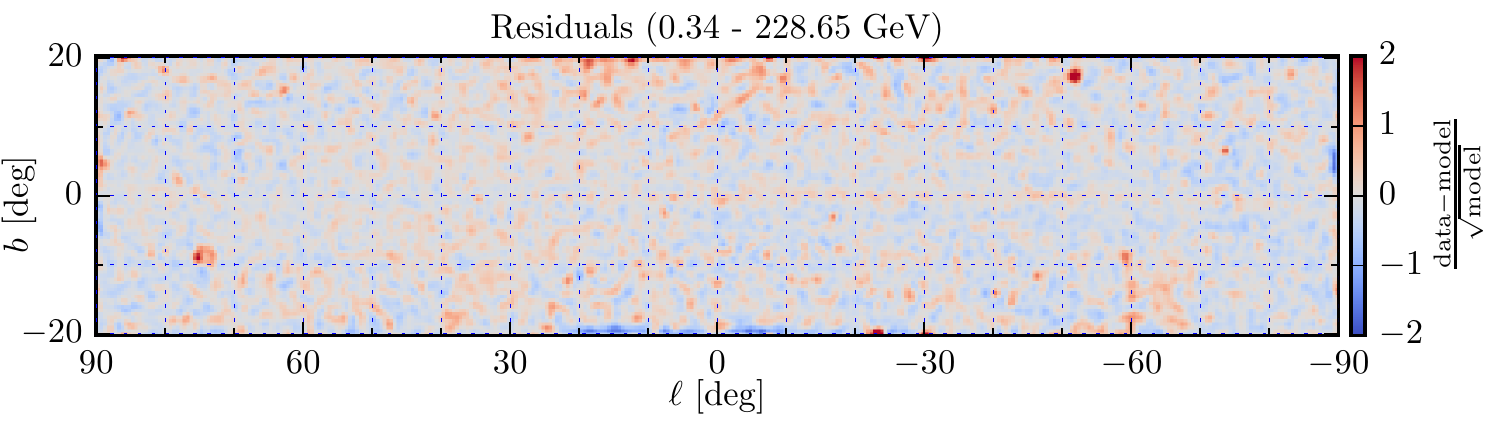}
    \includegraphics[width=0.49\linewidth]{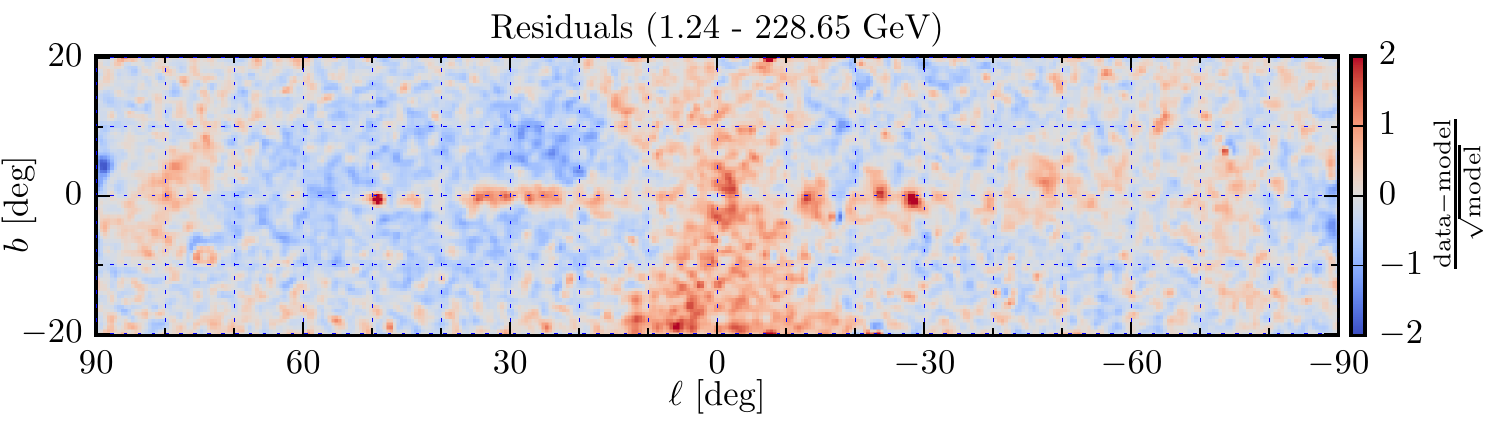}
    \includegraphics[width=0.49\linewidth]{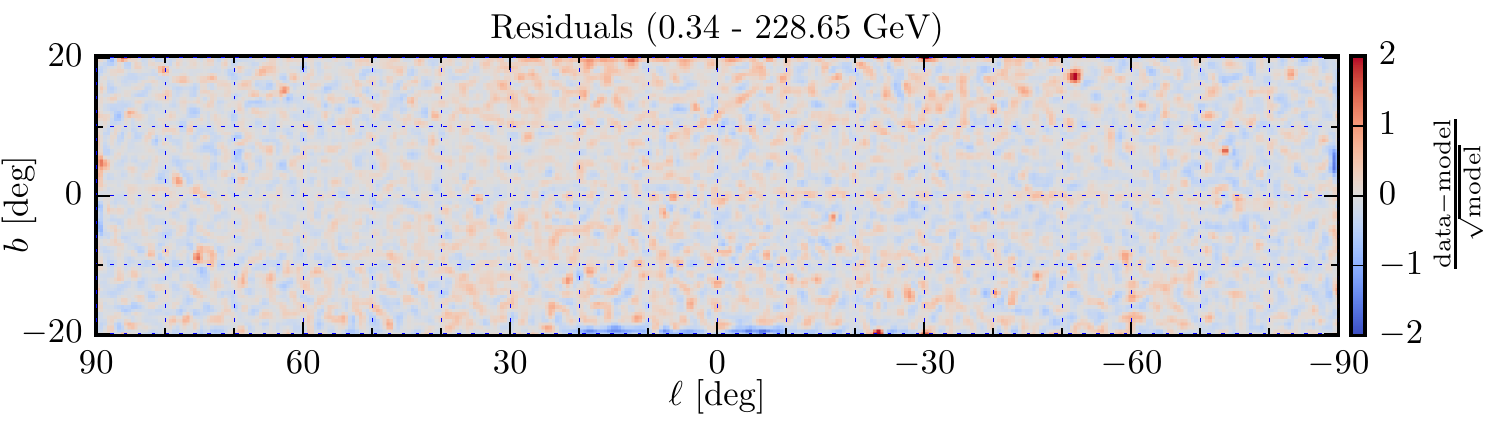}
    \includegraphics[width=0.49\linewidth]{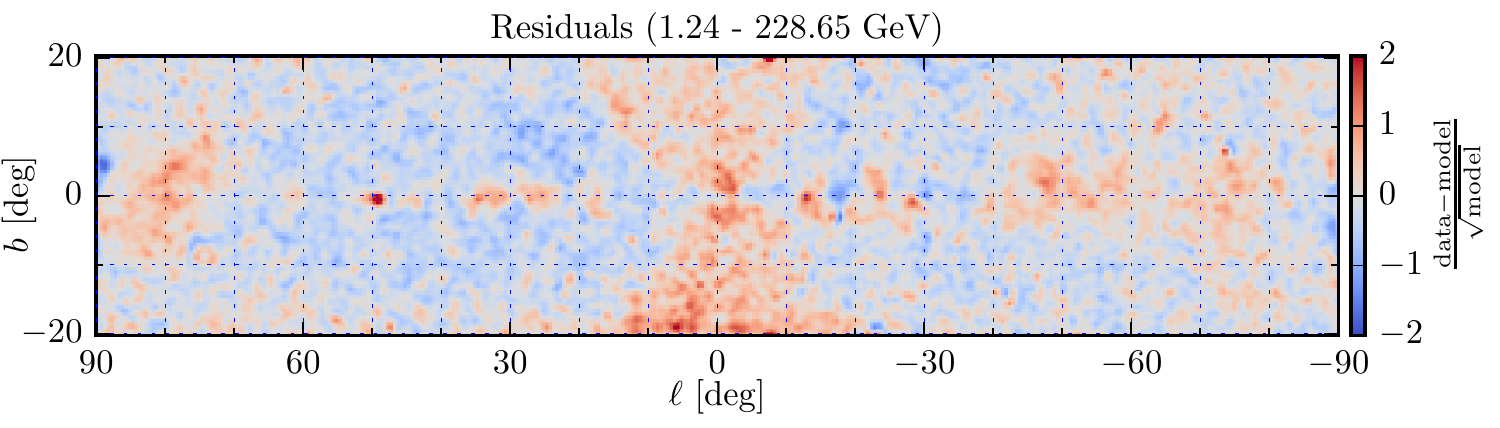}
    \includegraphics[width=0.49\linewidth]{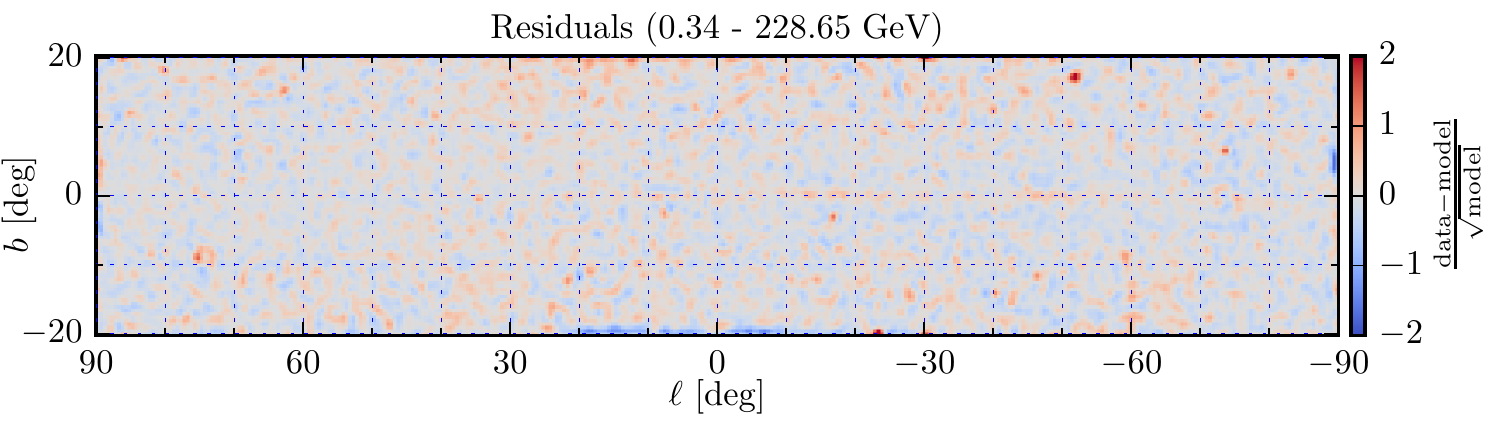}
    \includegraphics[width=0.49\linewidth]{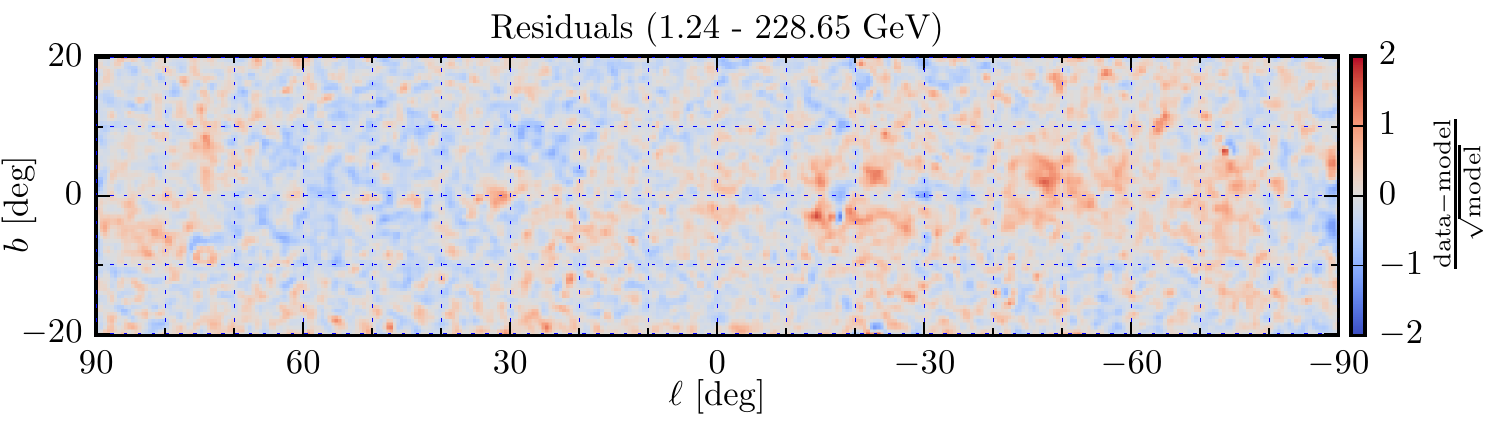}
    \includegraphics[width=0.49\linewidth]{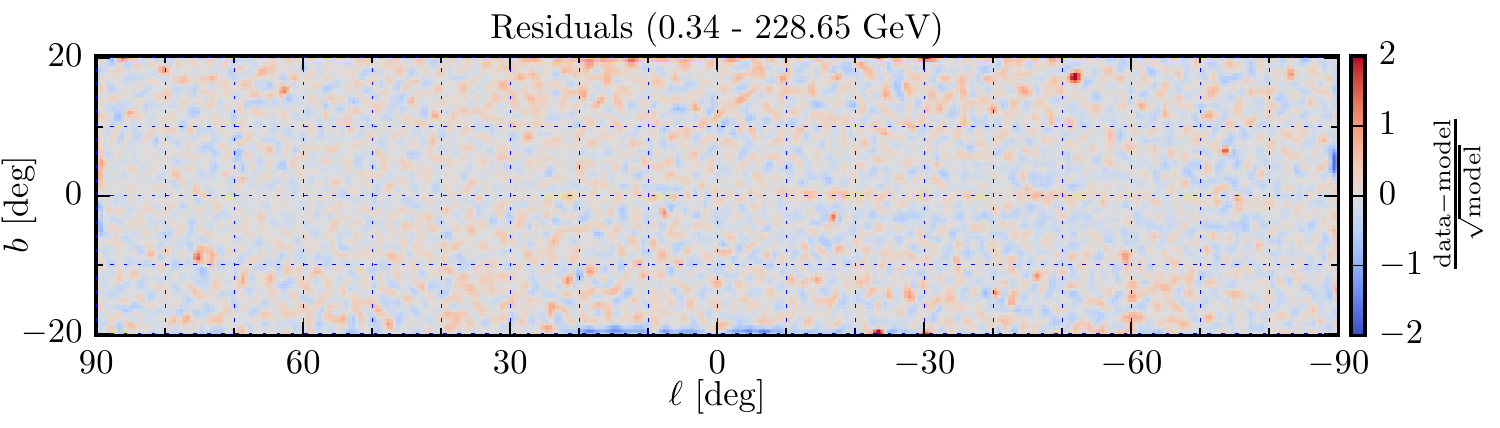}
    \includegraphics[width=0.49\linewidth]{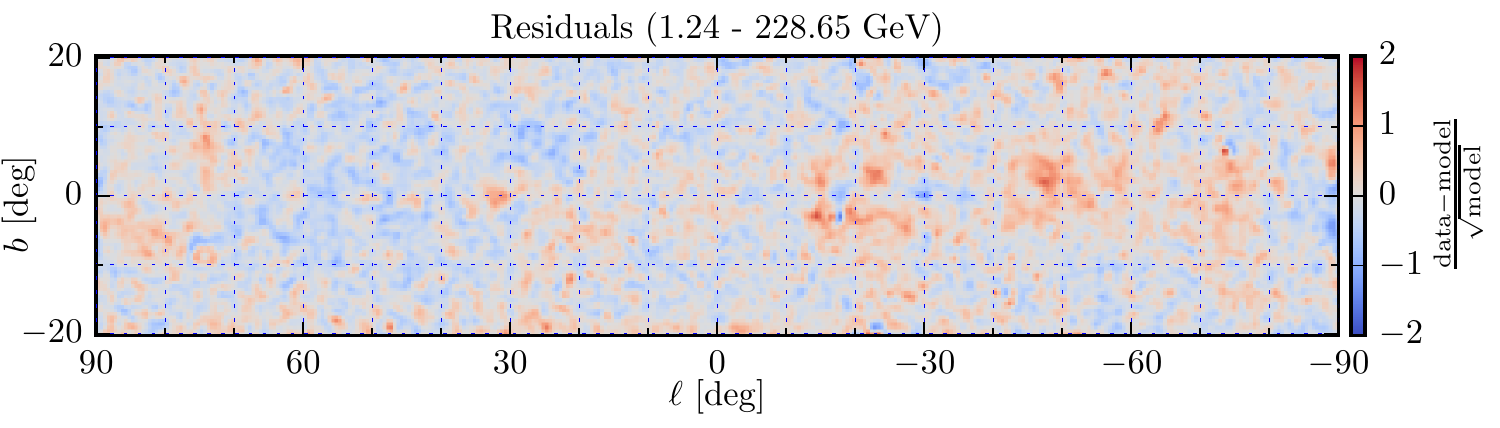}
    \caption{Significance of residuals for \run1 (top row) to \run5 (bottom row), for energies $>0.34$ GeV (left column) and $>1.24$ GeV (right column).}
    \label{fig:residuals_low}
\end{figure}

\begin{figure}[t]
    \centering
    \includegraphics[width=0.49\linewidth]{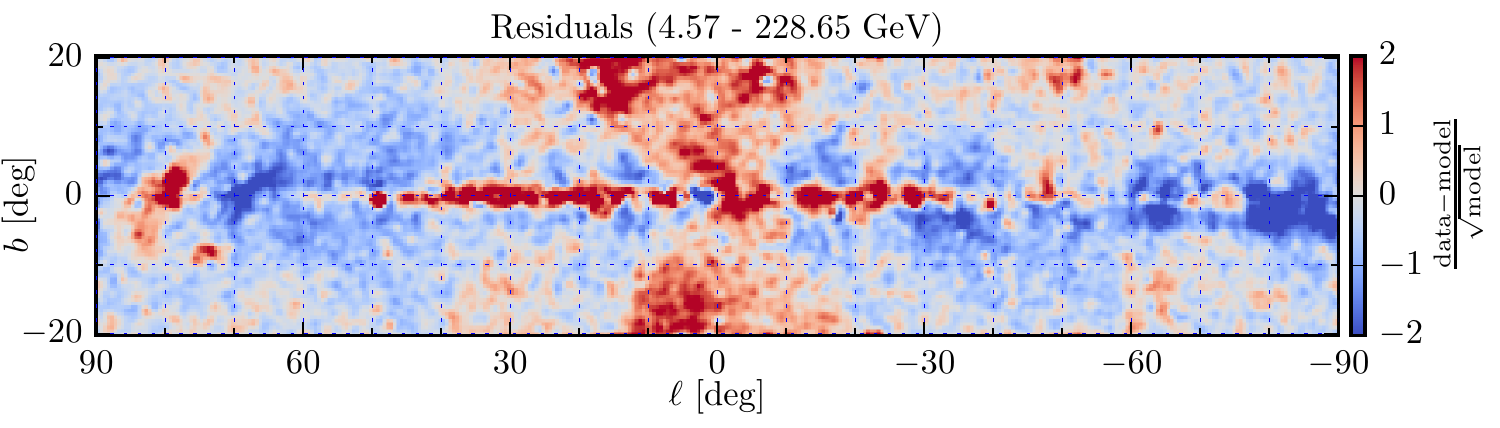}
    \includegraphics[width=0.49\linewidth]{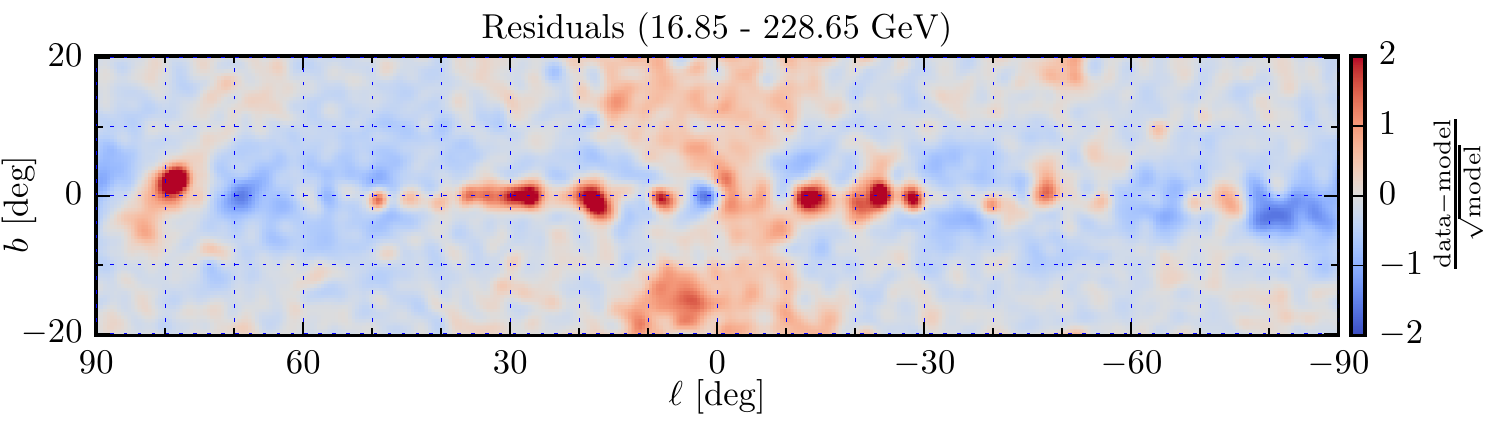}
    \includegraphics[width=0.49\linewidth]{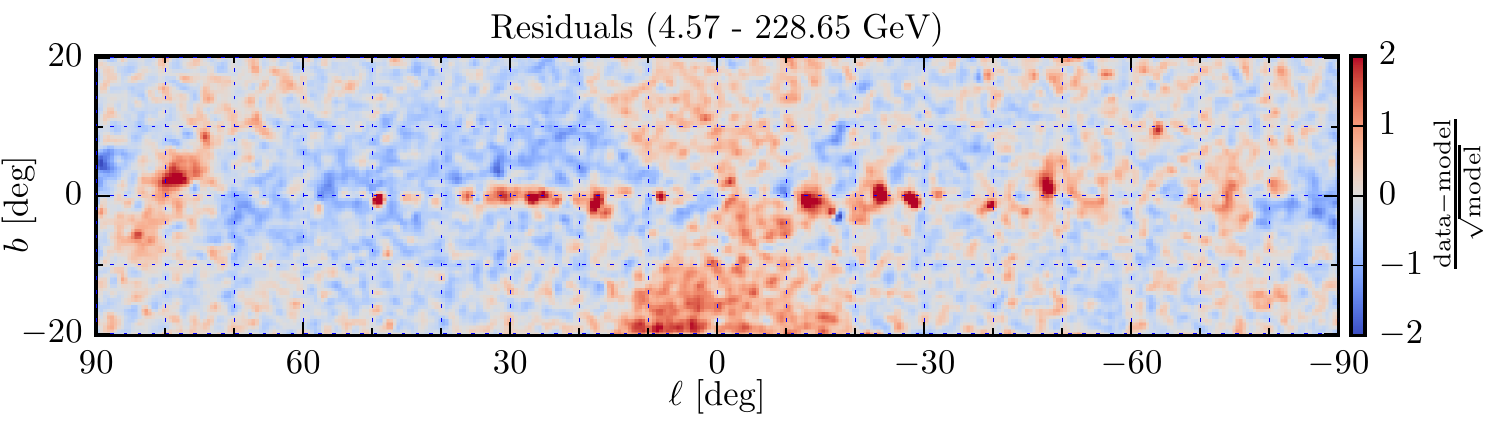}
    \includegraphics[width=0.49\linewidth]{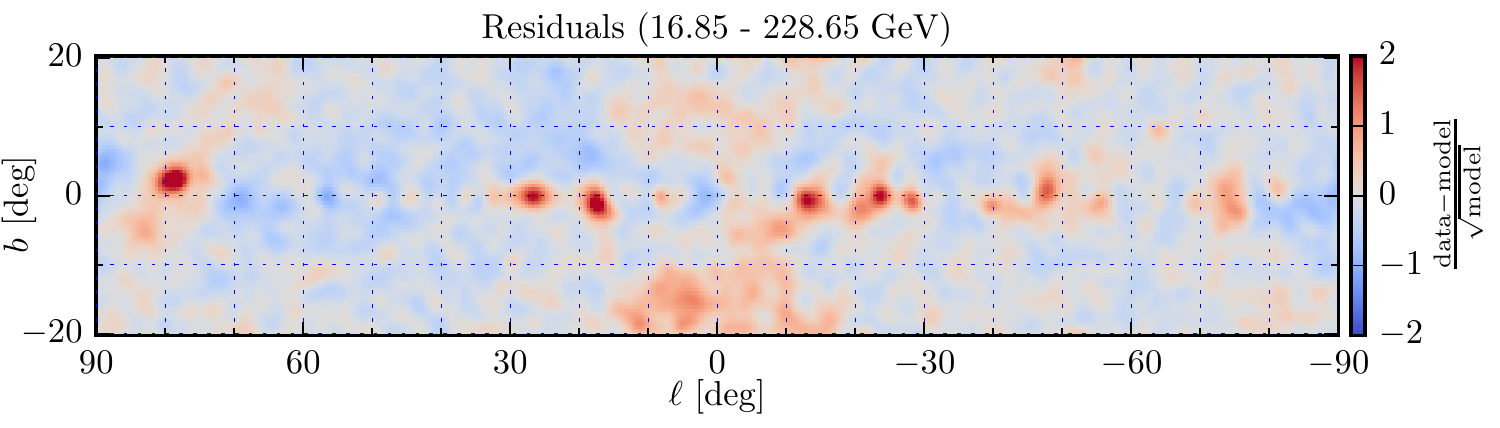}
    \includegraphics[width=0.49\linewidth]{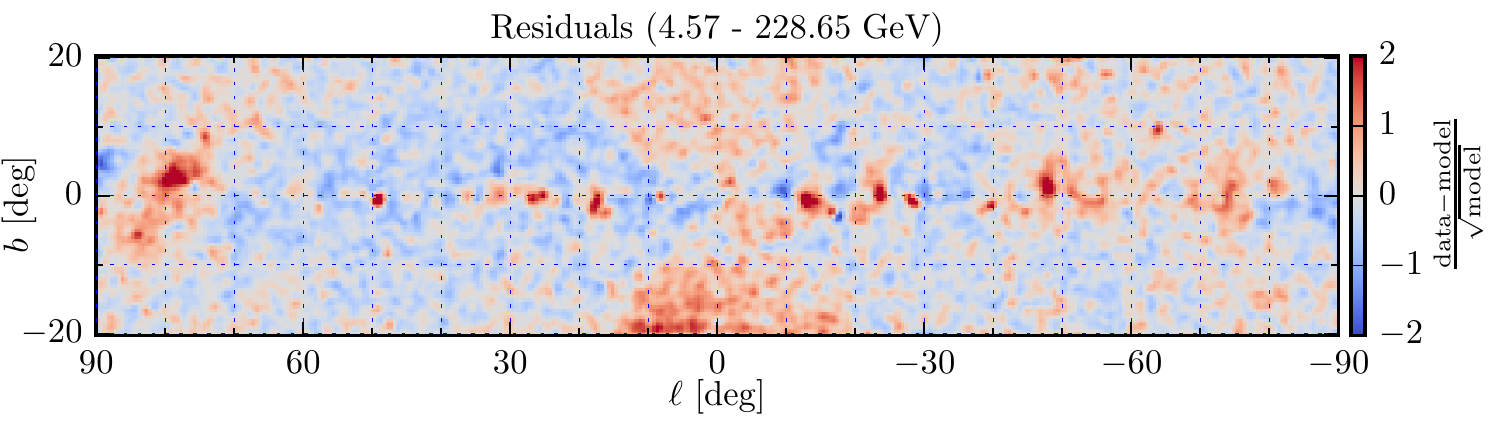}
    \includegraphics[width=0.49\linewidth]{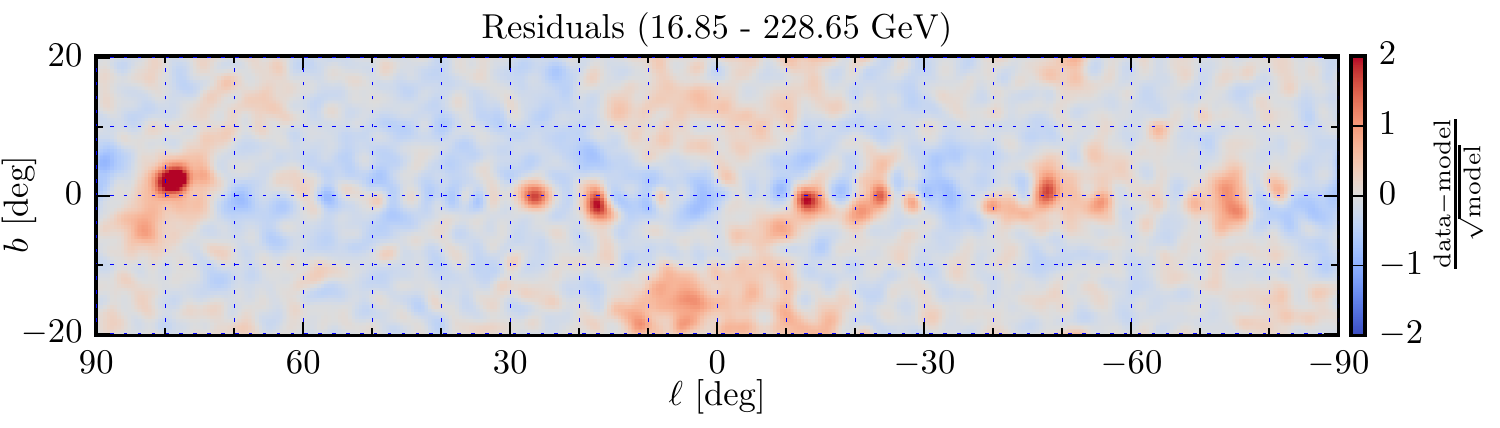}
    \includegraphics[width=0.49\linewidth]{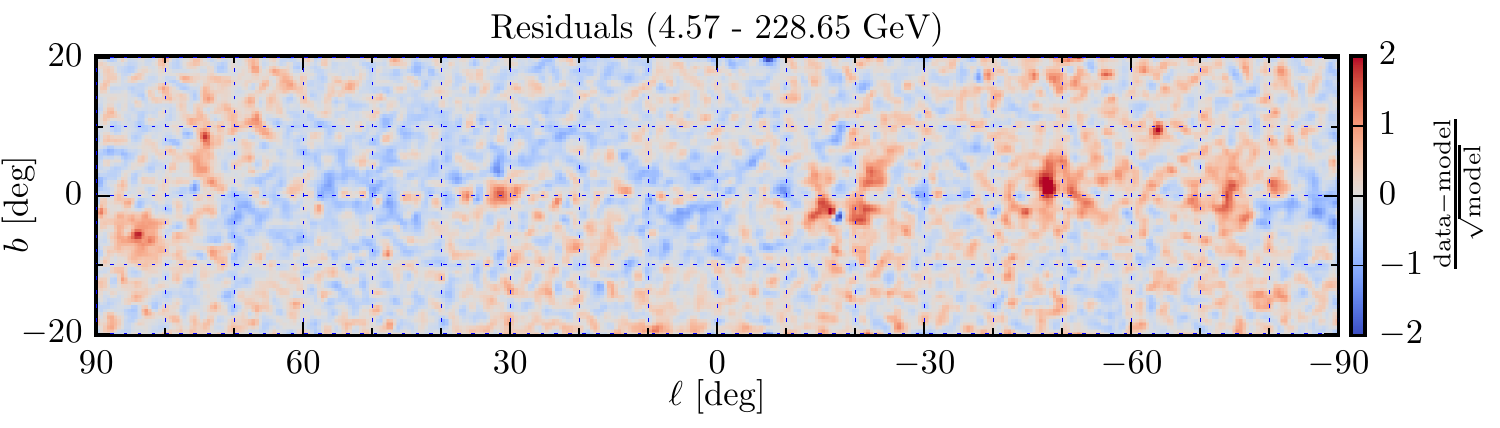}
    \includegraphics[width=0.49\linewidth]{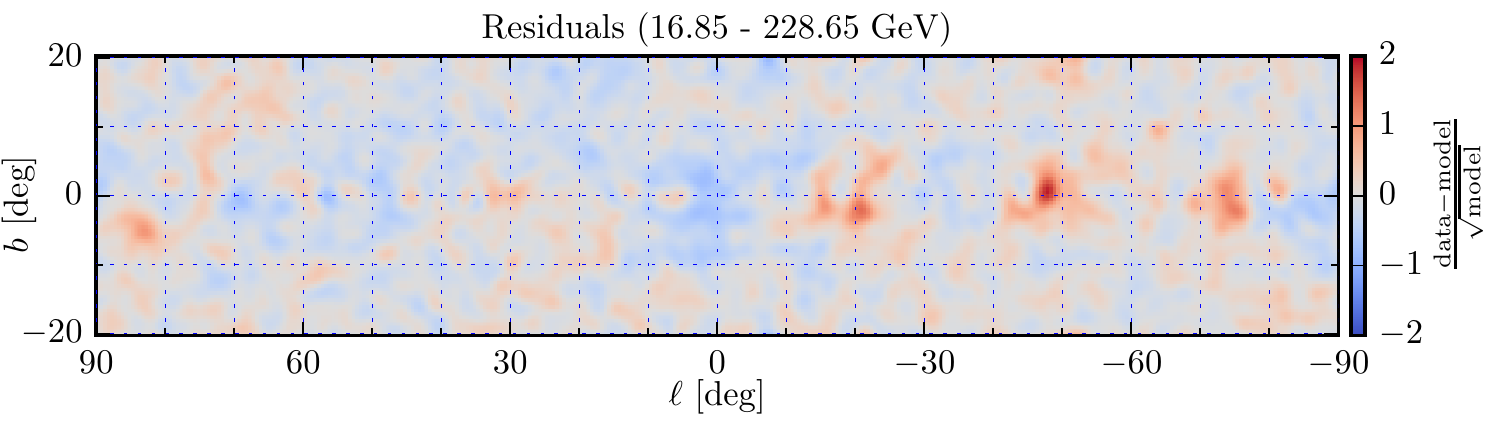}
    \includegraphics[width=0.49\linewidth]{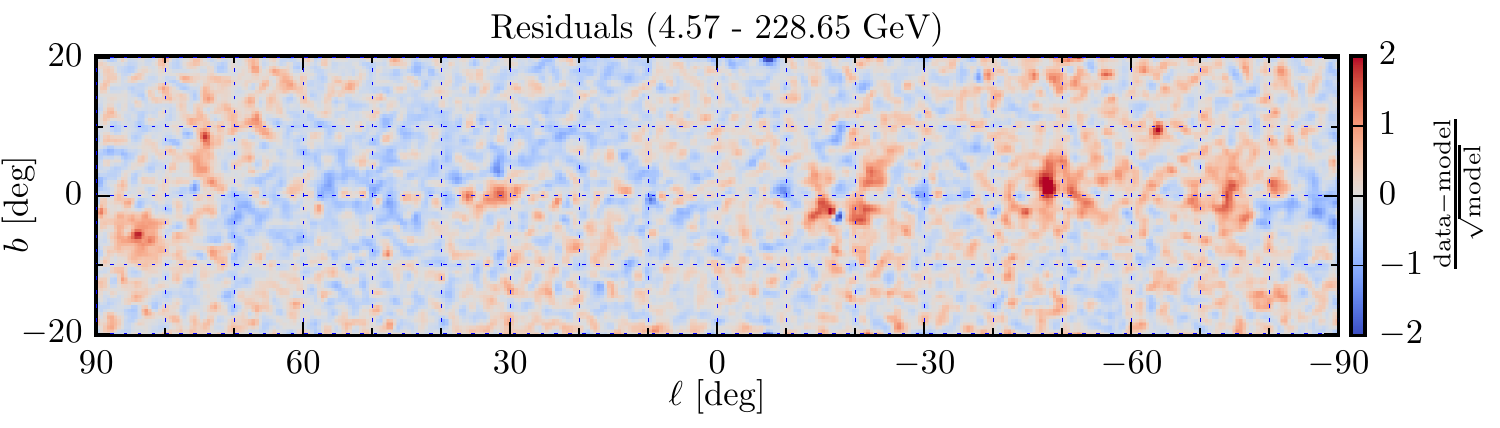}
    \includegraphics[width=0.49\linewidth]{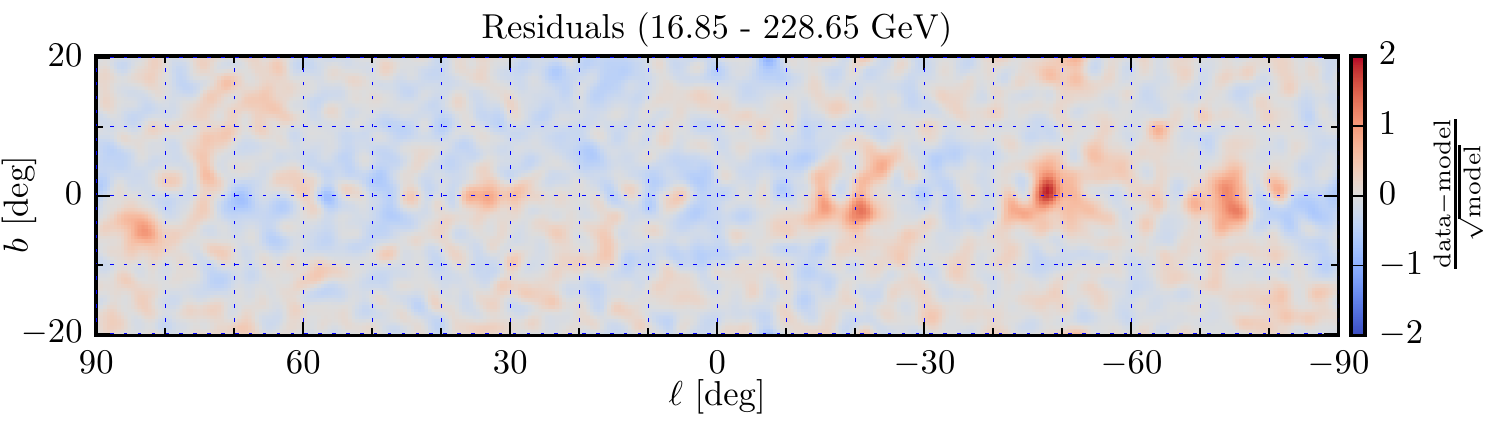}
    \caption{Significance of residuals for \run1 (top row) to \run5 (bottom row), for energies $>4.57$ GeV (left column) and $>16.85$ GeV (right column).}
    \label{fig:residuals_high}
\end{figure}

\begin{figure}[t]
  \centering
  \includegraphics[width=0.47\linewidth]{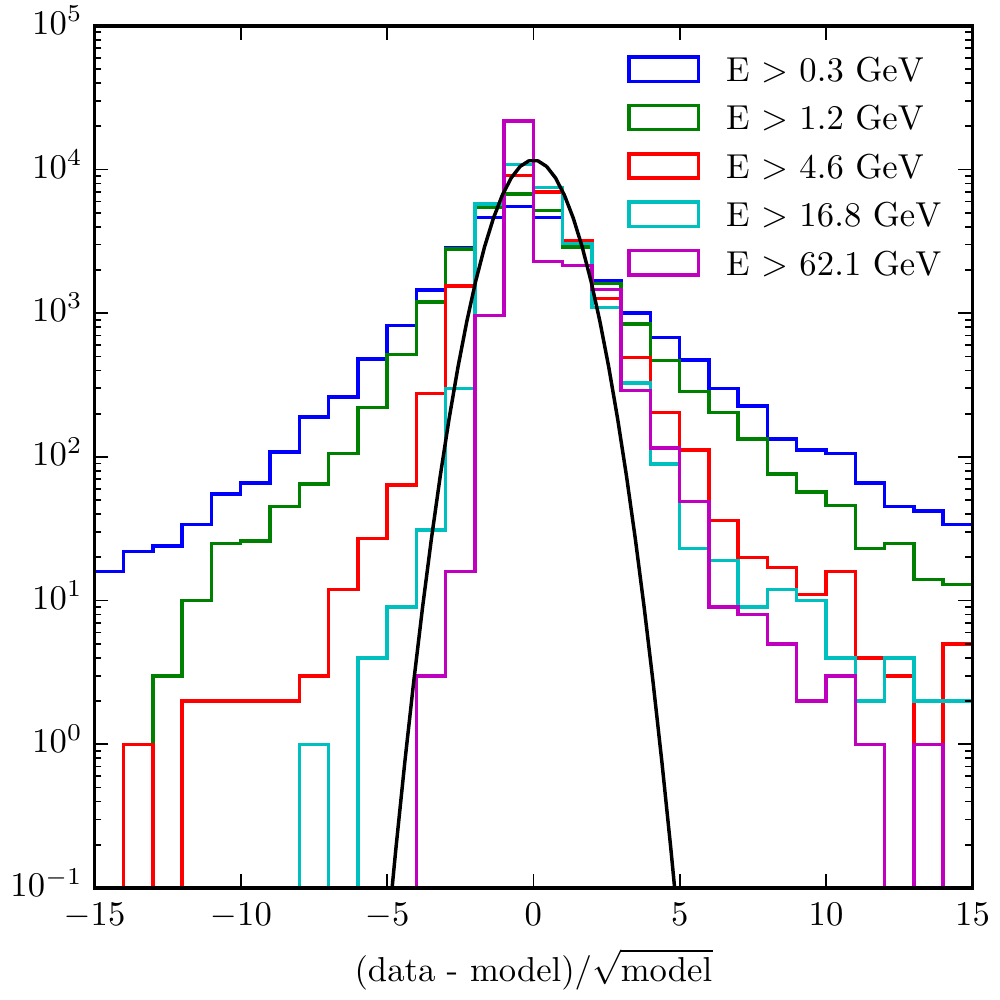}
  \hfill
  \includegraphics[width=0.47\linewidth]{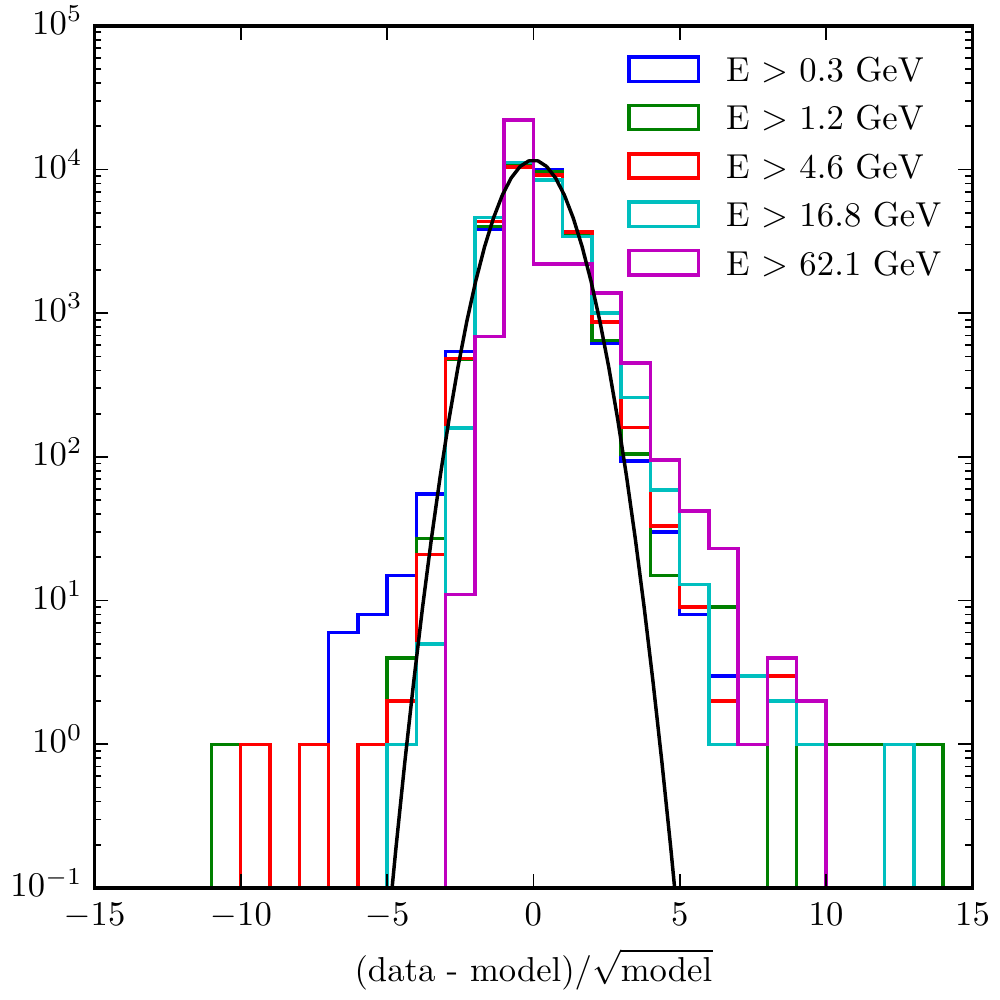}
  \caption{Histogram of residual significance for \run1 (left panel) and \run5 (right panel), for five different energy ranges. The black line shows the appropriately normalized PDF of a standard normal distribution for comparison.}
  \label{fig:residual_histogram}
\end{figure}

Furthermore, in figures~\ref{fig:residuals_low} and~\ref{fig:residuals_high}, we show in the top panels the significance of the residuals when subtracting the best-fit \run1 model from \Fermi-LAT data.  As expected, we find that this simple scenario reproduces observations, with fractional residuals around $\sim30\%$ in many regions of the sky. In particular, large and very significant residuals remain along the bright Galactic plane, and at higher energies in the \Fermi bubble regions and towards Cygnus X (at $\ell \sim 80^\circ$).  A histogram of the residuals is shown in figure~\ref{fig:residual_histogram}, and will be discussed further below in context of \run5.

\medskip

\run2.  As a next step, we introduce also spatial modulation for all components, except the IGRB, with values for the regularization hyper-parameters listed in table~\ref{tab:fits}. We introduce $33\%$ ($\lambda=10$) uncertainties for the morphology of the gas component, and $100\%$ ($\lambda=1$) uncertainties for the morphology of the ICS component.  We also now include smoothing regularization terms, with stronger smoothing for the ICS component than for the gas. The ICS component is smoothed with a hyper-parameter of $\eta=100$, which corresponds to roughly $10\%$ variations between neighboring pixels, while the gas component is smoothed with $\eta=25$ ($20\%$ variations).  The total number of parameters is now 78548. As in \run1, 20176 of these parameters correspond to point sources and their spectra. Most of the remaining parameters correspond to spatial modulation parameters for the ICS and gas components.

Looking at figures~\ref{fig:residuals_low} and~\ref{fig:residuals_high}, it is apparent that accounting for spatial and spectral nuisance parameters in the fit reduces the amount of residuals in the sky drastically.  Typical fractional residuals are now of the order $\lesssim 10\%$ at energies above 1.24 GeV.  Fractional residuals above the minimum energy of 0.34 GeV are practically zero (this can also be seen in figure~\ref{fig:residuals_low}), because there are enough degrees of freedom in the fit to remove the most significant residuals.  They happen to be at the lowest energies due to the larger number of low-energy photons.

One effect of the nuisance parameters is that \emph{reducible} residuals, in the sense that they can be absorbed by small changes of the local intensity of model components without significantly changing their spectra, are largely removed, whereas irreducible residuals remain.  Structures that become now evident are the bubble-shaped positive residuals that appear from the second energy band on, and, at high energies, numerous localized residuals along the Galactic plane.

\medskip

\run3.  This model is similar to the previous one, except that we split the gas component into three rings that cover the radial ranges 0--3.5 kpc, 3.5--6.5 kpc and 6.5--19.0 kpc.  Details about the construction of gas rings can be found in section~\ref{sec:gde}. The main effect is that residuals along the Galactic disk are further reduced, as can be seen in the third row of figures~\ref{fig:residuals_low} and~\ref{fig:residuals_high}.  However, localized residuals remain along the Galactic plane, both close to and at the Galactic center, as well as further away.  Most of these residuals grow stronger at higher energies, indicating rather hard fluxes.  Indeed, many of the residuals correspond to extended emission associated with the positions of various extended sources observed by \Fermi-LAT.  Below, we will be able to absorb most of these sources with extended source templates from the \Fermi 3FGL.

\begin{figure}[t]
    \centering
    \includegraphics[width=0.49\linewidth]{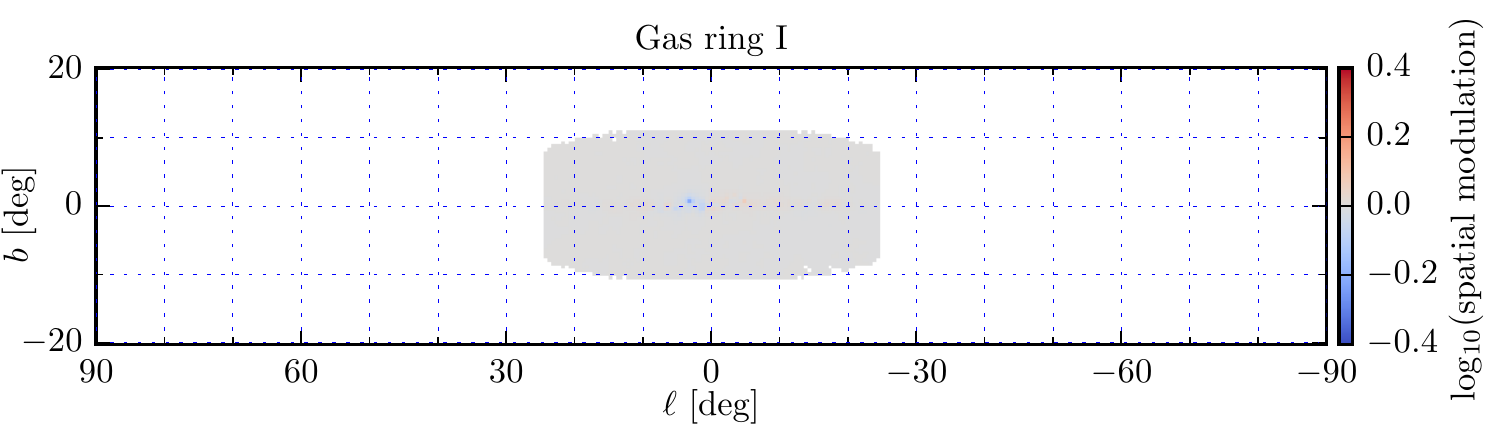}
    \includegraphics[width=0.49\linewidth]{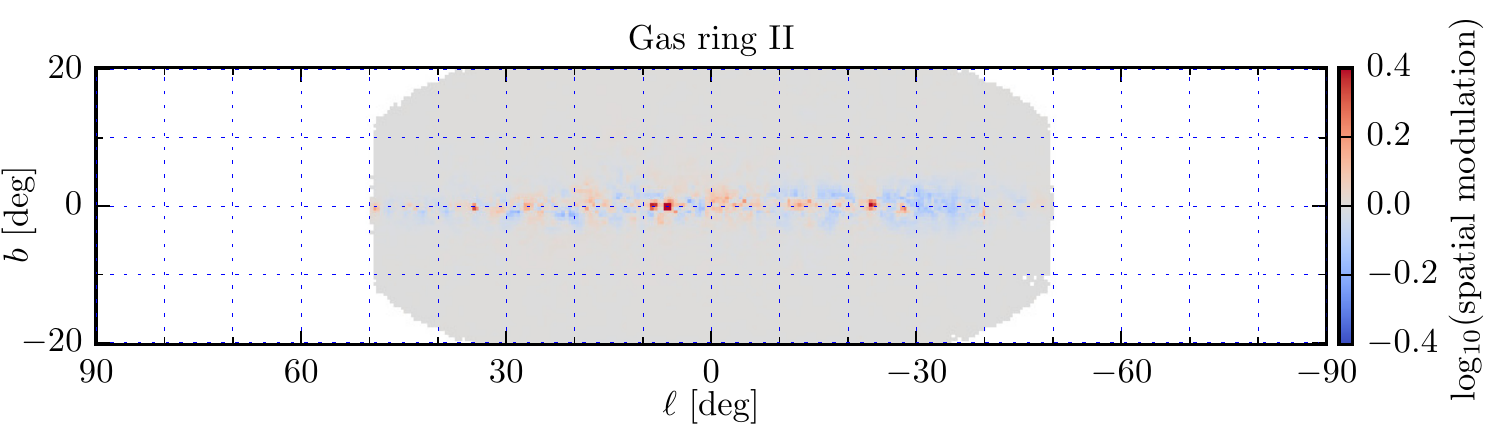}
    \includegraphics[width=0.49\linewidth]{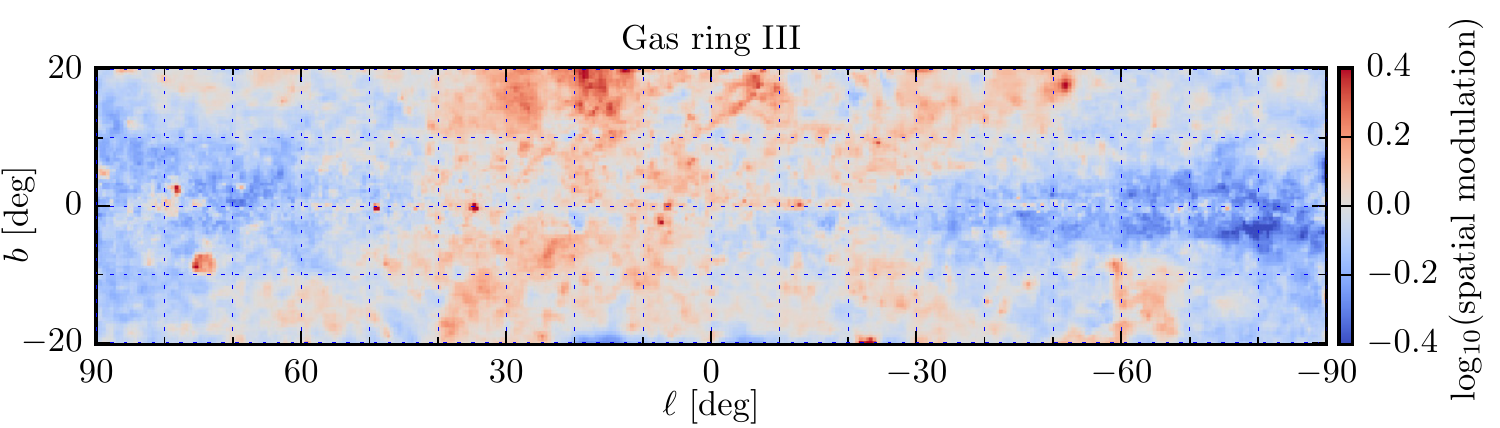}
    \includegraphics[width=0.49\linewidth]{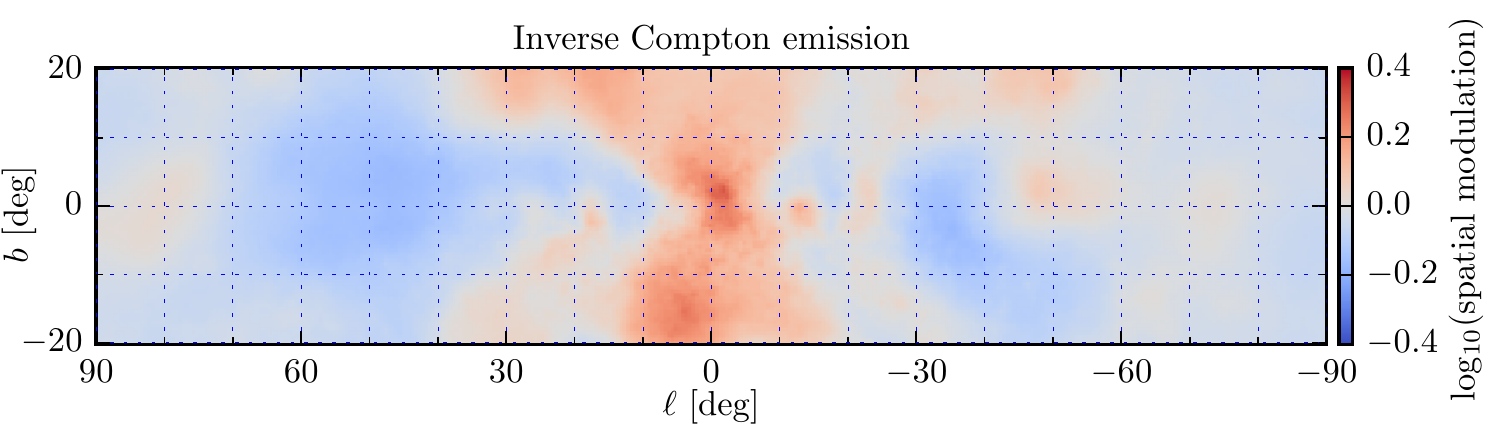}
    \caption{Spatial modulation parameters for the four diffuse components (three gas and one ICS) of \run3.}
    \label{fig:run3_rescaling}
\end{figure}

In figure~\ref{fig:run3_rescaling}, we show the modulation parameters that correspond to the four diffuse components in \run3.  For the gas ring I and gas ring II, the modulation parameters stay close to one except for a few bright spots along the Galactic disk. Most of these correspond to extended sources in the \Fermi 3FGL. However, in gas ring III, which reaches up to high latitudes, we can see the effects of `dark gas', which is not included in our gas templates and typically forms extended halos around CO and HI emission regions~\cite{2005Sci...307.1292G}.
This gas component is not captured by the usual CO line or 21 cm observations, but can be inferred from gas extinction maps to good accuracy.  We did not include such a correction in our initial maps shown in figure~\ref{fig:templates}.  Observing that the nuisance parameters are in fact recovering the dark gas from the gamma-ray data alone indicates that our method works as expected.
In the ICS template, we also clearly see an additional component in the modulation parameters: the \Fermi bubbles. The spectrum for the \Fermi bubbles is somewhat similar to the ICS spectrum, but it is, of course, a distinct component that we will include as such in subsequent fits.

\subsection{Additional spectral and spatial components}

\run4.  Up to now, we considered only the most conventional ingredients in the modeling of the GDE, although augmented with nuisance parameters.  We found in \run3 that various significant residuals still remain.  We will now add additional components that account for these residuals.  We note that this does not mean that this is the only valid way to decompose the observed gamma-ray emission from the Galaxy in individual components.

We first add a spectral template for the \Fermi bubbles.  The \Fermi bubbles component has an unconstrained morphology, while the spectrum is fixed, to within $5\%$ ($\lambda=400$), to the \Fermi bubble spectrum measured in ref.~\cite{Fermi-LAT:2014sfa}. We furthermore use a weak spatial smoothing regularization with $\eta=4$. In this case, we use non-zero smoothing regularization terms to reduce the effect of Poisson noise on the reconstructed morphology, although this is also expected to somewhat wash out potentially sharp features in the bubble edges.  We constrain the bubble template to be non-zero only in an hour-glass shaped region defined by $|\ell| + 4\cos(b/6.4^\circ)< 16^\circ$.

Furthermore, we add templates for the many spots of localized extended high-energy emission that we saw in \run3 along the Galactic disk. Many of these regions are colocated with the extended sources in the 3FGL. 18 such sources, out of 25 total in the 3FGL catalog, fall within our RoI. We therefore introduce new diffuse components that are only non-zero in regions that correspond to the extended sources in the 3FGL catalog that fall within our RoI. We initialize the spectra for these components to the extended source spectra from ref.~\cite{Acero:2015hja}, but leave considerable freedom for the spectra in the fit. Furthermore, the morphologies of the extended sources are left unconstrained, except for some smoothing. The templates of the extended sources are restricted to circular regions centered on their 3FGL locations with radii varying from 1.5--5$^{\circ}$, depending on the size of the residuals left by the source in \run3. We use a total of 16 templates for the extended sources. We exclude two sources, $\gamma$-Cygni and HESS J1616-508, because their templates overlap with the templates of other extended sources, and they are therefore completely degenerate with them, given the freedom allowed in both the spectral and spatial modulation parameters. Details about the adopted hyper-parameters can be found in table~\ref{tab:fits}.

We can see in figures~\ref{fig:residuals_low} and~\ref{fig:residuals_high} that the additional components remove much of the residual emission that remained in \run3.  Besides removing the \Fermi bubbles at high latitudes, most of the bright high-energy emission along the Galactic disk is removed. However, not all of the residuals disappear. The most notable one that remains at high energies is localized at $\ell\approx -48^\circ$ is extended by a few degrees. This residual is most likely associated with an extended source or sources not present the 3FGL catalog. In fact, there are two extended sources within $|b|<0.5^{\circ}$ near $\ell=-48^{\circ}$ in the recently released 3FHL catalog \cite{2017arXiv170200664T}, 3FHL J1420.3-6046e and 3FHL J1409.1-6121e  (see also the FGES catalog, built from a dedicated search for extended emission above 10 GeV along the Galactic plane: \cite{2017arXiv170200476T}).  Overall there are 19 extended emission sources in the 3FHL catalog that are within our RoI that are also not detected as extended emission in the 3FGL catalog. These sources may very well be contributing to some of the high-energy residuals we find along the disk in our fits. However, we restrict ourselves to the 3FGL catalog for now and save a full analysis of the extended emission sources along the disk for a future study.

\begin{figure}[t]
    \includegraphics[width=0.49\linewidth]{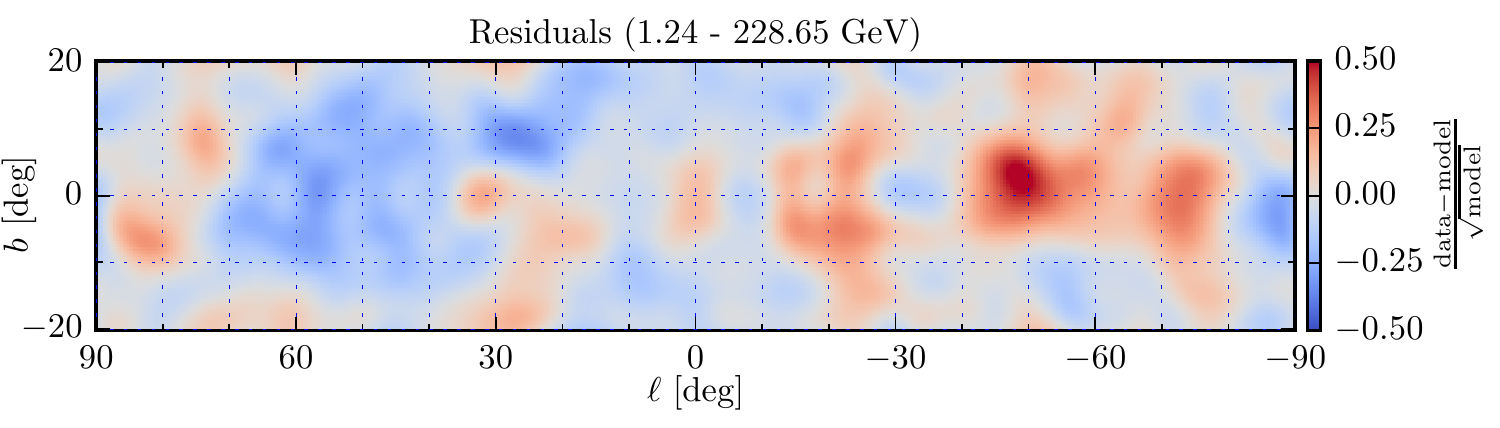}
    \includegraphics[width=0.49\linewidth]{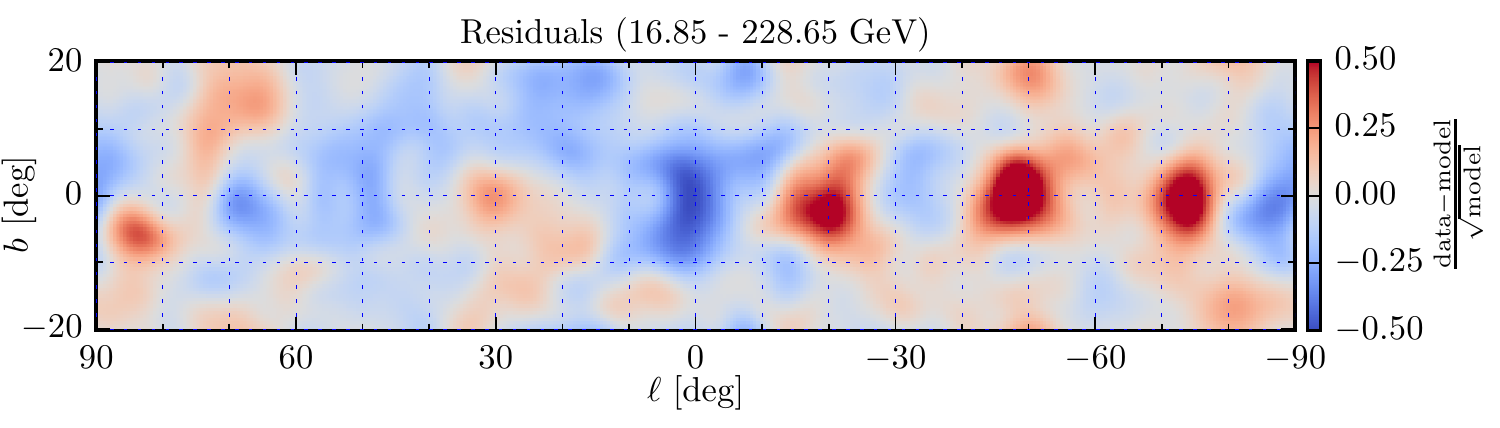}
    \centering
    \includegraphics[width=0.49\linewidth]{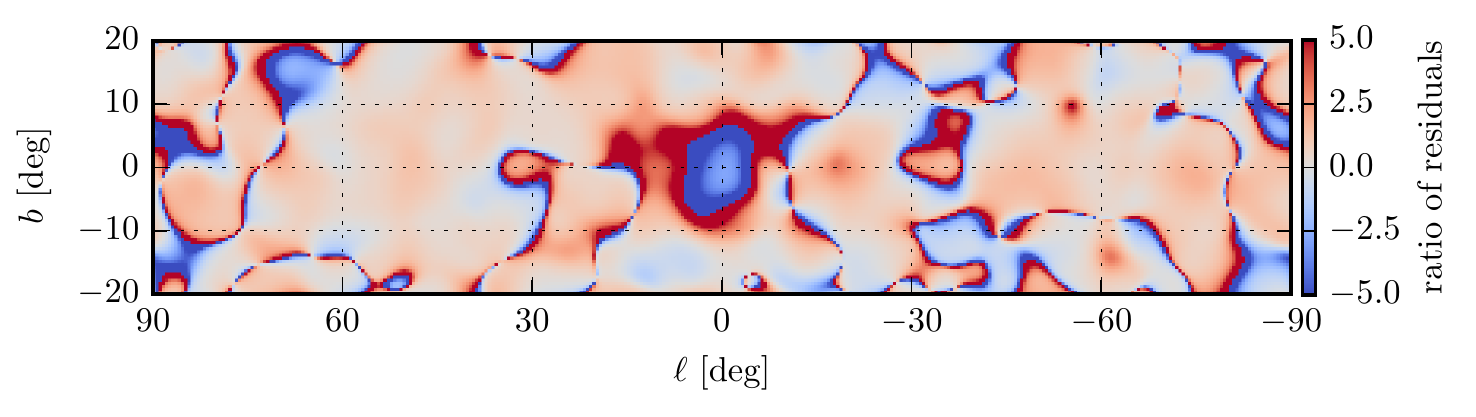}
    \caption{Top two figures: heavily smoothed residual maps for \run4 for two energy bands, $>~1.24$~GeV and $>~16.84$~GeV. Bottom: ratio of the heavily smoothed residual maps, where the ratio is the $>~16.84$~GeV map divided by the $>~1.24$~GeV map.}
    \label{fig:run4_residratio}
\end{figure}

Another residual that remains is some extended emission close to the Galactic center.  It has the peculiar feature that it is strongest in the second energy band, and appears almost as a negative residual in the third and fourth energy band. We show this feature in figure~\ref{fig:run4_residratio}. The top two figures are the same as in figures~\ref{fig:residuals_low} and \ref{fig:residuals_high}, but more heavily smoothed with a gaussian kernel to reduce the visual effect of noise and enhance the larger residual present close to the Galactic center. In the bottom panel of figure~\ref{fig:run4_residratio}, we show the ratio of the higher-energy residual map divided by the lower-energy map, after the maps were smoothed, to highlight the strength of this residual (the sharp positive-negative edges in the ratio map are the result of the smoothing). We address this residual in \run5.

\medskip

\run5.  This is the last and most complex model in the present analysis.  On top of the components of \run4, it includes a template that accounts for the central bulge excess that we found above.  It corresponds to the \Fermi Galactic center excess (GCE) extensively studied in the literature~\cite[\fex,][]{Goodenough:2009gk, Hooper:2010mq, Boyarsky:2010dr,Abazajian:2012pn,Macias:2013vya,Daylan:2014rsa,Gaggero:2015nsa,Calore:2014xka,deBoer:2016esu,TheFermi-LAT:2015kwa,Karwin:2016tsw}. As template for the GCE, instead of an analytic DM-inspired model as typically done in the literature, we use here, for the first time, the profile of the 511 keV flux as measured by INTEGRAL ~\cite[e.g.~][]{Leventhal_1978,Knodlseder:2005yq, Weidenspointner:2006nua, Siegert:2015knp}. The positron annihilation signal (for a comprehensive review see ref.~\cite{Prantzos:2010wi}) in the Galactic bulge shows a few common features to the GCE: both are almost spherically symmetric around the Galactic center, show a spectrum uniform in an extended RoI, and peaks in the direction of the center of the Galaxy. For a more thorough study about the connection between the two signals we refer the reader to a forthcoming publication~\cite{511keV}. We here report the main results of introducing such an additional template component. We only include the Galactic center and bulge component of the 511 keV flux, and neglect the disk for the present purpose. These components are modeled following ref.~\cite{Siegert:2015knp}.  Furthermore, we keep the morphology of the component constrained to within $20\%$, while the spectrum remains completely unconstrained, see table~\ref{tab:fits}.  The corresponding residuals in figures~\ref{fig:residuals_low} and~\ref{fig:residuals_high}, show that this component is enough to remove the remaining positive and negative residuals in the central region.  

In figure~\ref{fig:residual_histogram}, we show the distribution of residuals in units of standard deviations for \run1 and \run5. Comparing \run1 to \run5, we find that the introduction of nuisance parameters in the fit reduces the number of pixels with highly significant residuals (above/below $\pm3\sigma$) by over two orders of magnitude. In fact, the most drastic reduction of residuals appears from \run1 to \run2, which can be seen in the drop of the Poisson $-2\ln\mathcal{L}_P$ as shown in table~\ref{tab:fits}.

However, even in \run5 there are residuals above $\pm5\sigma$ that remain. On closer inspection, we find that at energies $\lesssim10\GeV$, they are usually related to regions around bright point sources which are over- or under-subtracted.  A likely cause is that we do not refit the positions of the 3FGL sources in our analysis; another may be due to potential, small numerical imperfections in our treatment of the \Fermi PSF.  At high energies, the remaining residuals often correspond to the regions that are also visible in the residual maps in figure~\ref{fig:residuals_high}: extended regions along the Galactic disk with hard emission.  The modeling of this emission component is still not optimal in the current analysis, and could be further improved by, \fex, increasing the size of the patches that we use for extended sources in our \run5 or including templates for additional known extended sources in the 3FHL.

Lastly, the distribution of residuals in figure~\ref{fig:residual_histogram} strongly deviates from a naive normal distribution, illustrated by the black line, which would be approximately expected if the residuals were due to statistical fluctuations only. Additionally, there is a strong tilt towards positive residuals, particularly for energies above 60 GeV, for both \run1 and \run5.  This can be understood by realizing that the smallest possible negative residual in a pixel with an expected number of $\mu$ photons is, in our adopted schema, $-\sqrt{\mu}$.  At high energies, $\mu$ can be of order one, leading to the apparent tilt.  In fact, we observe the same behaviour for our mock results that are discussed in appendix~\ref{apx:synthetic}.

\section{Characteristics of the reference model}
\label{sec:min}

\begin{figure}[t]
    \centering
    \includegraphics[width=0.49\linewidth]{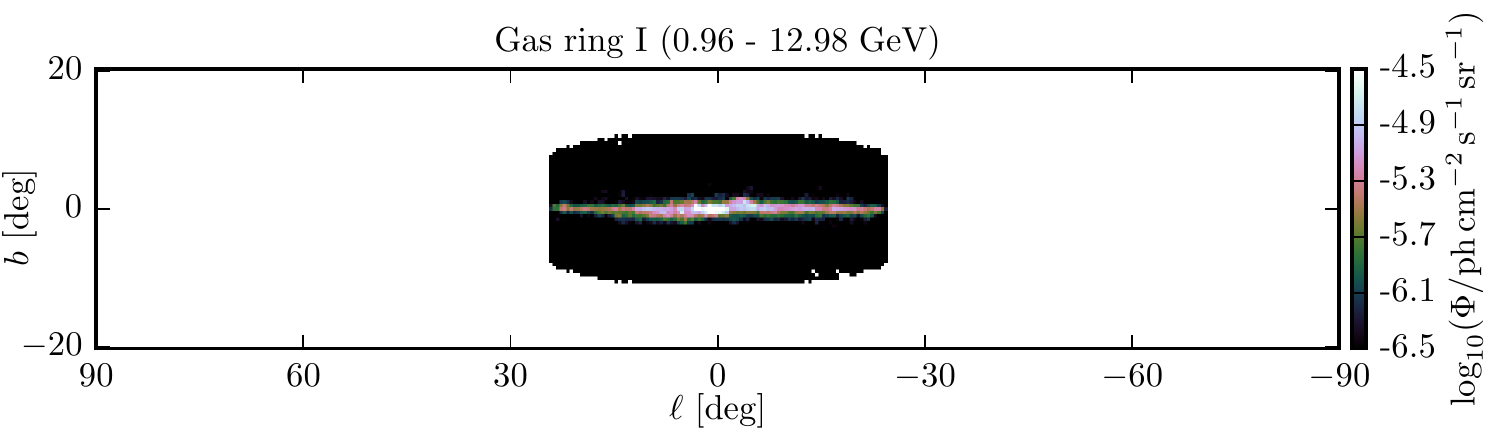}
    \includegraphics[width=0.49\linewidth]{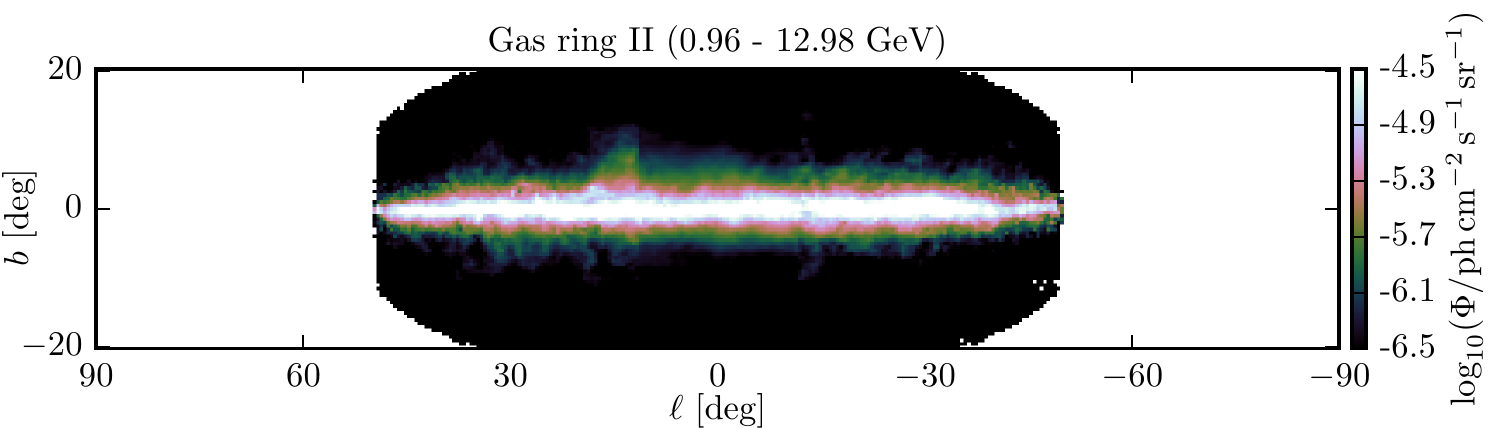}
    \includegraphics[width=0.49\linewidth]{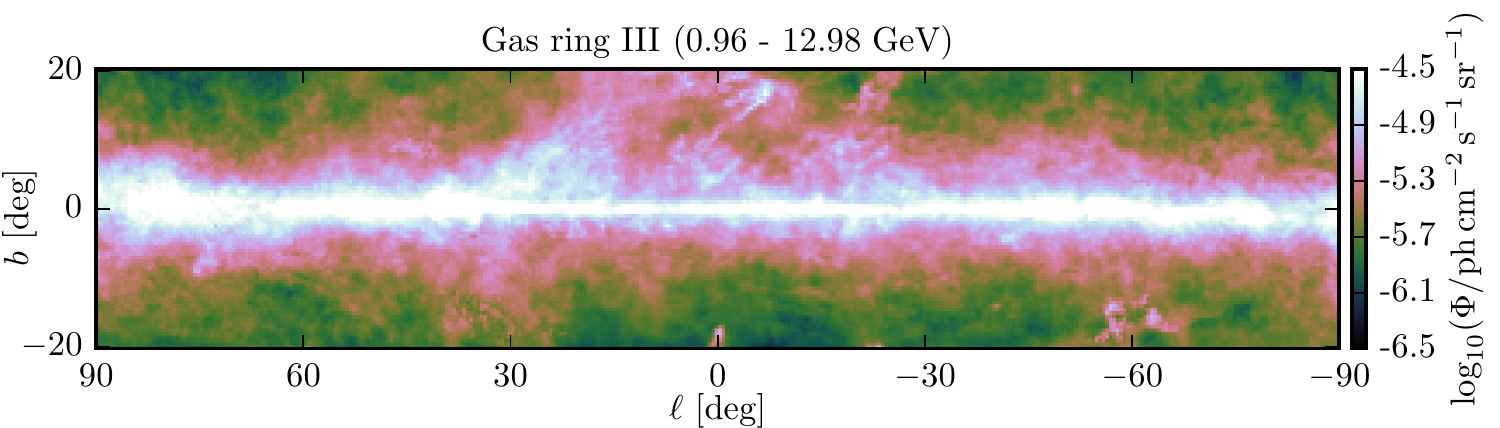}
    \includegraphics[width=0.49\linewidth]{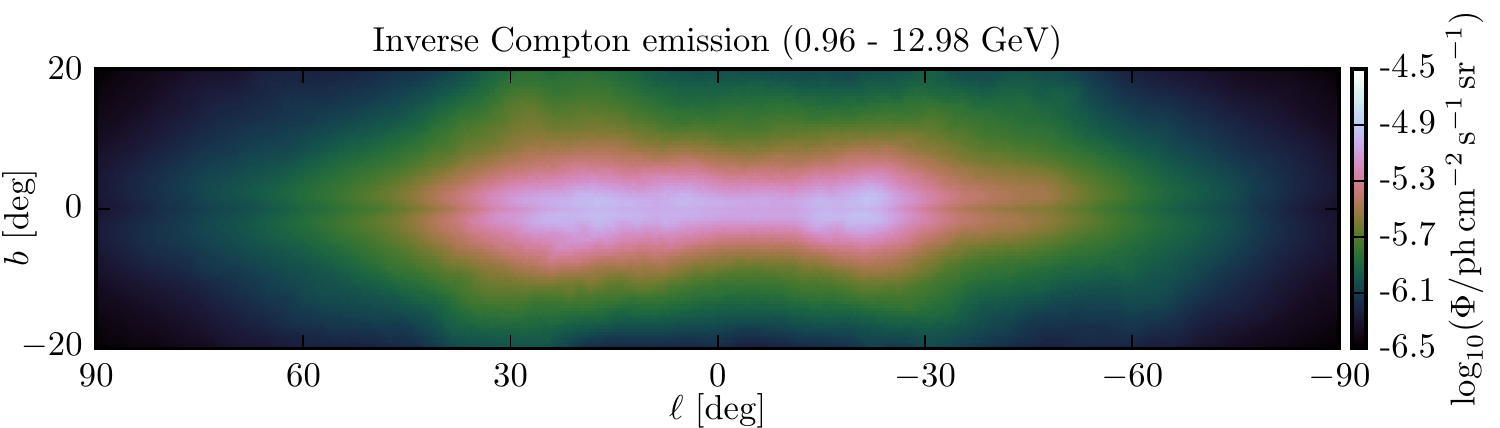}
    \includegraphics[width=0.49\linewidth]{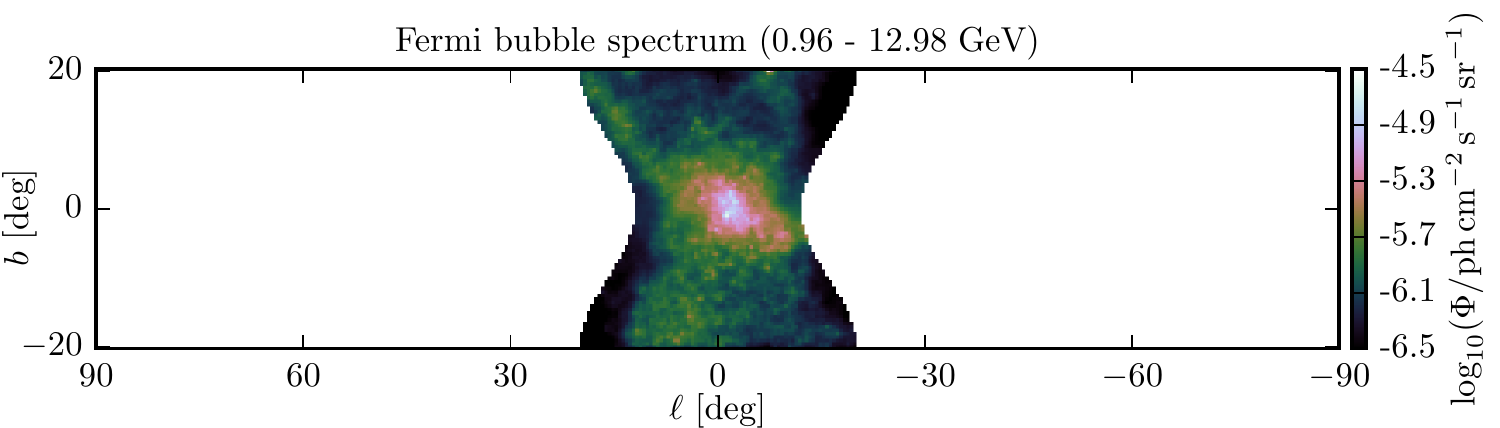}
    \includegraphics[width=0.49\linewidth]{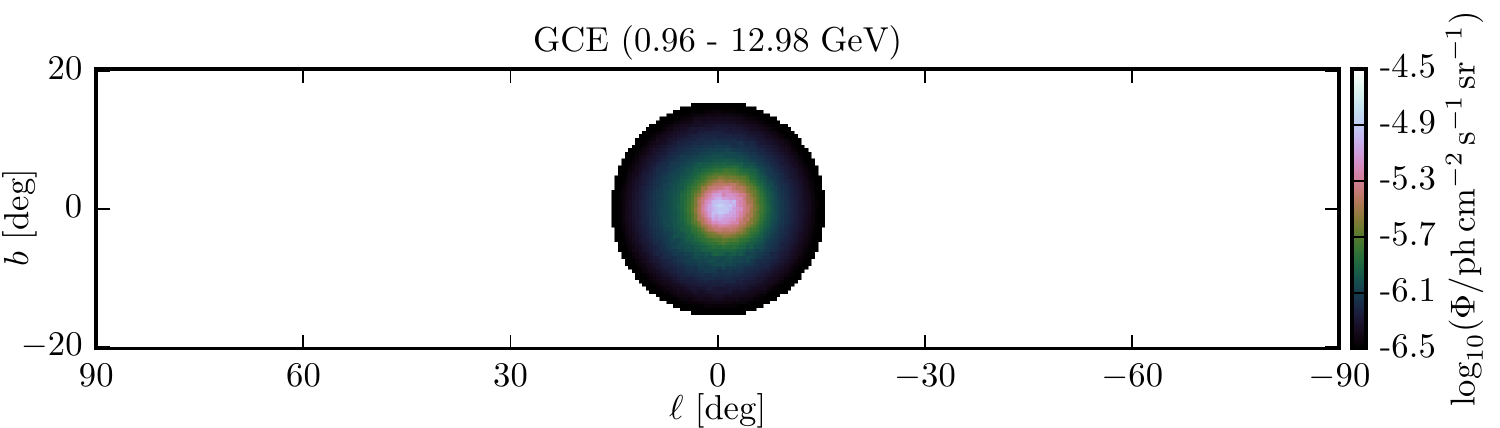}
    \includegraphics[width=0.49\linewidth]{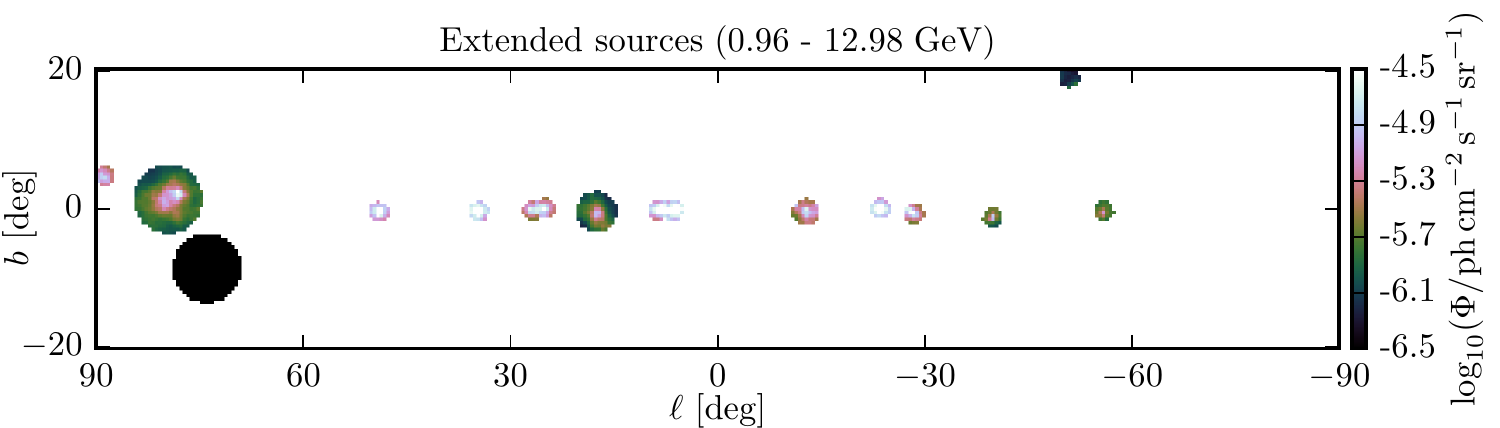}
    \includegraphics[width=0.49\linewidth]{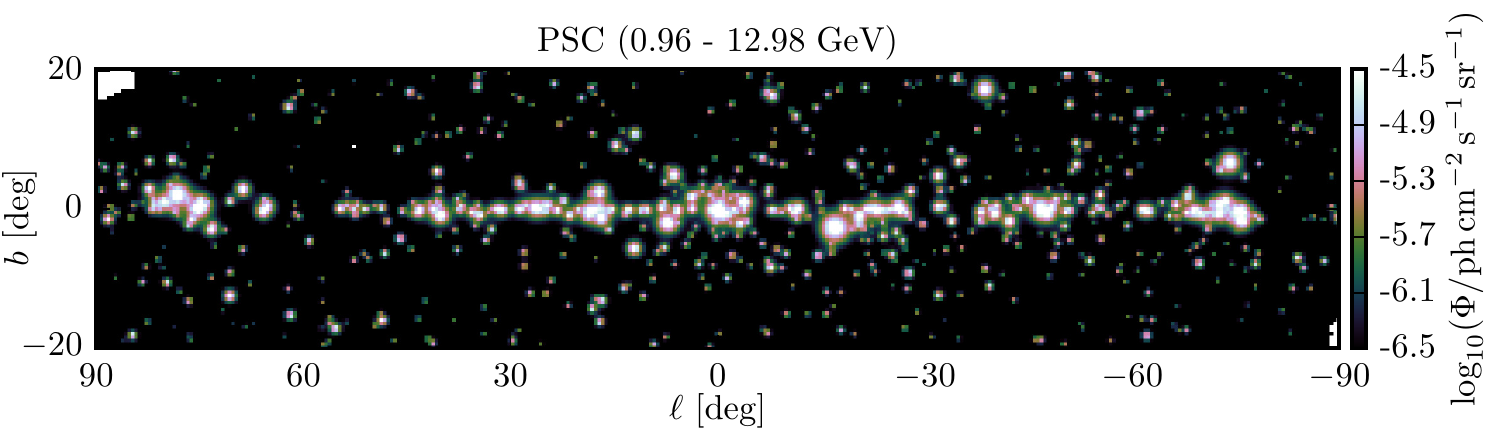}
    \caption{All extended emission components for the reference model \run5 (see table~\ref{tab:fits} for details about the model).  In the bottom row, we show the combined emission from extended sources (left panel), although their templates and spectra are kept independent in the fit. We also show the combined emission of point sources (right panel), which also have independent normalizations and spectra in the fits.}
    \label{fig:run5_components}
\end{figure}

In this section, we discuss in detail the aspects of \run5 that we motivated in the previous section.  In particular, we are interested in the plausibility of the values and variances of the modulation parameters in the analysis.

\subsection{Spatial modulation parameters}

\begin{figure}[t]
    \centering
    \includegraphics[width=0.49\linewidth]{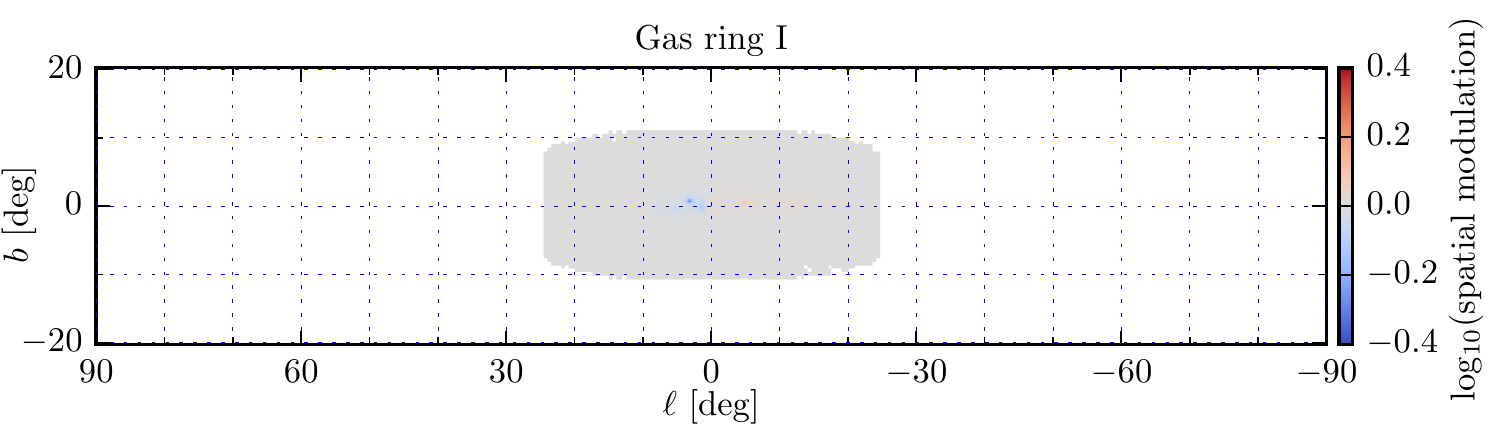}
    \includegraphics[width=0.49\linewidth]{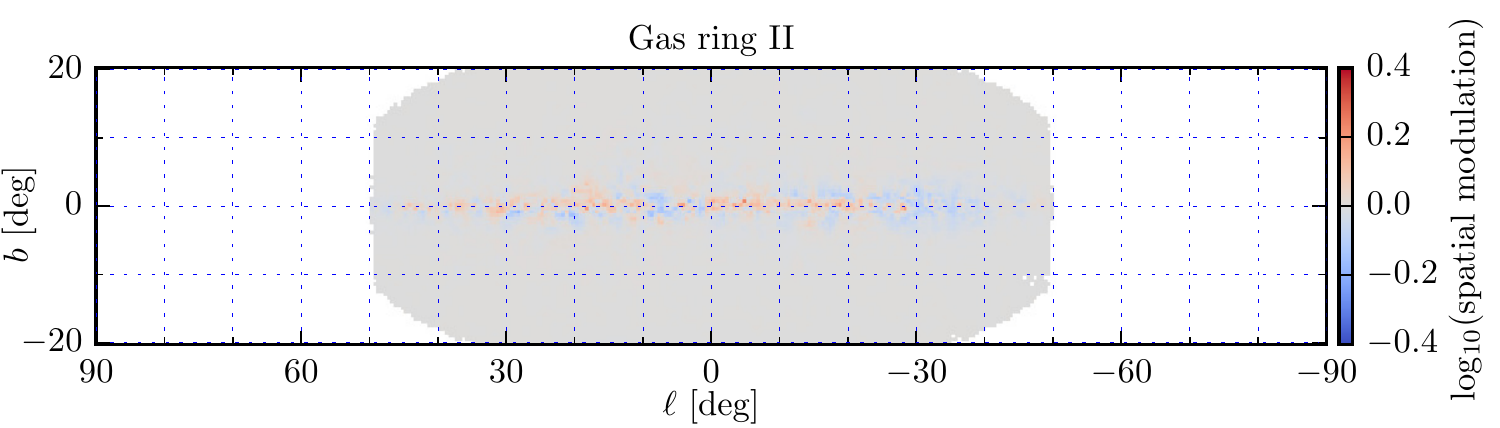}
    \includegraphics[width=0.49\linewidth]{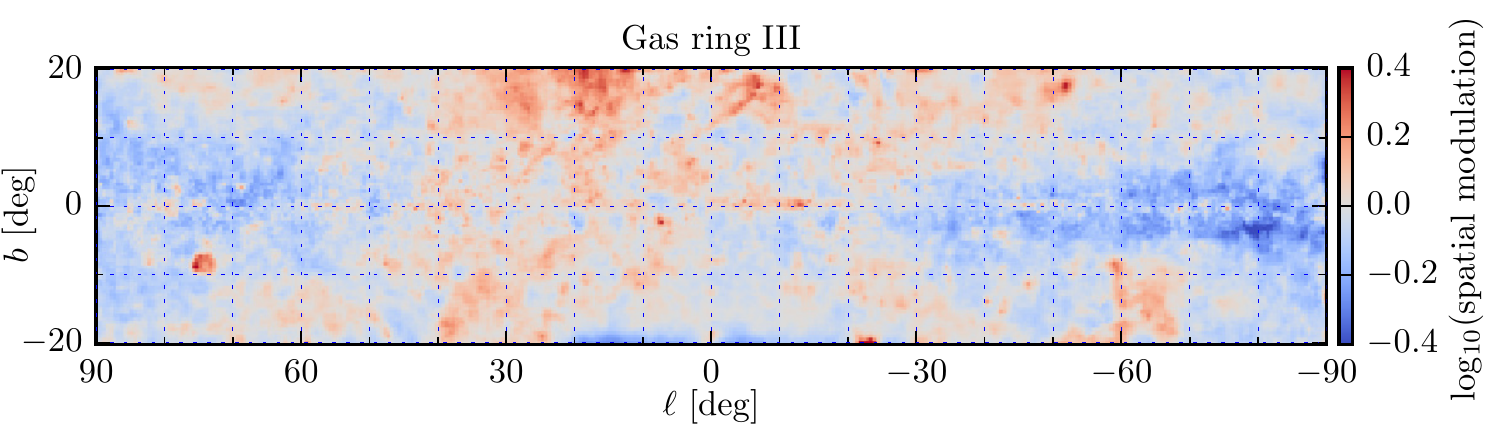}
    \includegraphics[width=0.49\linewidth]{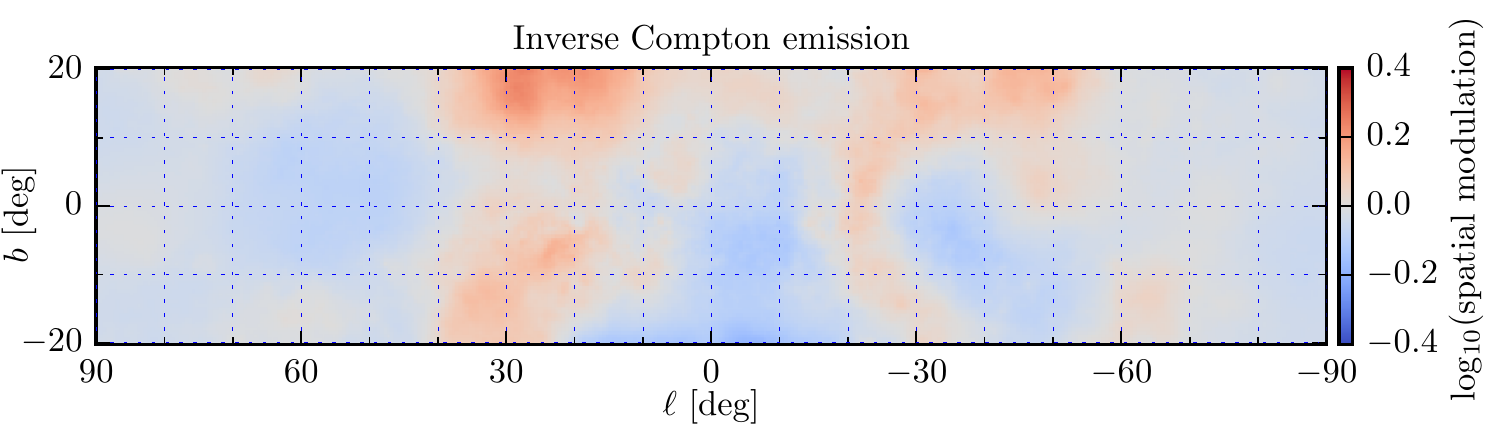}
    \caption{Modulation parameters for the conventional diffuse emission components of \run5, which includes multiple disk components.}
    \label{fig:run5_rescaling}
\end{figure}

\begin{figure}[t]
  \centering
  \includegraphics[width=0.49\linewidth]{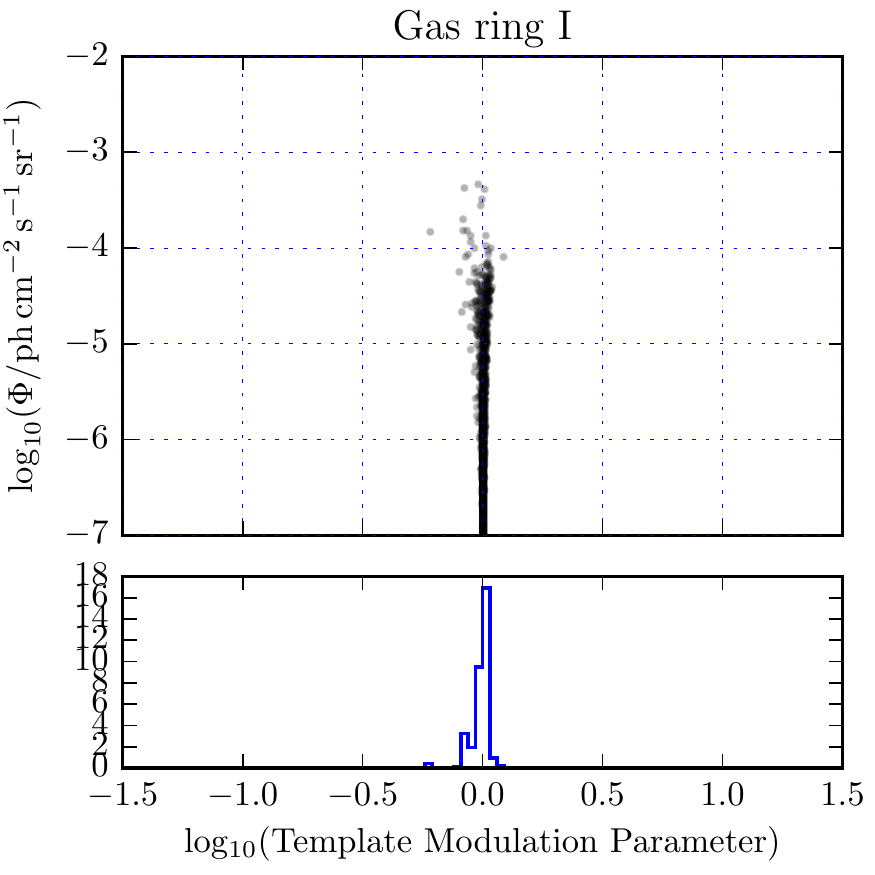}
  \includegraphics[width=0.49\linewidth]{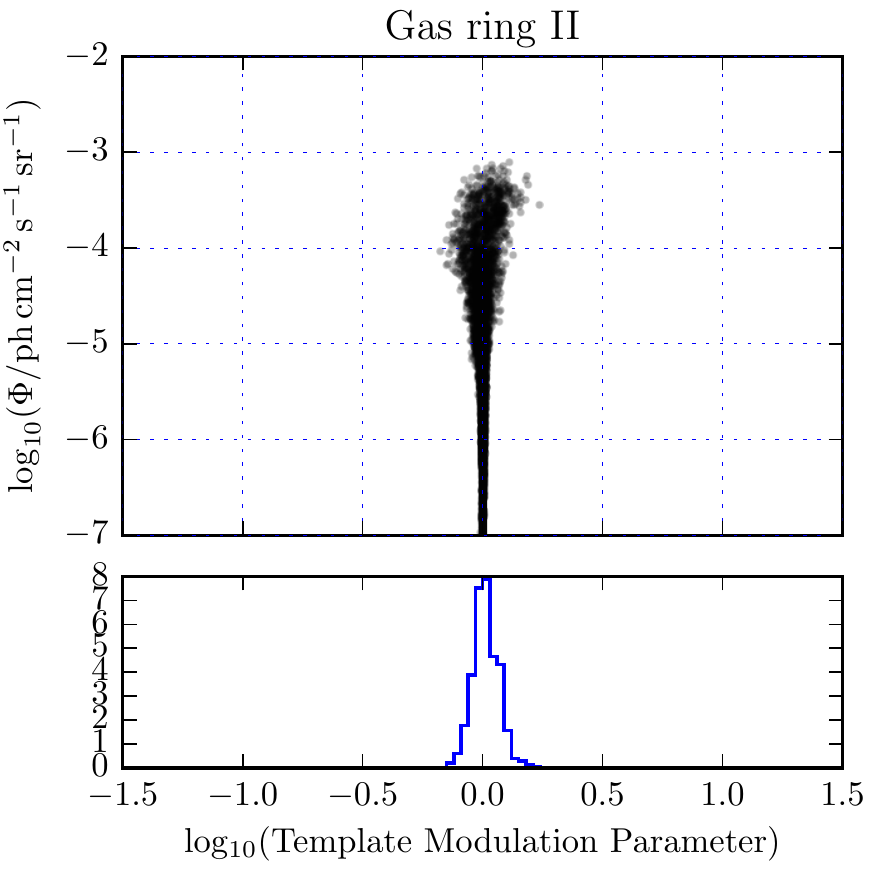}
  \includegraphics[width=0.49\linewidth]{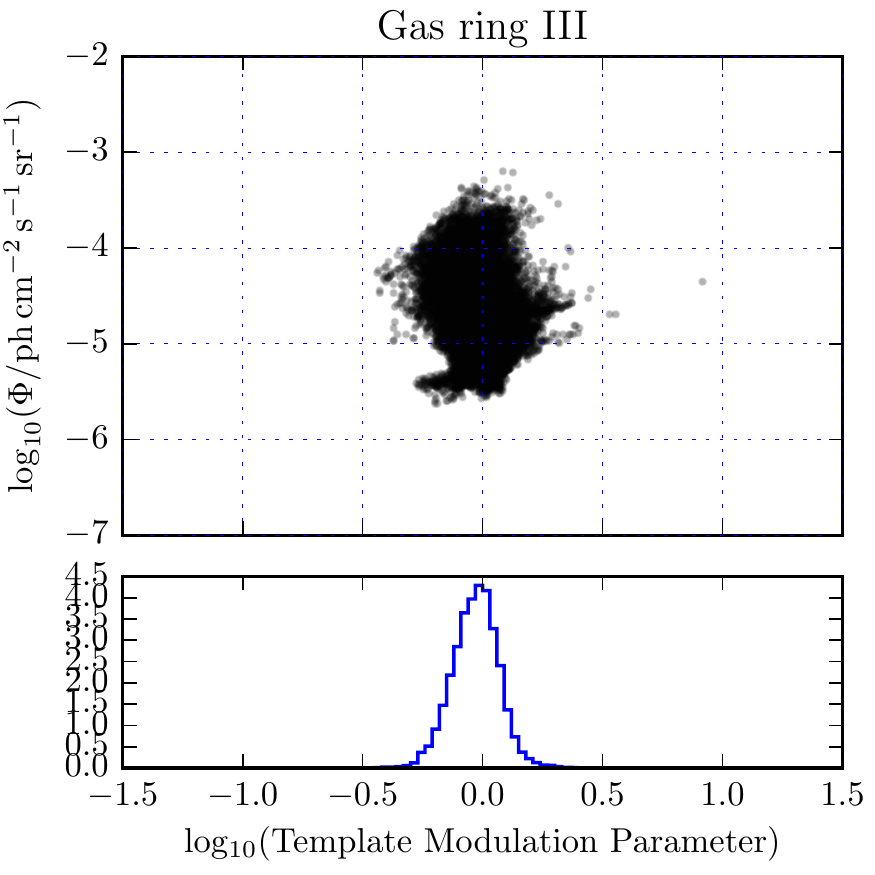}
  \includegraphics[width=0.49\linewidth]{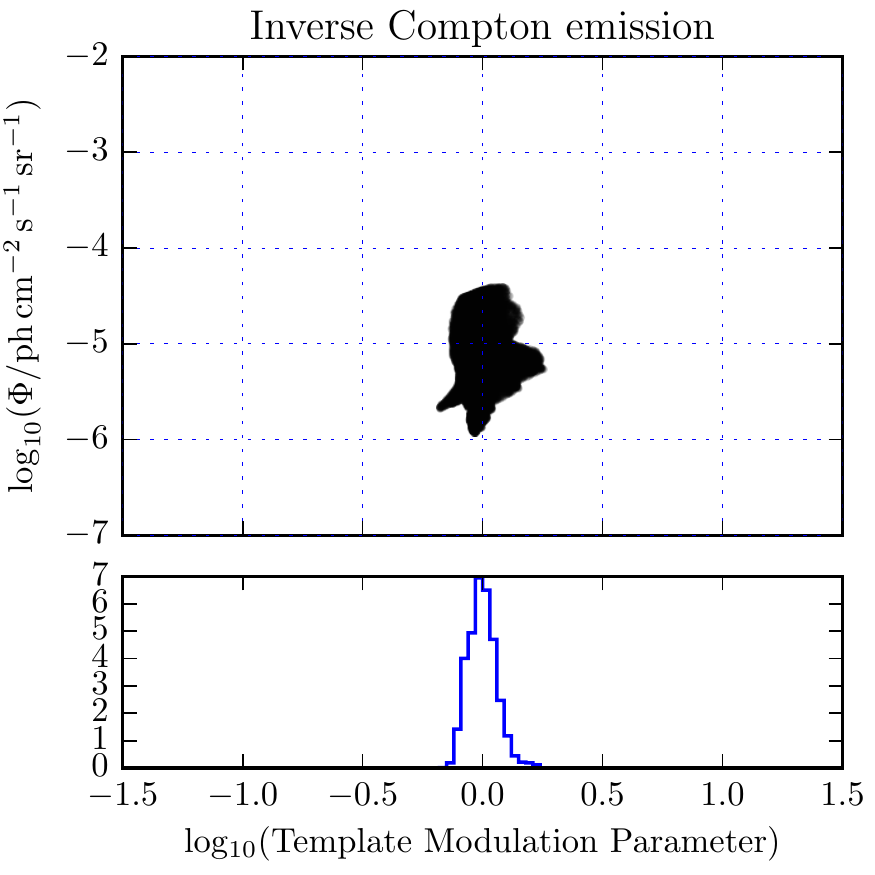}
  \caption{Scatter plot and histogram of rescaling factors of the spatial modulation parameters of the conventional diffuse emission components of \run5 (see also figure~\ref{fig:run5_rescaling}).  The scatter plot shows the rescaling factor vs.~the template flux in each pixel.  The histogram in the lower panel is weighted by the pixel flux.}
  \label{fig:run5_rescaling_hist}
\end{figure}

The various best-fit fluxes for the extended components of \run5 are shown in figure~\ref{fig:run5_components}.  The spatial modulation parameters of \run5 are shown in figure~\ref{fig:run5_rescaling}, for the four conventional diffuse emission components. Histograms for the same modulation parameters are shown in figure~\ref{fig:run5_rescaling_hist}, weighted by pixel intensity.  The original morphology of the four components is shown in figure~\ref{fig:templates}, where the spatial modulation parameters are set to one.

We find that for the innermost gas ring I, the nuisance parameters remain rather close to one. The weighted $90\%$ central quantile covers the range [0.88, 1.04], indicating typically less than $10\%$ deviations from the original template. Weighting these parameters is important, since a large number of pixels of the template correspond to very small fluxes, which are not constrained by the data and hence not rescaled.  This does not, however, necessarily reflect that the original template was very accurate, but more likely that measuring this template is difficult in the bright Galactic disk.

The gas rings II and III have modulation parameters with a broader distribution.  Their weighted $90\%$ central quantiles cover [0.80, 1.46] and [0.70, 1.45], respectively.  Note that there are also a few outliers, which likely correspond to isolated point sources that get incorrectly absorbed by the template.  These do not affect the quantiles, which are relatively resilient towards outliers.  The larger variations indicate that the templates are well constrained by the data, and give an estimate for the quality of gas maps in describing the data.

The gas ring III, which includes the position of the Sun and is responsible for the structured emission at mid-latitudes, exhibits an interesting pattern in the nuisance parameters.  The spatial modulation parameters are in some regions as large as 2--3 (in the tail of the distribution); in other regions we observe a suppression of similar magnitude.  As for \run3, the emerging structures are very similar to the `dark gas' corrections~\cite{2005Sci...307.1292G} (see discussion for \run3).

In figure~\ref{fig:run5_rescaling}, we also show the rescaling of the ICS emission w.r.t.~the initial template from figure~\ref{fig:templates}.  The weighted $90\%$ quantile covers here the values [0.84, 1.21], which is a smaller range than for the gas emission.  However, there are some similarities between the large-scale features of the gas ring III and the ICS rescaling map.  This suggests that both components are not completely separable, and probably some of the ICS emission is falsely absorbed by the gas map and vice versa.  This is expected if, for instance, the spectra of the two components are not perfectly uniform throughout the entire RoI (this effect could be mitigated by further splitting up these diffuse components).  On the other hand, looking at the total ICS emission shown in figure~\ref{fig:run5_components} suggests that the emission absorbed by our ICS template might actually have a bi-modal ring-like structure, with enhancements towards $\ell\approx\pm25^\circ$.  This corresponds to the edges of the molecular ring and could be potentially explained by the 3D distribution of CR sources, or yet by the asymmetry expected in the case of 3D interstellar radiation field.

\subsection{Spectral modulation parameters}

\begin{figure}[t]
    \centering
    \includegraphics[width=0.49\linewidth]{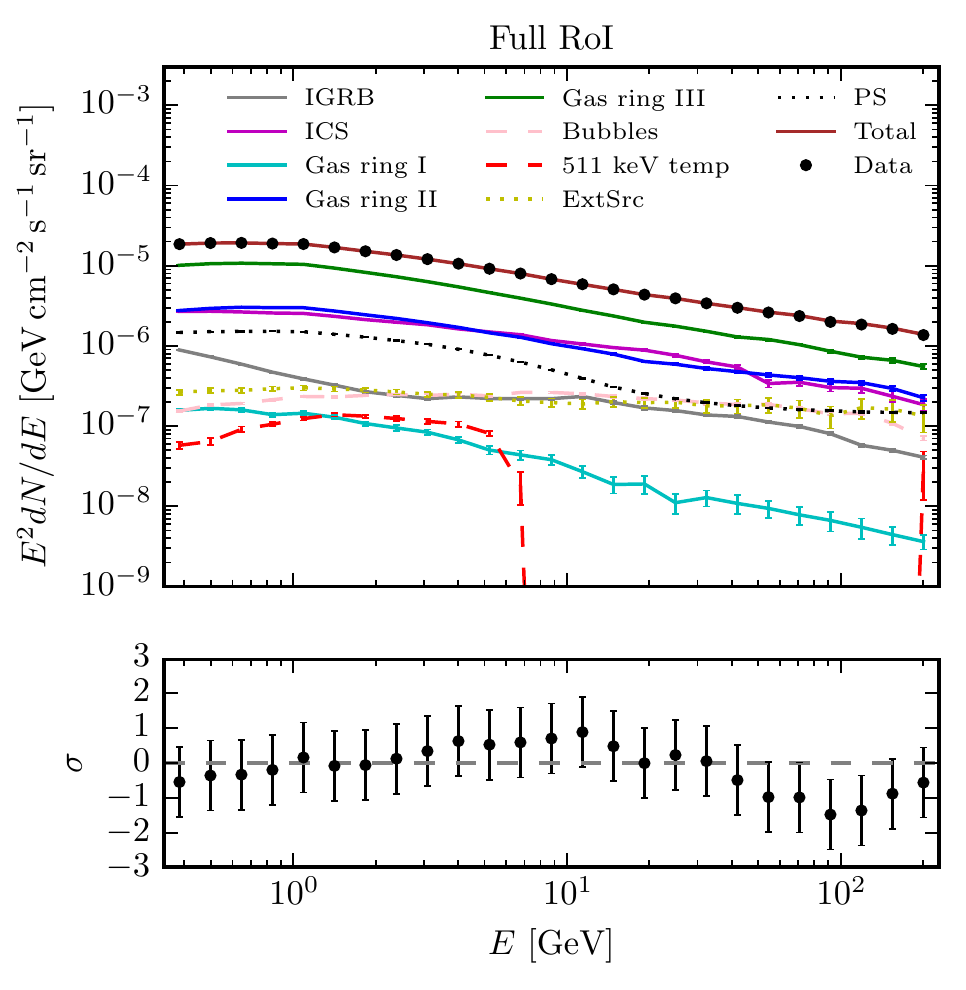}
    \includegraphics[width=0.49\linewidth]{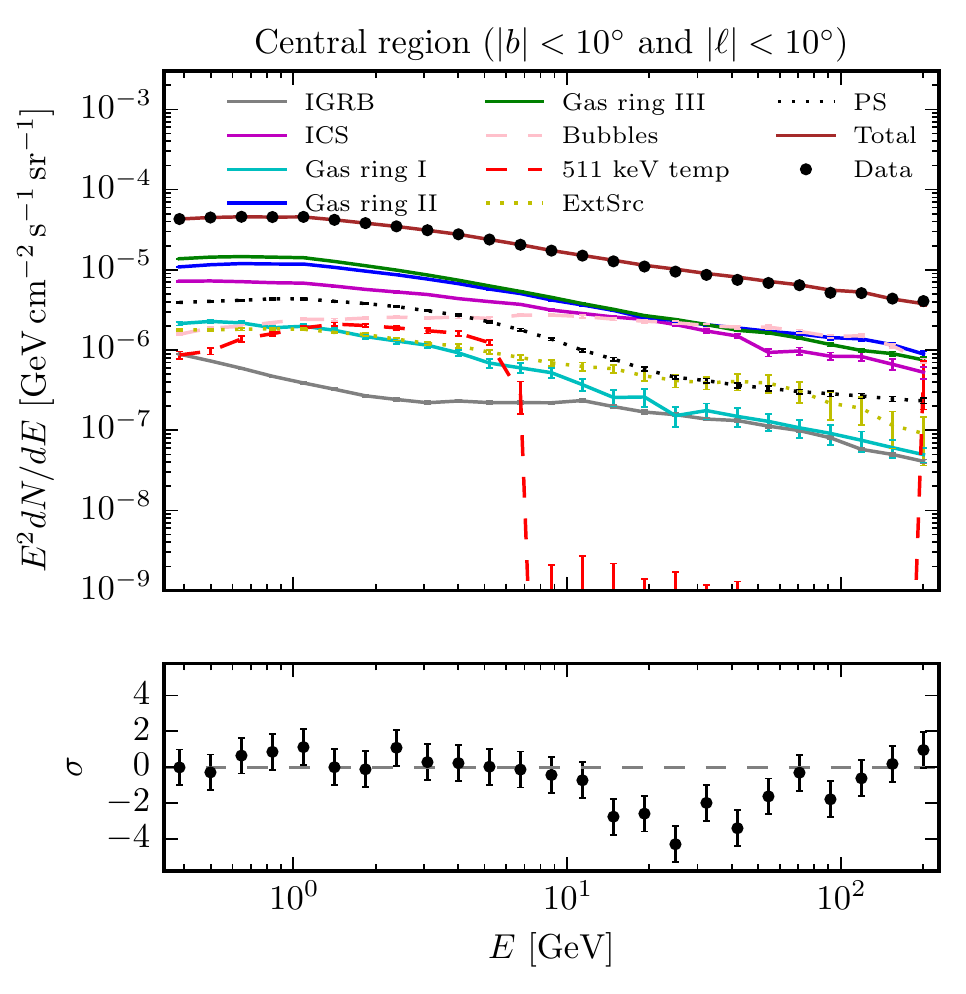}
    \caption{Spectra from \run5, in the same regions as used in figure~\ref{fig:run1_spectra}.  In the upper panels, we show the various components of the fit, as well as their sum compared to the data from that region.  The lower panels show the residuals in units of standard deviations, where $\sigma = (\text{data}-\text{model})/\sqrt{\text{model}}$.}
    \label{fig:run5_spectra}
\end{figure}

The best-fit spectra from \run5 are shown in figure~\ref{fig:run5_spectra}, both for the full RoI considered in this work as well as for a smaller central region that covers the Galactic bulge.  In both regions, the measured spectrum is well reproduced. The error bars on the ICS, gas ring II, and gas ring III components are very small, which is a consequence of the large sky region that they cover.  The same is true for the combined emission of the 776 point sources in our analysis, although some $20\%$ uncertainties becomes visible at high energies in the central region.  Gas ring I, however, has sizable uncertainties, as it only covers a small sky region and overlaps along the line-of-sight with the other gas rings. We remark that our results are relatively stable w.r.t.~choices of the adopted smoothing parameters, as discussed in appendix~\ref{apx:run5}.

\begin{figure}[t]
  \centering
  \includegraphics[width=0.6\linewidth]{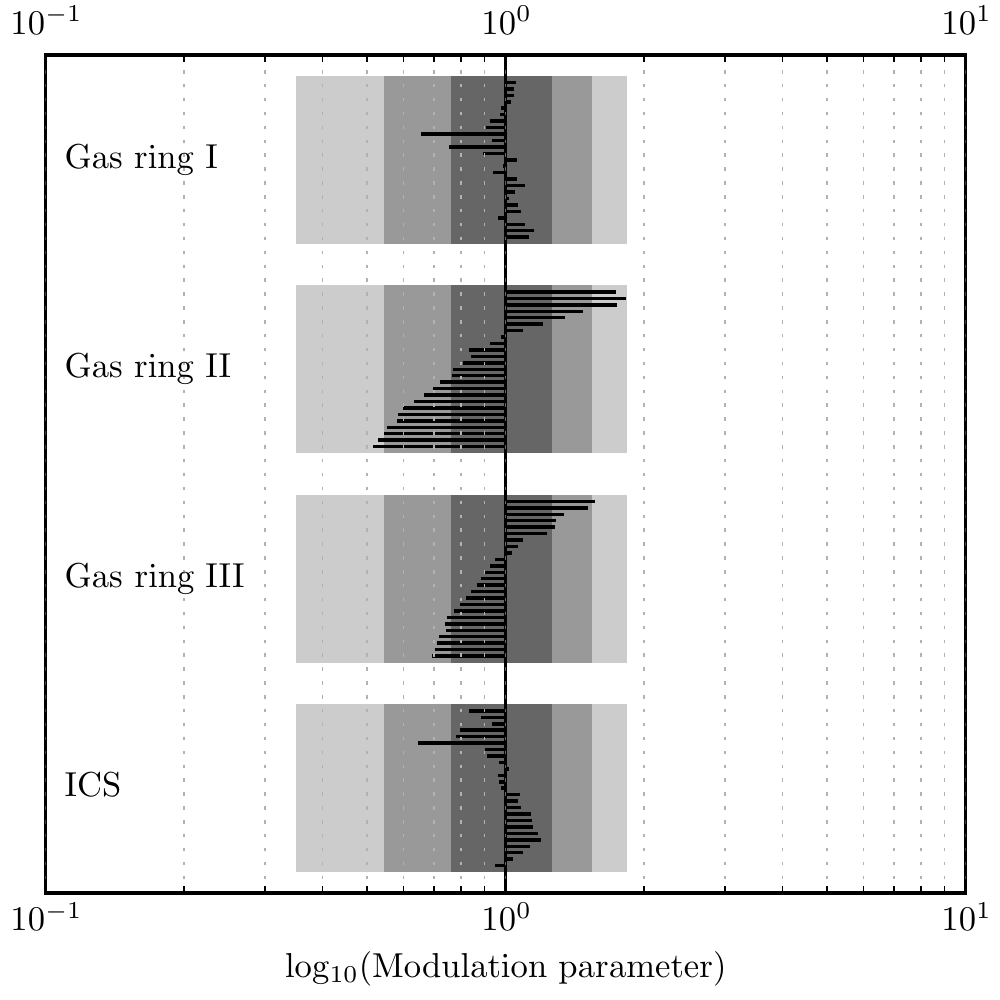}
  \caption{Spectral modulation parameters, $\sigma_i^{(k)}$, for the four `conventional' diffuse emission components in \run5.  The rescaling is w.r.t.~the nominal input spectra, which are taken from ref.~\cite{Ackermann:2012pya}. The different gray shaded regions show the 1, 2 and 3-$\sigma$ ranges corresponding to the MEM regularization (namely the values of $\sigma_i^{(k)}$ where $\lambda=16$).  Lines correspond respectively, from bottom to top, to the energy bins from $0.3$ to $230$ GeV.}
  \label{fig:run5_tree}
\end{figure}

We find that the inner gas ring II has a somewhat harder spectrum at higher energies than the outer gas ring III (the effect is not visible is the poorly constrained ring I). This hardening can also be observed in figure~\ref{fig:run5_tree}, where we show the strength of the spectral modulation parameters for the diffuse emission components in \run5. There is no strong drag for gas ring I and the ICS components, which indicates self-consistency in the results. However, the spectra of the gas rings II and III harden by more than 3$\sigma$. Although (part of) the indicated spectral change could be also caused by some unmodeled extended sources, this hardening of the spectrum towards the inner Galaxy could be a real feature of the interaction of CR protons with the gas, and point towards a concentration of CR accelerators or specific propagation effects in that region~\cite{Gaggero:2014xla}.  Although a more systematic analysis would be required in this respect, we find results qualitatively in agreement with Refs.~\cite{Yang:2016jda, Acero:2016qlg}.

\subsection{The Galactic Center Excess}
Lastly, the 511 keV template, which we use to model the \Fermi Galactic center excess (GCE), exhibits a spectrum that peaks, as usual, at 1--3 GeV, and falls quickly off at lower and higher energies.  The error bars are here at the level of $10-20\%$ at the peak.  As can be seen in the right panel of figure~\ref{fig:run5_spectra}, at energies above 10 GeV, the template does not absorb any significant emission, which shows that the high-energy emission found in previous studies~\cite{Calore:2014xka,Linden:2016rcf} is here absorbed by other components, in particular the \Fermi bubbles. Here, the \Fermi bubbles absorb the high-energy emission as a consequence of not having imposed a uniform brightness bubble template. Very marginal negative residuals remain at energies between 10 and 50 GeV, most likely because of the spectral constraint on the \Fermi bubbles, which could be reduced by relaxing the constraints on that component. 

The only change in our model between \run4 and \run5 is the addition of the GCE component. We can in theory determine its significance straightforwardly with the standard likelihood ratio test; the difference in the poisson likelihood between \run4 and \run5 is approximately $208$. However, the difference in the degrees of freedom between the two models cannot be read off from table~\ref{tab:fits}, because the difference between \run4 and \run5  is smaller than the uncertainty in the number of effective degrees of freedom (see section~\ref{sec:gof}).

  We therefore performed a test where the difference in degrees of freedom is obvious: instead of allowing some small spatial modulation in the GCE component, we fix the spatial template completely, but allow the spectral modulation to be completely free. We kept all other parameters the same as in \run5. For this run, the difference in the degrees of freedom between \run4 and \run5 is equal to $25$, or the number of spectral bins. The Poisson likelihood decreases such that the signifcance for the GCE is about $12\sigma$. The overall fit is for this run is quantitatively very similar to our original version of \run5, as expected. We can therefore conclude that the GCE component is highly significant.

\section{Discussion}
\label{sec:discussion}
\subsection{The goodness-of-fit}\label{sec:gof}

We are interested in estimating the goodness-of-fit for our final \run5 model.  The traditional Pearson's chi-squared goodness-of-fit test cannot be applied to the current problem for two reasons.  First, the data is Poisson and not normally distributed.  Second, given that we introduced a number of penalization constraints on the model parameters, as well as smoothing constraints, it is not clear what the effective number of free model parameters really is.

We use an approach that fully takes into account the Poisson nature of the problem as well as the unclear effective number of degrees of freedom (DOF), and which is based on the mock results obtained in appendix~\ref{apx:run5}.  First, we consider here only the Poisson part of the likelihood function, and define $\chi^2\equiv-2\ln\mathcal{L}_P$.  We assume that the $\chi^2$ function is chi-squared distributed under repeated data realizations (this is true to very good approximation, since for the very high number of DOF the central limit theorem dominates the shape of the distribution).  We estimate the effective number of DOF from fits to mock data that are based on the best-fit models for each of the five runs in table~\ref{tab:fits}.  It can be estimated to be $k \approx \langle -2\ln\mathcal{L}_P\rangle_{\rm mock}$; its associated uncertainty is simply the square root of the variance of the chi-square distribution, $\sqrt{2k}$, about $\sim1100$ for our five runs. The number of effective DOF along with the final Poisson likelihood and regularization terms, $-2\ln\mathcal{L}_P$ and $-2\ln\mathcal{L}_R$, can be found in table~\ref{tab:fits} for each fit.

Naively, the number of model parameters is just given by the number of free parameters in the fit, as listed in table~\ref{tab:fits}. The number of independent data points would normally be simply equal to the number of pixels times the number of energy bins,
\begin{equation}
  N_\text{data} = N_\text{pix}\times N_\text{ebin}\;.
\end{equation}
For the current analysis, we have $N_\text{data} = 360\times81\times25 = 729000$.  However, these estimates are only valid if we are in the Gaussian regime in all data bins, and if all model parameters are free to vary and can independently and without degeneracies improve the fit to the data. However, these conditions are not satisfied in our current setup.  We therefore estimate the \textit{effective} number of data bins, model parameters, and DOF in a way that makes these estimates useful for goodness-of-fit calculations.

\begin{figure}[h]
  \centering
  \includegraphics[width=0.6\linewidth]{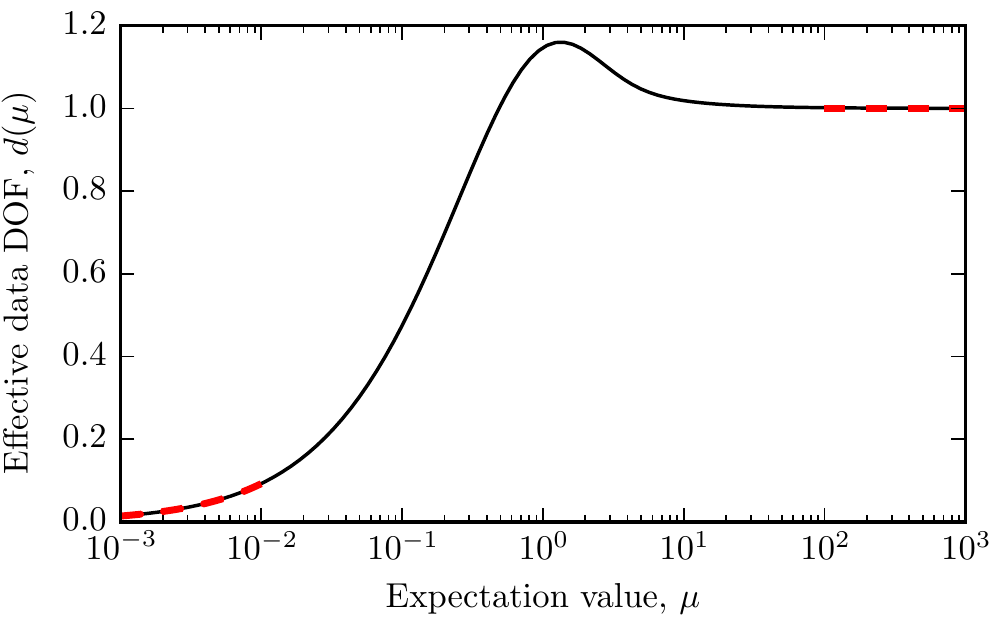}
  \caption{Effective number of DOF per data bin, $d(\mu)$, as function of the Poisson expectation value $\mu$, see Eq.~\eqref{eqn:Neffdata}.  In the Gaussian regime, $\mu\gg1$, the function converges to one; in the deep Poissonian regime, $\mu\ll1$, it is well approximated by $2\mu\ln(1/\mu)$ (red dashed lines).}
  \label{fig:d}
\end{figure}

The effective number of data bins is calculated by averaging the $-2\ln\mathcal{L}_P$ over mock realizations of the best fit model, without refitting model parameters.  This can be written as
\begin{equation}
  N_\text{data}^\text{eff} \equiv \left\langle -2\ln\mathcal{L}_P(\vect\theta)\right\rangle_{\mathcal{D}(\vect\theta)}
  = \sum_{i=1}^{N_\text{data}} 2\left\langle c \ln\frac{\mu_i(\vect\theta)}{c}\right\rangle_{c\sim P(\mu_i(\vect\theta))}
  = \sum_{i=1}^{N_\text{data}} d(\mu_i(\vect\theta))
  \;,
  \label{eqn:Neffdata}
\end{equation}
where we explicitly show that the average can be written as a sum over all data bins. We furthermore defined a function $d(\mu)$, which can be interpreted as the effective number of ``degrees of freedom'' that correspond to a data bin with expectation value $\mu$, assuming Poisson noise. The sum of this function over all data bins gives the total number of \textit{effective} data bins, $N_\text{data}^\text{eff}$. We show that function in figure~\ref{fig:d}.  For $\mu\ll1$, one can show that it follows $2\mu\ln(1/\mu)$; for $\mu\gg1$ it converges to one.  We use a tabulated version of the function $d(\mu)$ to calculate $N_\text{data}^\text{eff}$ for each of the five runs.  The results are similar for all runs, with values around 620000, and shown in table~\ref{tab:fits}.

The number of effective model parameters can now be estimated in a similar way.  We define it as
\begin{equation}
  N_\text{param}^\text{eff}  \equiv
  N_\text{data}^\text{eff}
  -\left\langle -2\ln\mathcal{L}_P(\vect{\hat\theta})\right\rangle_{\mathcal{D}(\vect\theta)}\;,
\end{equation}
where $\vect{\hat\theta}$ are the maximum-likelihood estimators for the model parameters, and different for each data set the average is taken over. Note that $  \left\langle -2\ln\mathcal{L}_P(\vect{\hat\theta})\right\rangle_{\mathcal{D}(\vect\theta)}$ is simply equal to the DOF, $k$, which we define above. For each of the five runs, instead of performing the average, we in practice use only one mock realization to obtain $N_\text{param}^\text{eff}$ (the expected standard deviation is $\sim1100$ and not relevant for our discussion).  The resulting values are always significantly smaller than the nominal number of free model parameters, $N_\text{param}$, and summarized in table~\ref{tab:fits}.

We estimate the fit-quality of the five runs by using some `model fidelity' parameter,
\begin{equation}
  \mathcal{F} = \sqrt{-2\ln\mathcal{L}_P(\vect{\hat\theta})-k}\;.
\end{equation}
It can be interpreted as the expected statistical significance (in standard deviations) of any additional \emph{single-parameter} component that would bring the model at hand in perfect agreement with the data.  Since the expected significance is lower for a multi-parameter component, our model fidelity parameter $\mathcal{F}$ provides an upper limit on the significance at which additional unmodeled components could be identified in the data by improvement of the model.  We consider this quantity to be more useful than the standard $p$-values, which remain extremely small for all models considered in this work ($p \lll 10^{-315}$ for \run1 and $p=10^{-77}$ for \run5).

The model fidelities $\mathcal{F}$ for \run1--5 are summarized in table~\ref{tab:fits}. We find that the most drastic improvement is between \run1 and \run2, where the fit fidelity drops from 627 to 164. Subsequent refinements of the model appear to lead only to a relatively mild improvement of the fit quality.  The still relatively large value for $\run5$, 144, is likely dominated by residuals around bright point sources and boundary effects related to our treatment of the PSF.  We stress that the statistically significance for including additional model components with many parameters cannot be directly read off by comparing the improvement in $\mathcal{F}$ alone.

\subsection{Future directions and potential applications}
\label{sec:potential}

In this paper we present a novel method for fitting the gamma-ray sky that can
help us in deepening our understanding of gamma-ray emission mechanisms.  We
discuss here a number of possible or planned extensions of the current framework,
and some of the science questions that it can be used for to address.

\medskip

On the \emph{scientific} side, we are interested in full parameter
scans over GALPROP/DRAGON predictions for cosmic-ray diffusion and gamma-ray
emission.  To this end, we would leave both the propagation model parameters as
well as the relevant template modulation parameters free to vary.  One could
then study constraints on GALPROP/DRAGON parameters (like the halo height, the
diffusion coefficient, source distributions, etc) while simultaneously
accounting for, e.g., hard-to-model components like the \Fermi bubbles.
However, such a simultaneous fit is currently hindered by the long run-times of
GALPROP/DRAGON when generating high-resolution gamma-ray predictions for large
regions of the sky.

A combined analysis could be used, e.g., to study potential cosmic-ray
gradient, or substructure in the inverse Compton emission. This is for instance
relevant to understand whether there is really a hardening of the proton
spectrum suggested by the data, as previously claimed~\cite{Yang:2016jda, Acero:2016qlg}.
One particularly interesting aspect is the potential
observability of spiral arm structures in the gamma-ray emission of the Milky
Way.  It is well known that the distribution of CR sources as modeled in
numerical codes might not be the most realistic one. For example, we do know
that CR sources are expected to be distributed in spiral arms.  The impact of
spiral arms modeling has been discussed in the literature in the context of CR
observables~\cite{PhysRevLett.103.111302,Gaggero:2013rya,2015APh....64...18W}.
Also, the spiral arm dynamic can have an impact of secondary-to-primary
ratios~\cite{Benyamin:2016xcq}.  However, so far, the implications for gamma
rays observables remain largely unexplored, with a pioneering
work~\cite{2017MNRAS.466.3674N} showing that the spiral arm dynamic might have
a clear signature on gamma-ray pion spectrum.  Fitting both propagation model
parameters and nuisance parameters simultaneously would be likely important to
make sure that a detection of spiral arm structures is not just an effect of
mismodeling of gas.

Finally, an obvious application is the characterization of \Fermi bubbles at
low latitudes and of possible degeneracies with the Galactic center excess.  As
we saw above, we are able to reconstruct the bubbles template given their spectral
distribution, without including a specific spatial template. A more
detailed analysis of the degeneracy between the spectral and spatial characteristics
of the bubbles and the GCE would enable us to characterize very large
scale emission down to the center of the Galaxy and might provide a way to
disentangle various emission processes.

\medskip

On the \emph{technical} side, a full Bayesian extension of the current analysis
would be interesting and desirable.  To this end, Hamiltonian Monte Carlos are
particularly promising, since they make use of the available gradient
information.  In this framework, our sampling from the inverse Fisher
information to estimate errors could be replaced by directly drawing from the
proper posterior distributions of the model parameters.  However, a
sufficiently fine sampling of the posterior would require speeding up the code
significantly and/or using variational inference for the posterior
distributions. Another interesting technical extension would be the
parallelization of the code, which is, as mentioned above, currently hindered
by the difficulties of
parallelizing sparse matrix multiplications.

\section{Conclusions}
\label{sec:conclusions}

We presented a new hybrid approach for studying, modeling and decomposing diffuse gamma-ray emission, which we dubbed \textsc{SkyFACT} (\textbf{Sky} \textbf{F}actorization with \textbf{A}daptive \textbf{C}onstrained \textbf{T}emplates).  Our approach combines methods of image reconstruction and adaptive spatio-spectral template regression in one coherent framework, based on penalized Poisson likelihood regression.

We discussed in detail the implementation of our approach, as well as various solutions to technical challenges related to the high dimensionality of the optimization problem.  In particular, we showed how the L-BFGS-B algorithm can be used to find unique solutions in high-dimensional parameter spaces.  The largest example in this work has $1\times10^5$ parameters.  We showed that the optimization problem is, for scenarios considered here (and for sufficiently soft smoothing), convex, and hence has no non-global minima.  With conventional desktop computers, convergence can be reached in a few dozens of core hours.  We demonstrated how Cholesky decomposition of the Fisher matrix, which we use here as approximation to the inverse covariance matrix, can be used to sample parameter errors.  From these, uncertainties and correlations of various model components can be inferred.  This is an important step towards going beyond the typically adopted strategy of `bracketing uncertainties' with discrete model choices, although model assumptions still have to be made.  

We applied our approach to the gamma-ray emission from the inner Galactic disk, $|\ell|<90^\circ$ and $|b|<90^\circ$, using 7.6 years of \Fermi-LAT data binned in $0.5^\circ$ pixels and covering the energy range 0.34--228.65 GeV.  We presented a series of five increasingly complex models that show the challenges and typical residuals for conventional fits to the data, and the reduction of residuals when realistic nuisance  parameters are included in the fit.  The final, most complex model includes three gas rings, an ICS template, the IGRB, a spectral template for the \Fermi bubbles, a spatial template for the \Fermi Galactic center excess (based on INTEGRAL measurements of the 511 keV line), as well as \Fermi-LAT extended and point sources.  As a result, we can remove most of the residuals over much of the RoI and across all energies in the analysis.

We also performed a series of tests with synthetic data to explore the performance of the regularization used in our fits; see appendix~\ref{apx:synthetic}. In general, we found that moderate regularization in the form of either modulation or smoothing can effectively recover model components without signficant overfitting.

The main physics results of our analysis will be presented elsewhere.  However, we summarize a few findings:  
\begin{itemize}
\item We do not include dark gas corrections in the gas maps, but instead recover the missing dark gas from the nuisance parameters in the fit.  In general, the spatial nuisance parameters rescale the \textit{a priori} gas and ICS templates that we use from DRAGON by less than $\pm40\%$, which is enough to provide good fits to the data, and within the uncertainties of these components.  
\item We furthermore do not include a spatial template for the \Fermi bubbles, but instead recover their flux at high and lower Galactic latitudes.  We find that the bottom of the \Fermi bubbles is not coincident with the Galactic center, but lies at around $\ell\approx-2^\circ$, which was also observed by \cite{TheFermi-LAT:2017vmf}.  However, despite being bright, the emission is not very significant, and we cannot exclude that this is an extended unmodeled component of the foreground.  
\item For the Galactic center excess we use a spatial template based on the morphology of the Galactic center and bulge emission of the 511 keV line measured by INTEGRAL.  We find that the template is capable of removing the inner \Fermi Galactic center excess and is detected at about the $12\sigma$ level, and that it picks up a spectrum compatible with the combined emission of (young or millisecond) pulsars.

\end{itemize}

\paragraph*{Acknowledgements.}
We are indebted to D.~Gaggero for providing the package GammaSKY and helping with the GDE templates.  
F.C. thanks E.~Charles for valuable discussion.  C.W. and E.S.~acknowledge funding from an NWO Vidi fellowship.

\clearpage

\appendix

\section{Tests with synthetic data}
\label{apx:synthetic}

\subsection{Image reconstruction, MEM and smoothing}
\label{apx:circles}

\begin{figure}
  \centering
  \includegraphics[width=0.75\linewidth]{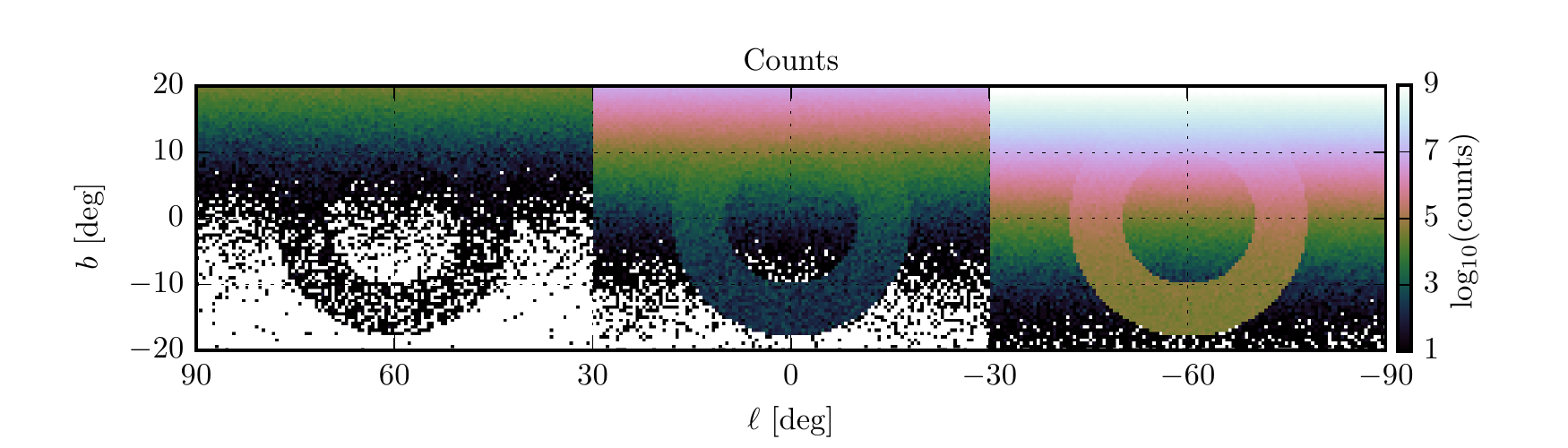}
  \caption{Counts map for the mock data set built from flat rings and an exponentially-varying background. The number of counts in each section, from left to right, total $10^5$, $10^6$, and $10^7$.}
  \label{fig:ring_counts}
\end{figure}

\begin{figure}
  \centering
  \includegraphics[width=0.74\linewidth]{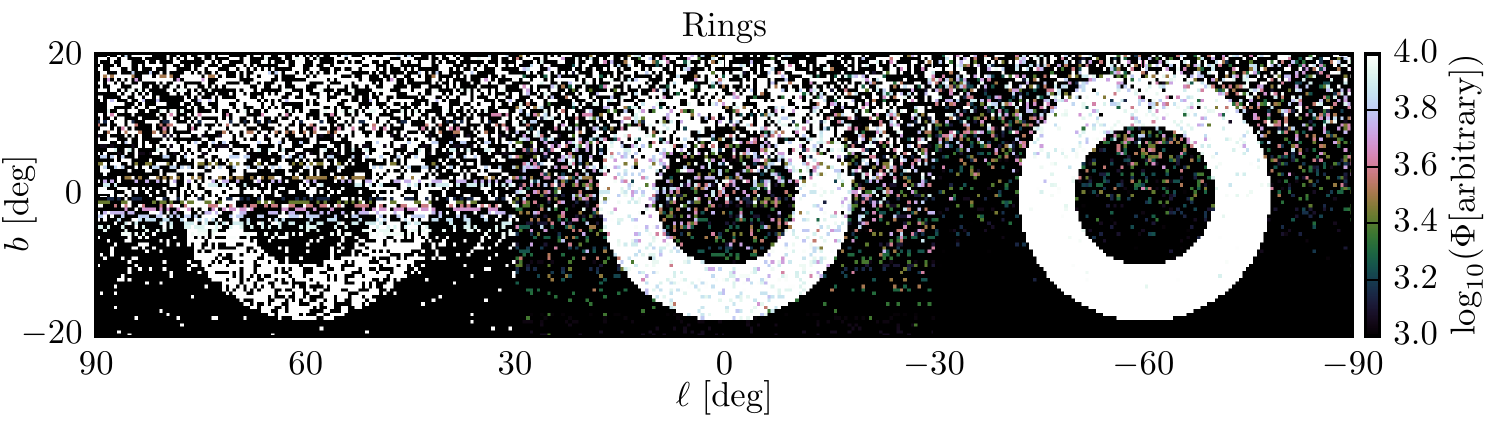}
  \includegraphics[width=0.25\linewidth]{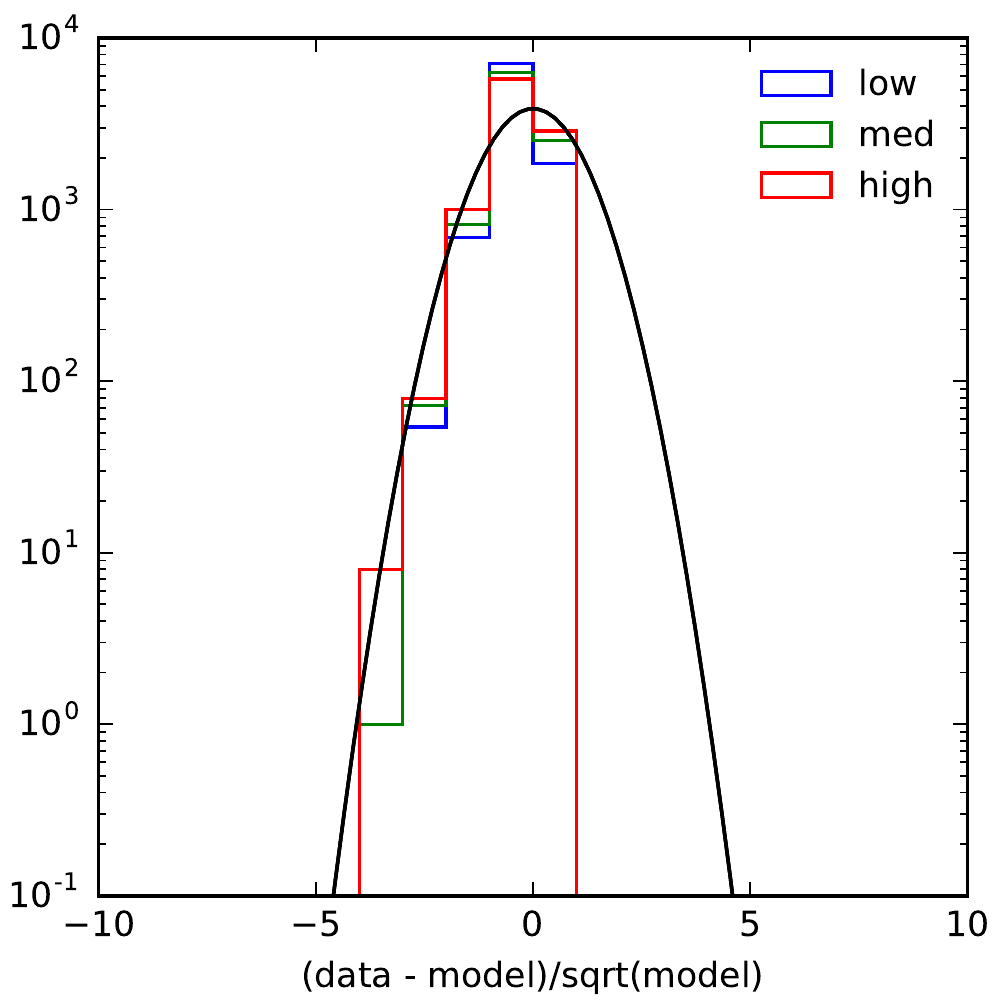}
  \includegraphics[width=0.74\linewidth]{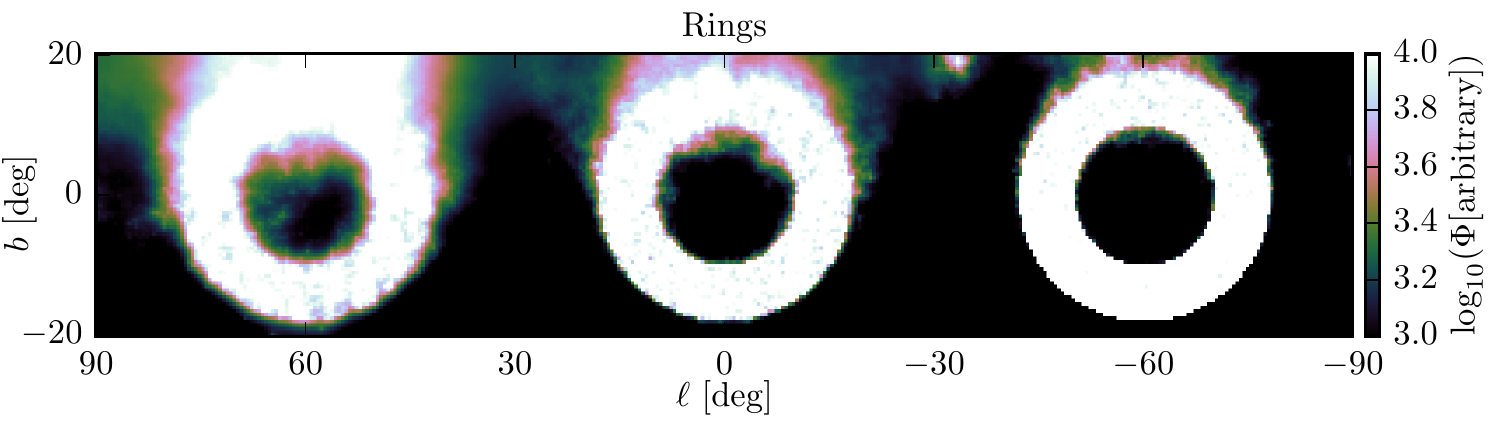}
  \includegraphics[width=0.25\linewidth]{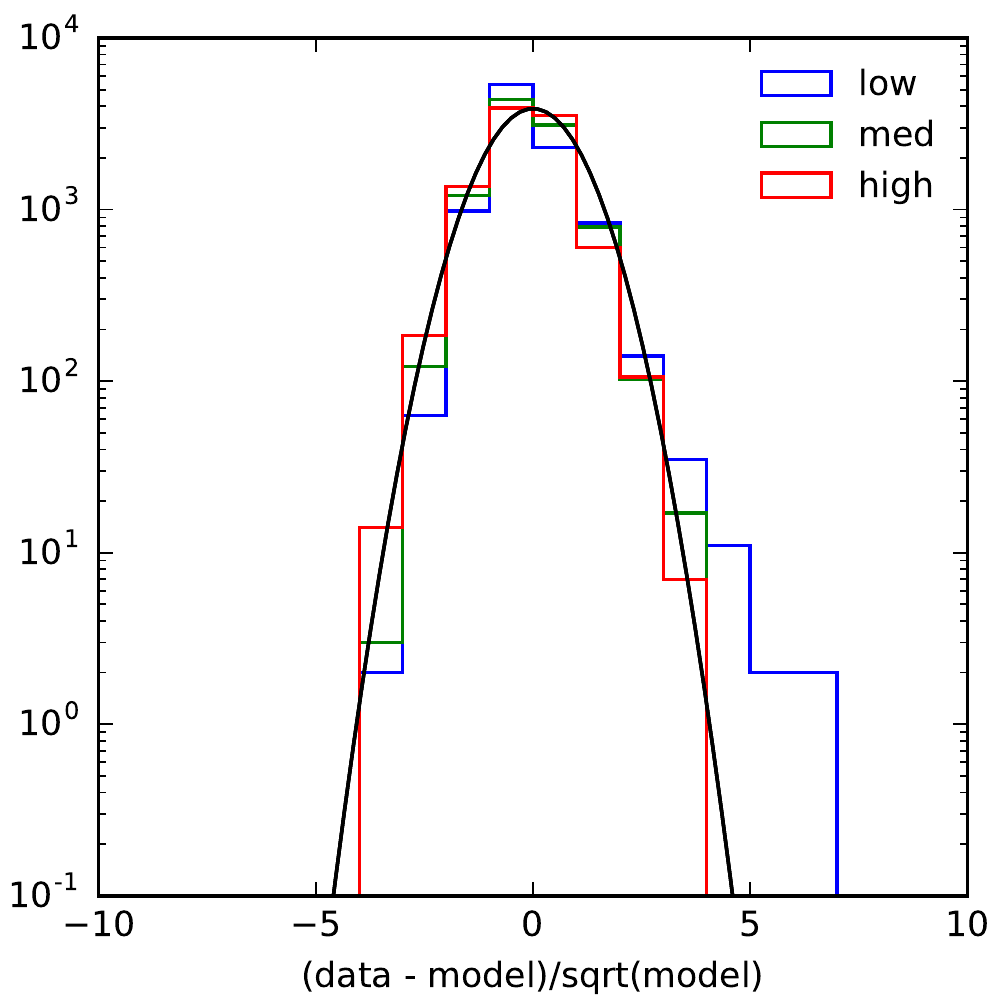}
  \includegraphics[width=0.74\linewidth]{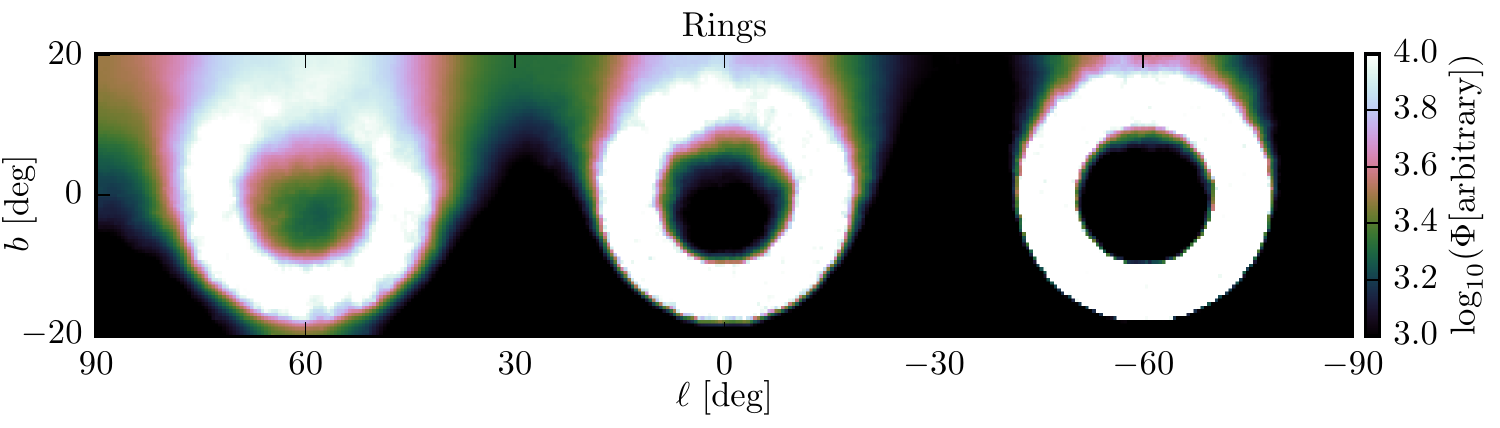}
  \includegraphics[width=0.25\linewidth]{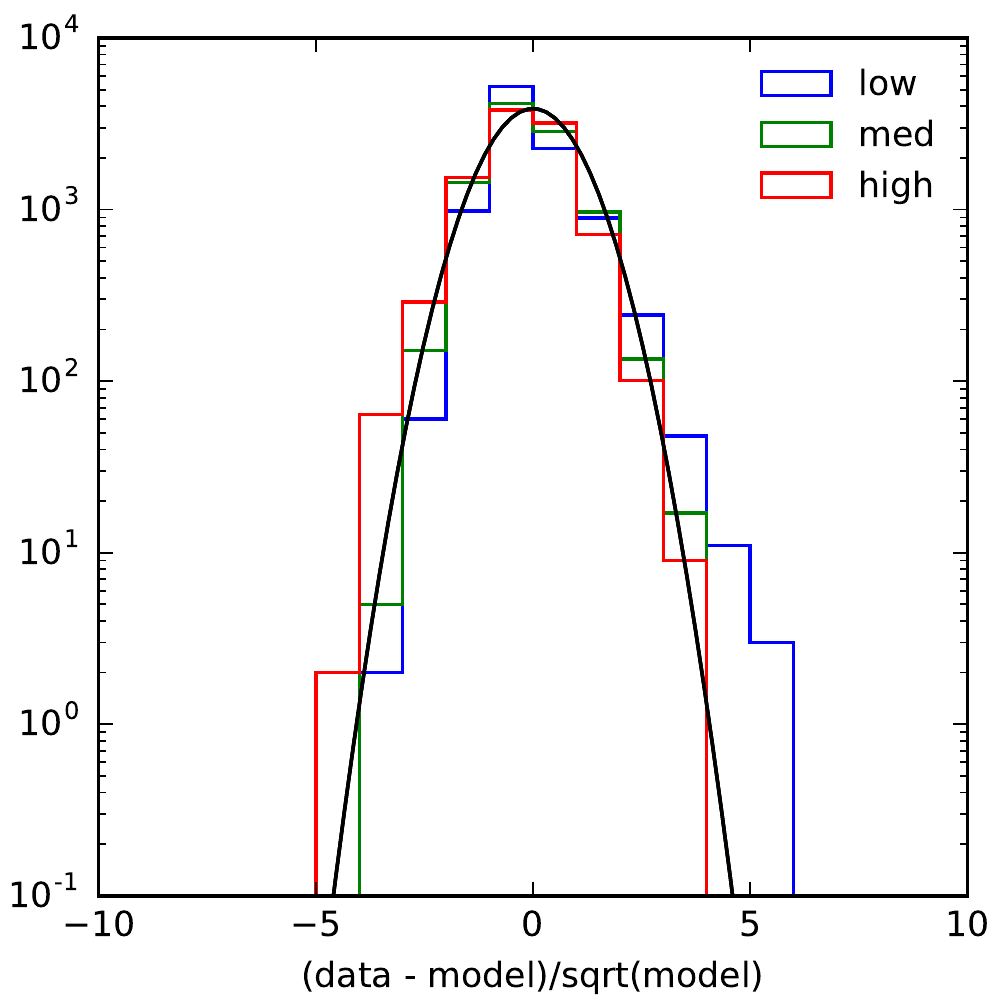}
  \includegraphics[width=0.74\linewidth]{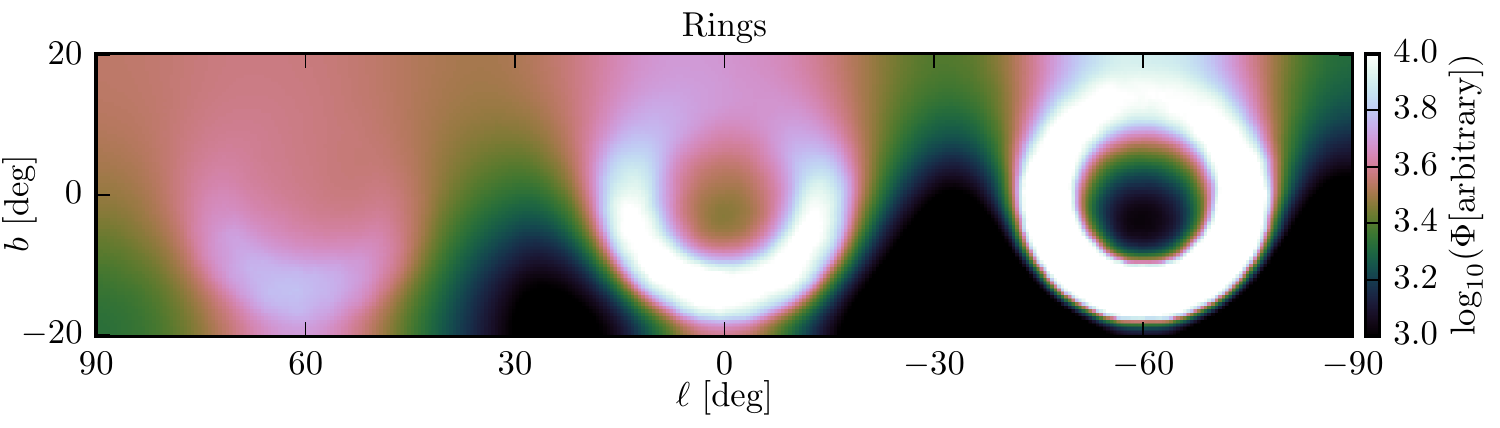}
  \includegraphics[width=0.25\linewidth]{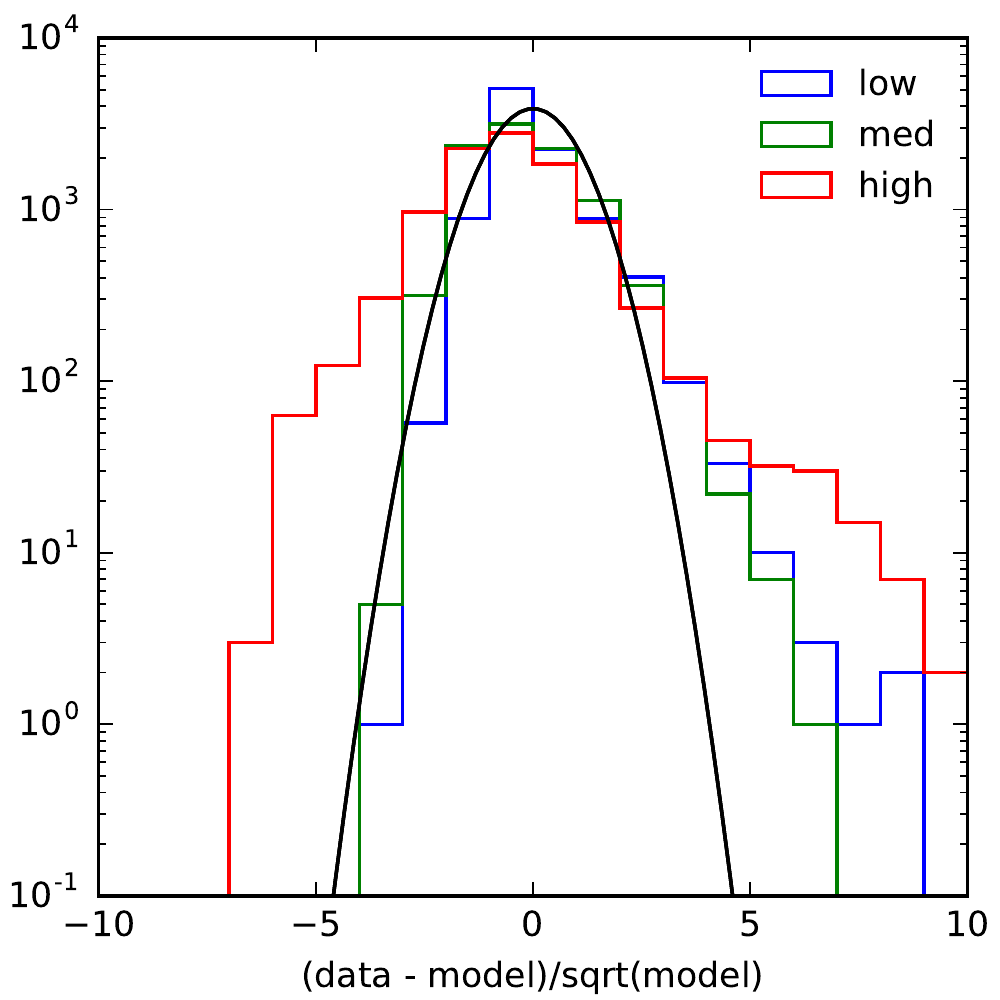}
  \caption{Results of fits to rings with three different total counts in each map, and four levels of smoothing applied to the ring template (from top to bottom: $\eta = 0$, $1$, $4$, $100$; $\lambda$ is set to $0$ for these runs.) The left 4 panels are the recovered ``flux'' maps for the ring template, in arbitrary units. The right four panels are histograms of the residuals. Each histogram plot contains three histograms that represent the residuals for each of the three exposures. The histograms marked ``low'' correspond to the left-most panel with $10^5$ counts, ``med'' correspond to the middle panel with $10^6$ counts, and ``high'' correspond to the right-most panel with $10^7$ counts.}
  \label{fig:ring_mean_hist}
\end{figure}

\begin{figure}
  \centering
  \includegraphics[width=0.74\linewidth]{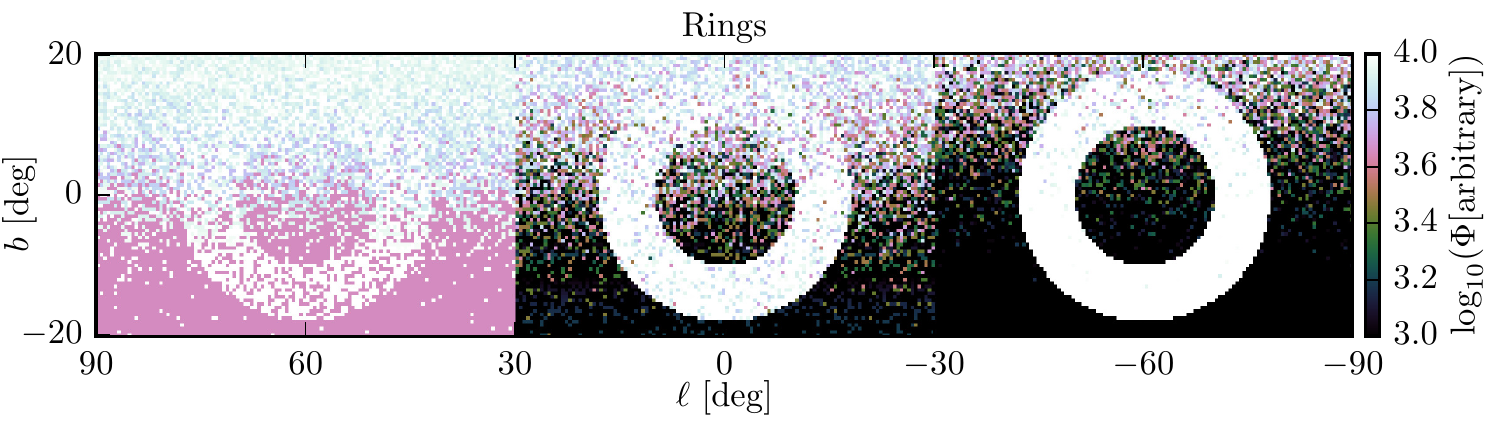}
  \includegraphics[width=0.25\linewidth]{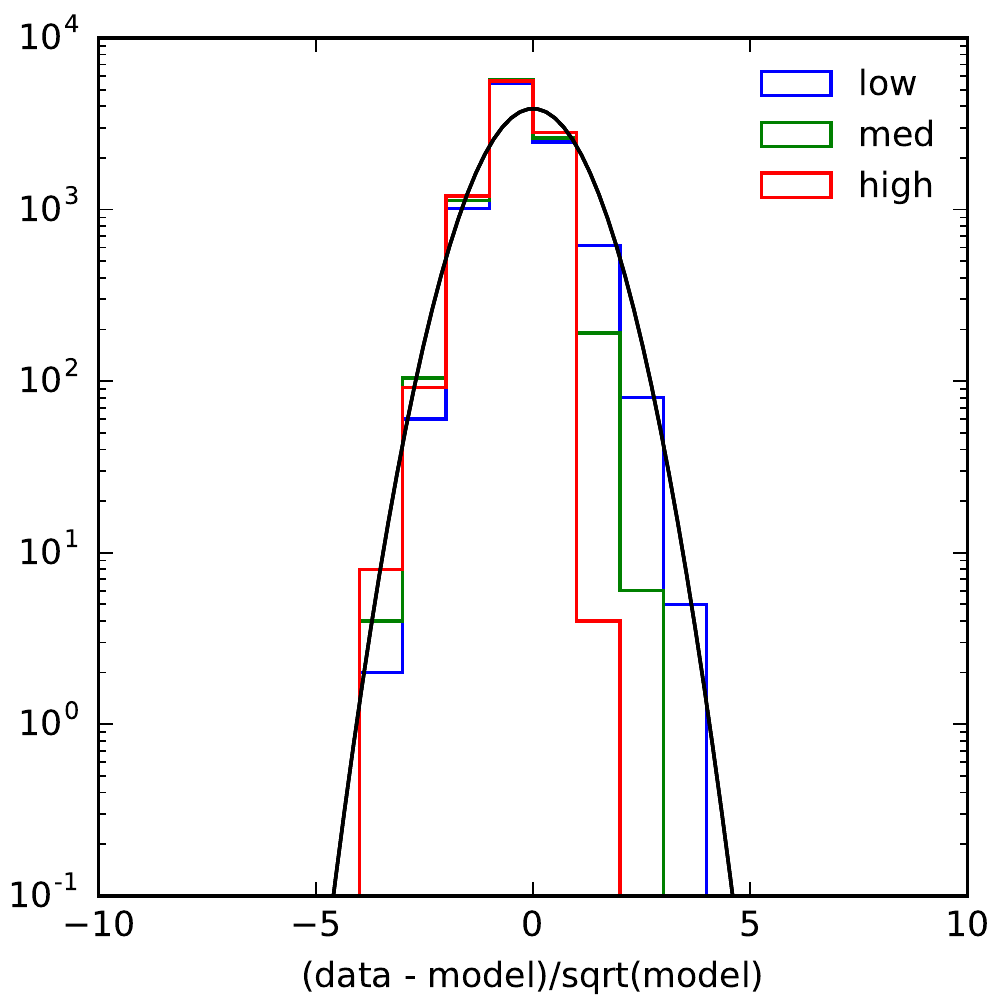}
  \includegraphics[width=0.74\linewidth]{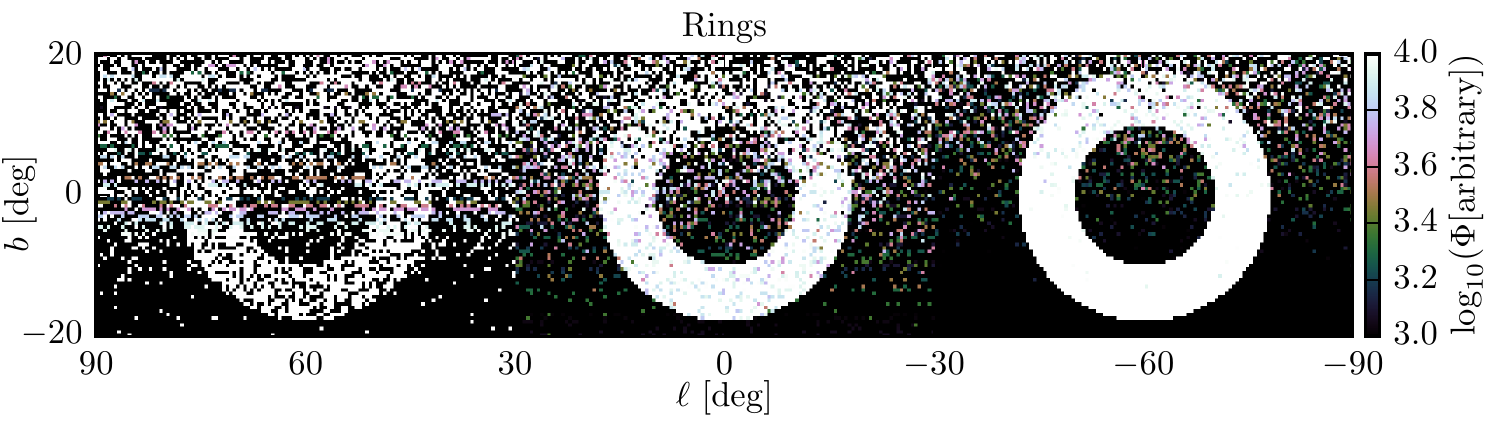}
  \includegraphics[width=0.25\linewidth]{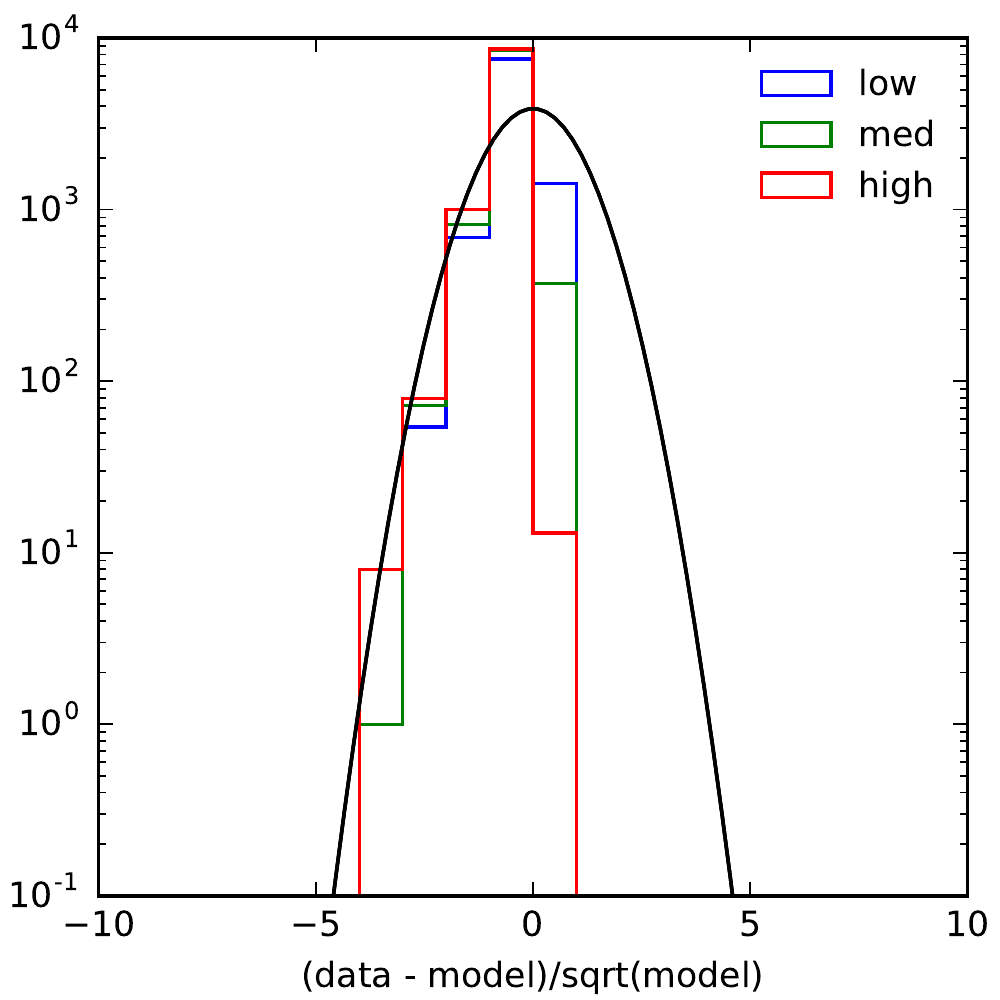}
  \includegraphics[width=0.74\linewidth]{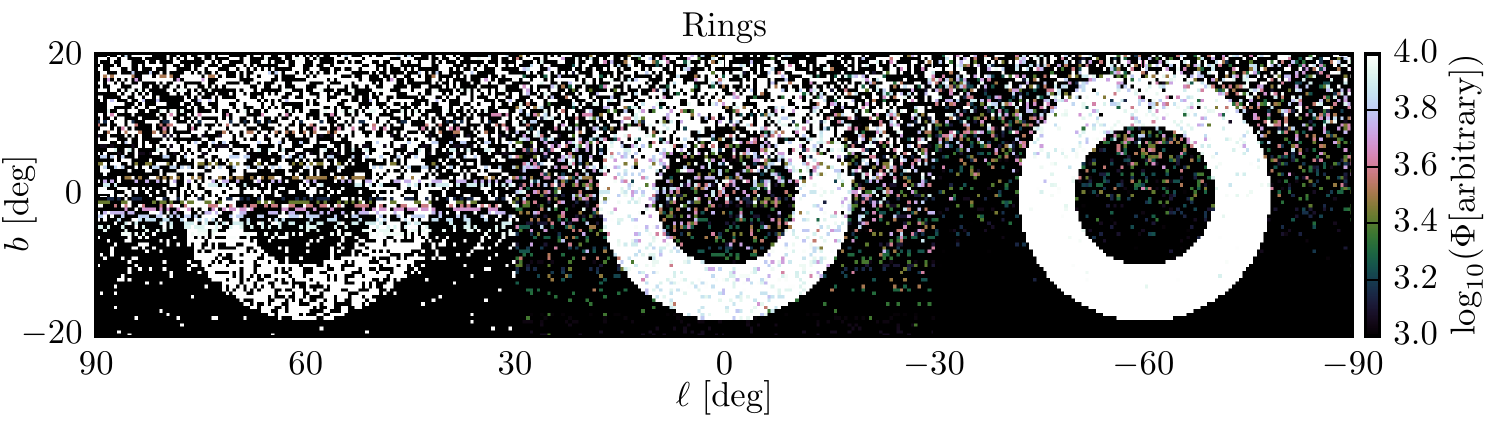}
  \includegraphics[width=0.25\linewidth]{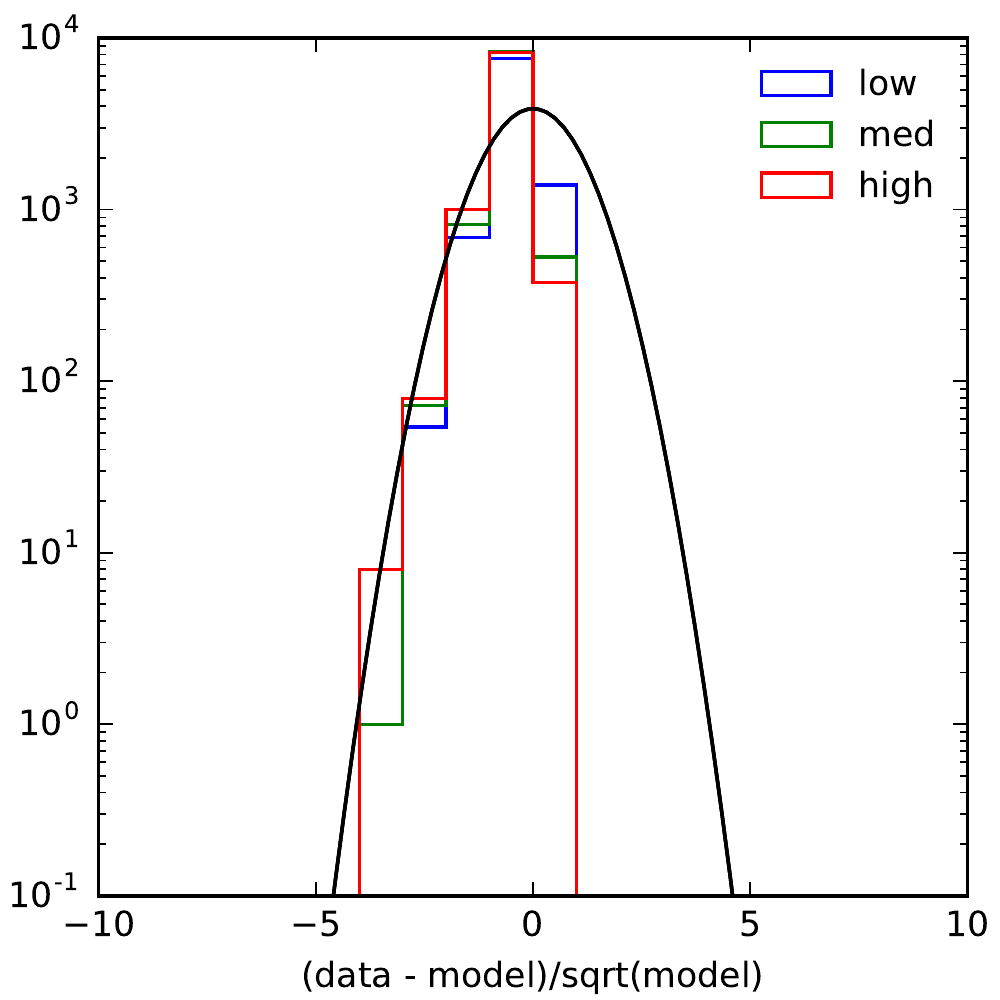}
  \caption{Results of fits to rings with three different total counts in each map, and three values of the template modulation parameter for ring template (from top to bottom: $\lambda = 1$, $10^{-2}$, $10^{-4}$; $\eta$ is set to $0$ for these runs.) The left 3 panels are the recovered ``flux'' maps for the ring template, in arbitrary units. The right four panels are histograms of the residuals. Each histogram plot contains three histograms that represent the residuals for each of the three exposures. The histograms marked ``low'' correspond to the left-most panel with $10^5$ counts, ``med'' correspond to the middle panel with $10^6$ counts, and ``high'' correspond to the right-most panel with $10^7$ counts.}
  \label{fig:ring_mean_hist_MEM}
\end{figure}

We first test both the effect of smoothing and MEM on image reconstruction for varying signal-to-background ratios and number of counts. We build a mock data set that consists of two spatial templates: a background that varies exponentially with latitude and is constant in longitude, and a set of three rings with flat, uniform intensity. No energy dependence is included in the data or fits. The ring and background templates are normalized such that the signal-to-background ratio is roughly equal to $10^{-2}$ at the top of the map ($b=20.25^{\circ}$), $10^{-1}$ at $b=10.125^{\circ}$, $1$ at $b=0^{\circ}$, $10$ at $b=-10.125^{\circ}$, and $10^{2}$ at the bottom ($b=-20.25^{\circ}$).

Poisson noise is added and the data is divided into three equal-longitude sections of $60^{\circ}$. The exposure in each section varies such that the total number of counts is different in each section. There are approximately $10^5$ in the left section, $10^6$, in the middle, and $10^7$ in the right section. In the real \Fermi-LAT data we use in our fits, there are roughly $5 \times 10^6$ photons in the lowest energy bin. At the top of the map, where the signal-to-background is small, the average number of background counts is $10^2$ in the left section, $10^3$ in the middle, and $10^4$ in the right section. Correspondingly, at the bottom where the signal-to-background is large, the average number of background counts is 0.01, 0.1, and 1 in the left, middle, and right sections. We show the counts map used for this analysis in figure~\ref{fig:ring_counts}.

To perform fits, we fix the background component completely. For the ring component, we initialize the template to be completely flat, and set the template modulation parameter to be essentially free ($\lambda = 0$). We perform four fits with increasing amounts of smoothing applied to the ring template ($\eta = 0$, $1$, $4$, and $100$).

We show the results of these runs in figure~\ref{fig:ring_mean_hist}. The top-left image shows the recovered ``flux'' map where no smoothing was used in the template reconstruction. It is obvious even by eye that some amount of background is being recovered in the ring template; the histograms of residuals for this run also indicate this. In general, the rings are more accurately recovered in regions of high signal-to-background and large number of counts, as seen in the bottoms of right-most sections of the ``flux'' maps. For regions with fewer photons, and low signal-to-background, the rings are less well-recovered, seen in the tops of the middle and left sections of the flux maps. However, moderate amounts of smoothing can help to recover the rings and suppress too much fitting of the Poisson flucuations; compare, for example, the left sections of the middle two flux maps. If the smoothing parameter is too large, the ring shape is washed out, as seen in bottom flux map, especially for low photon counts. The strong positive and negative residuals seen in the histogram for this run with the strongest smoothing are mostly the result of the edge of ring getting smeared out: there are strong positive residuals on the inside edge of the ring and strong negative residuals on the outside edge.

We perform a second set of runs without smoothing, but with varying template modulation parameters, to illustrate how MEM regularization alone influences image reconstruction. Again, we fix the background component and initialize the ring template to be flat. We vary the template modulation parameter in each fit ($\lambda = 1, 10^{-2}, 10^{-4}$). If $\lambda>1$, there is insufficient variation permitted in the template parameters to recover the ring flux.

These results are shown in figure~\ref{fig:ring_mean_hist_MEM}. In the top panel, where the template modulation is strongest, there are still significant remaining residuals. Essentially, this template modulation is almost too strong, especially in the low-photon section, to allow a flat template to completely reconstruct the ring flux. For weaker template modulation, there are very few positive residuals, indicating that the ring flux is fully recovered (this is evident from the residual maps, not shown, as well.) The negative residuals that remain are the result of the flat template not being able to go completely to zero; the rescaling for most of those pixels is at their smallest allowed value ($10^{-10}$; this is also true for the runs with smoothing). However, also evident is some amount of Poisson noise being picked up by this template, seen especially in the high-background-count upper regions of the flux maps. As seen in the previous set of runs, even a weak smoothing constraint applied here would drastically reduce the amount of fitted Poisson noise, especially in low-photon regions.

\subsection{Component separation and Poisson-induced bias with two components}
\label{apx:two}

\begin{table}[t]
    \centering
    \begin{tabular}{cccc}
        \toprule
        Components & free & constrained & smoothed  \\
                   & \multicolumn{2}{c}{Regularization hyper-parameters: \sconf{\lambda}{\lambda'}{\lambda''}{\eta}{\eta'} }\\
        \midrule
        Gas             & \sconf{1}{1}000
                        & \sconf{4}{4}000
                        & \sconf{1}{1}0{100}0
        \\[0.1cm]
        Inverse Compton & \sconf{1}{1}000
                        & \sconf{4}{4}000
                        & \sconf{1}{1}0{100}0 \\
        \bottomrule
    \end{tabular}
    \caption{Hyper-parameters for mock 2-component runs.}
    \label{tab:pi0_ICS_mock}
\end{table}

\begin{figure}
    \centering
    \includegraphics[width=0.6\linewidth]{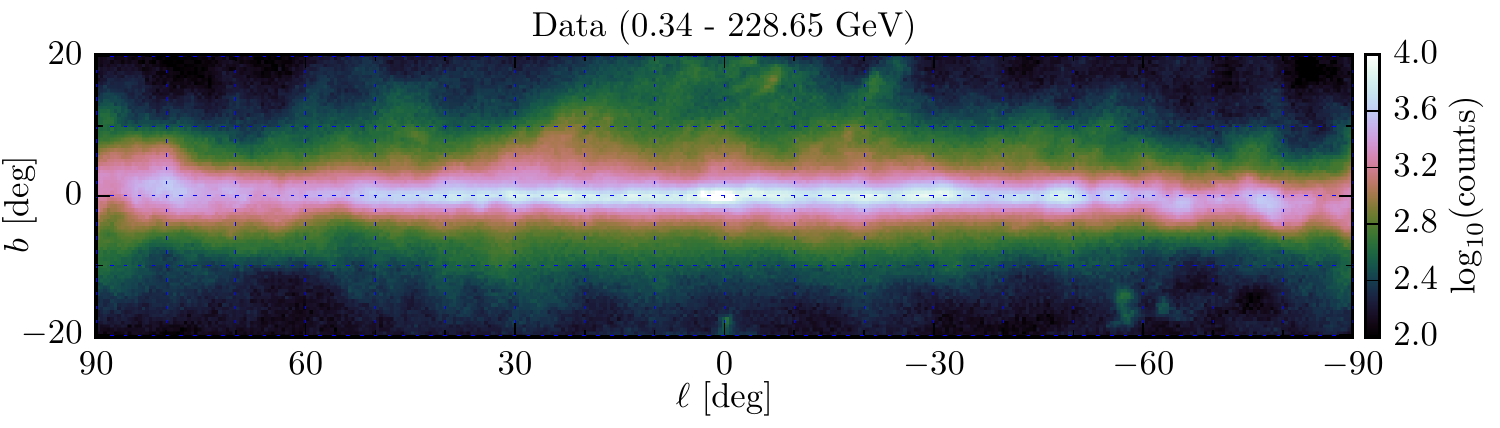}
    \includegraphics[width=0.4\linewidth]{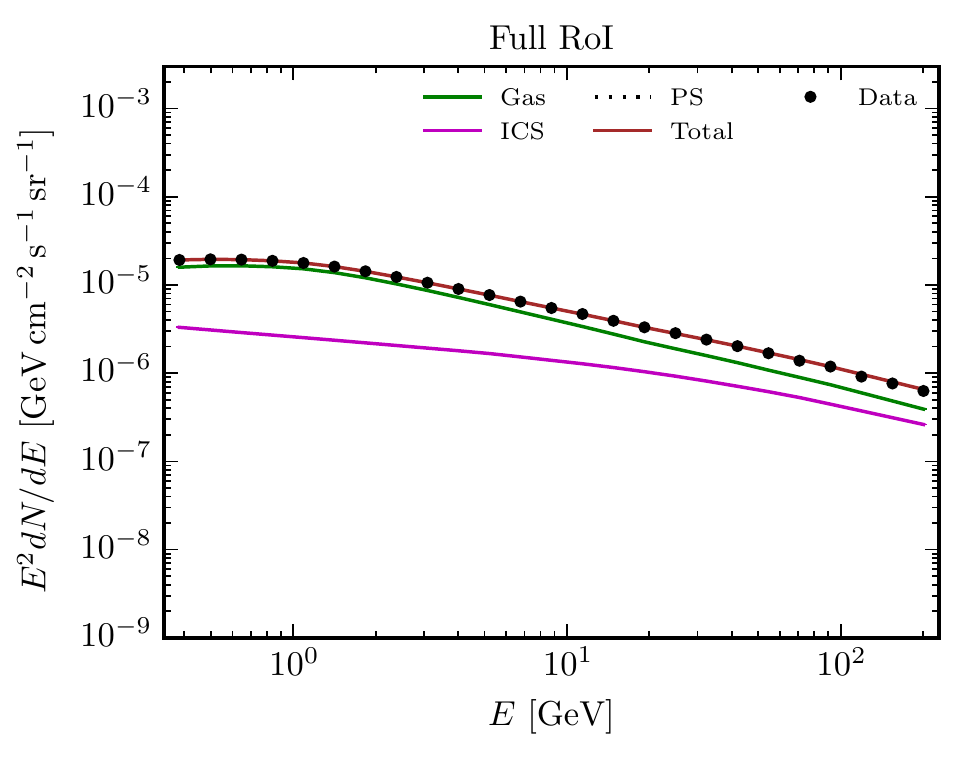}
    \caption{Top: counts map, summed over all energies. Bottom: spectral model components.}
    \label{fig:2comp_initial}
\end{figure}

For the next set of mock data runs, we wish to test how fits perform at separating model components (both templates and spectra) with different values for smoothing and template modulation parameters. For these tests, we build a mock data set with two components: the gas and ICS components described in section~\ref{sec:gde}. Here we also include energy dependencies for both components, using the spectra described in section~\ref{sec:gde} for the gas and ICS components. The energy range we fit over is the same as for the real LAT data: $0.34-228.65$~GeV. We choose these two components because they are roughly similar in overall brightness in our RoI, have similar spectral shapes at high energies, and share some morphological characteristics. The diffuse emission from the gas and ICS also dominate the real LAT data we consider. We show the mock data counts map and the initial spectral energy distribution in figure~\ref{fig:2comp_initial}. 

\begin{figure}
  \begin{minipage}{0.45\textwidth}
    \includegraphics[width=0.8\linewidth]{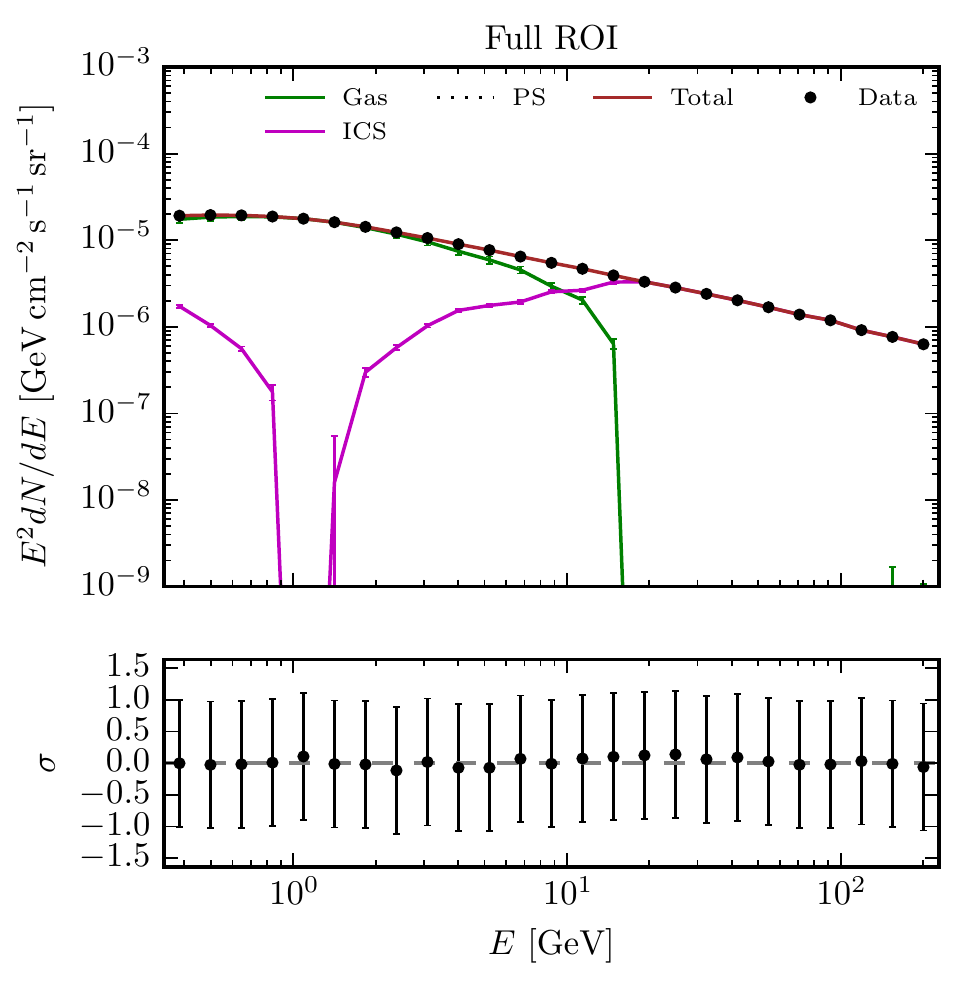}
  \end{minipage}
  \begin{minipage}{0.55\textwidth}
    \includegraphics[width=0.99\linewidth]{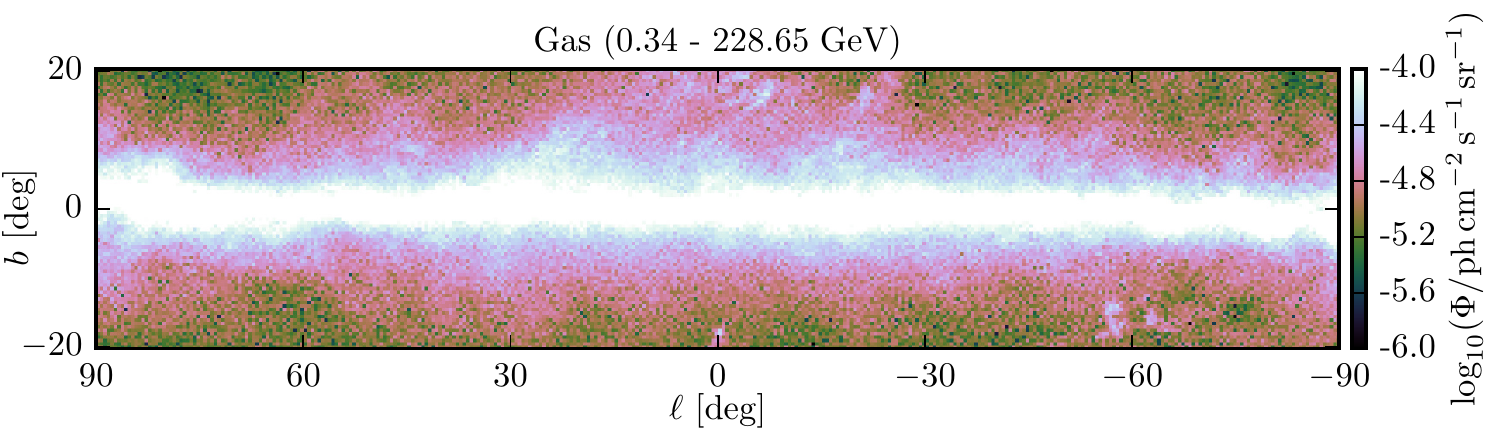}
    \includegraphics[width=0.99\linewidth]{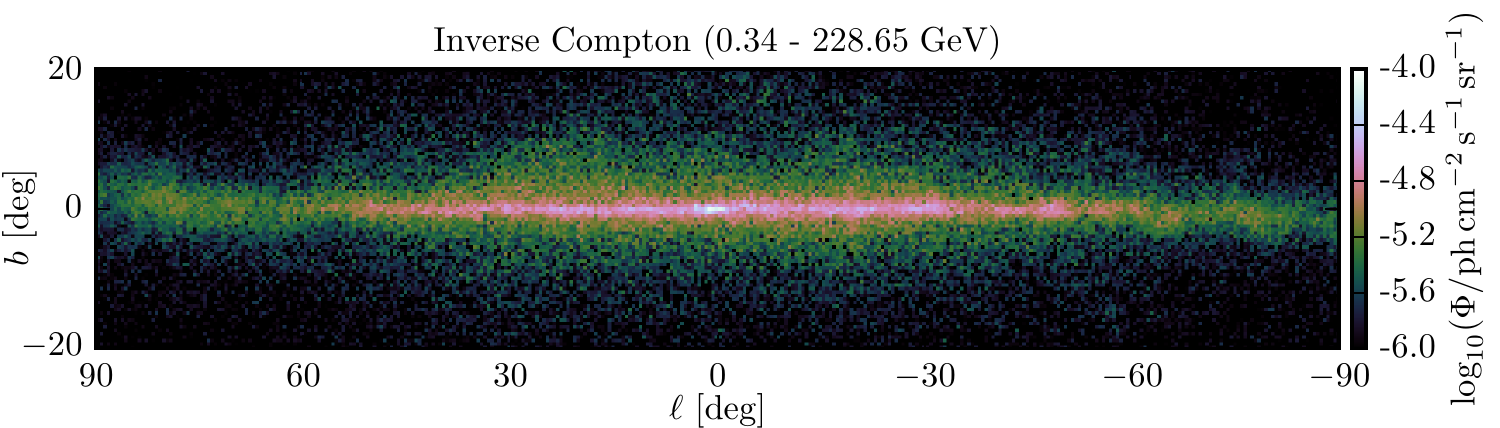}
  \end{minipage}
  \caption{Results for mock 2-component run (labeled ``free'' in table~\ref{tab:pi0_ICS_mock}) with moderate template and spectra regularization terms and no smoothing ($\lambda,\lambda' = 1$, $\eta = 0$).}
  \label{fig:2comp_sm0p0_t1_s1}
\end{figure}

\begin{figure}
  \begin{minipage}{0.45\textwidth}
    \includegraphics[width=0.8\linewidth]{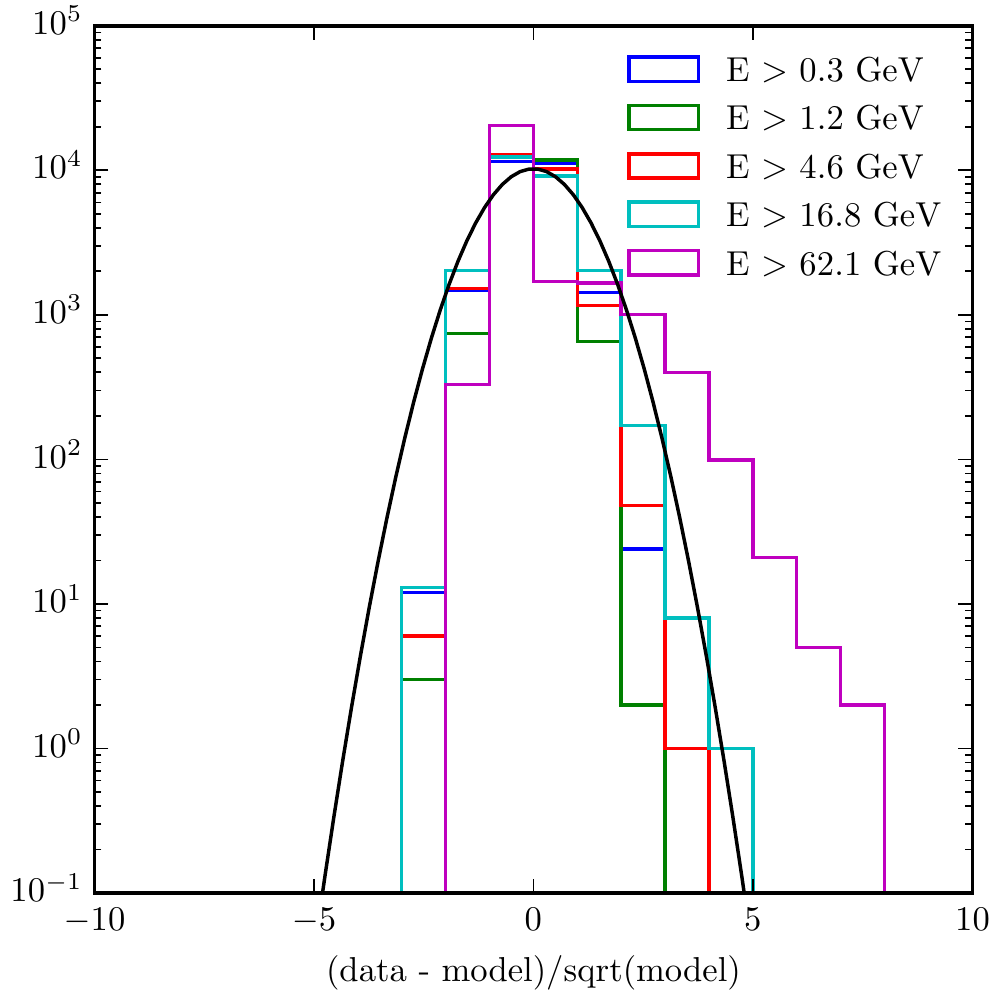}
  \end{minipage}
  \begin{minipage}{0.55\textwidth}
    \includegraphics[width=0.99\linewidth]{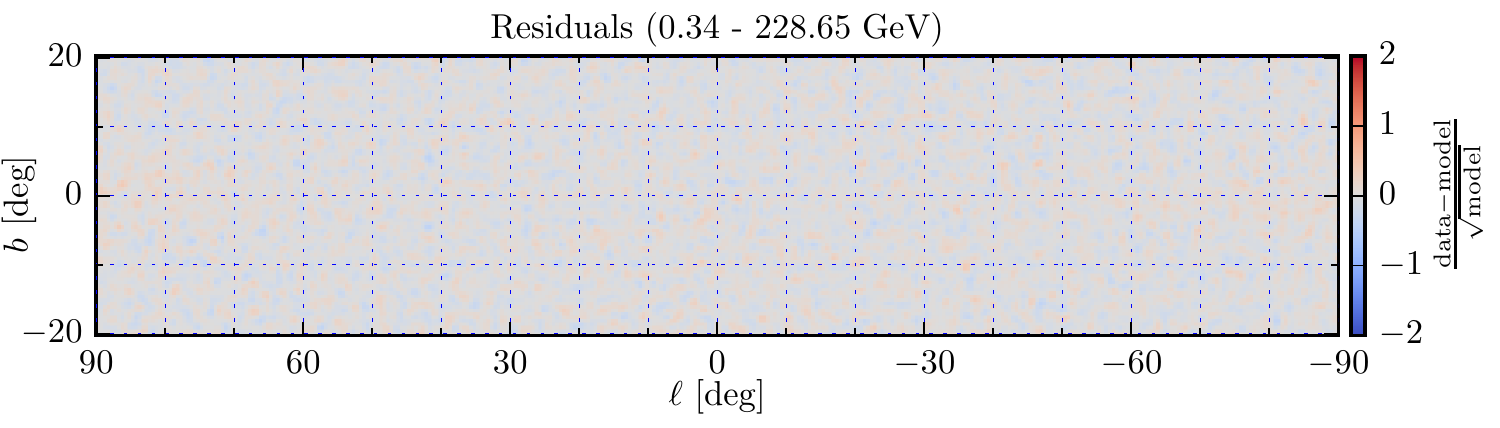}
    \includegraphics[width=0.99\linewidth]{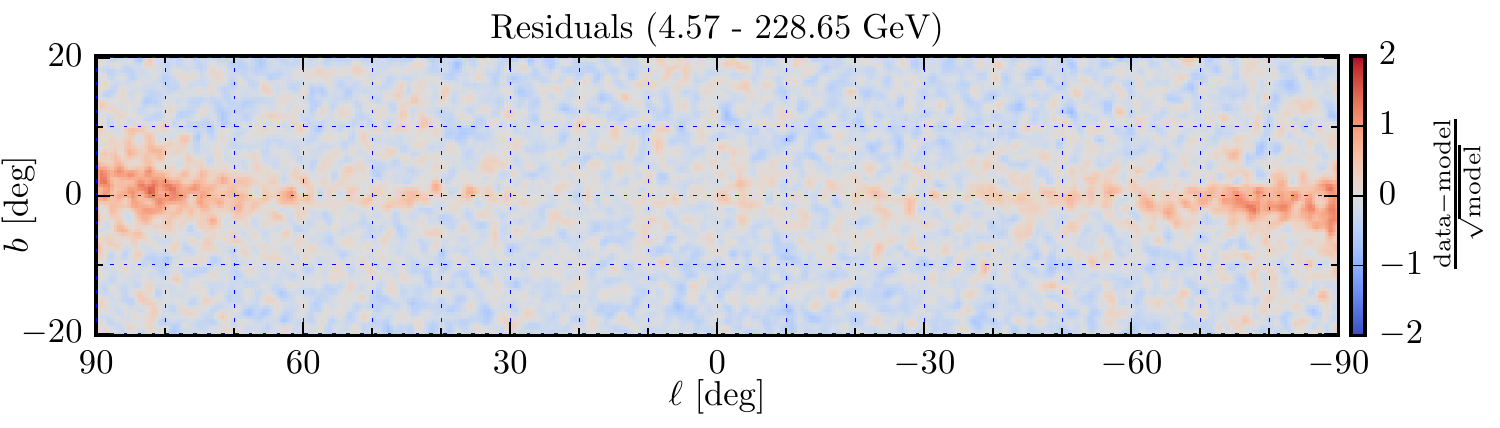}
  \end{minipage}
  \caption{Residuals for mock 2-component run (labeled ``free'' in table~\ref{tab:pi0_ICS_mock}) with stronger template and spectra regularization, and no smoothing ($\lambda,\lambda' = 1$, $\eta = 0$). Residual maps and histograms are in units of signficance.}
  \label{fig:2comp_sm0p0_t1_s1_res}
\end{figure}

\begin{figure}
  \centering
  \includegraphics[width=0.32\linewidth]{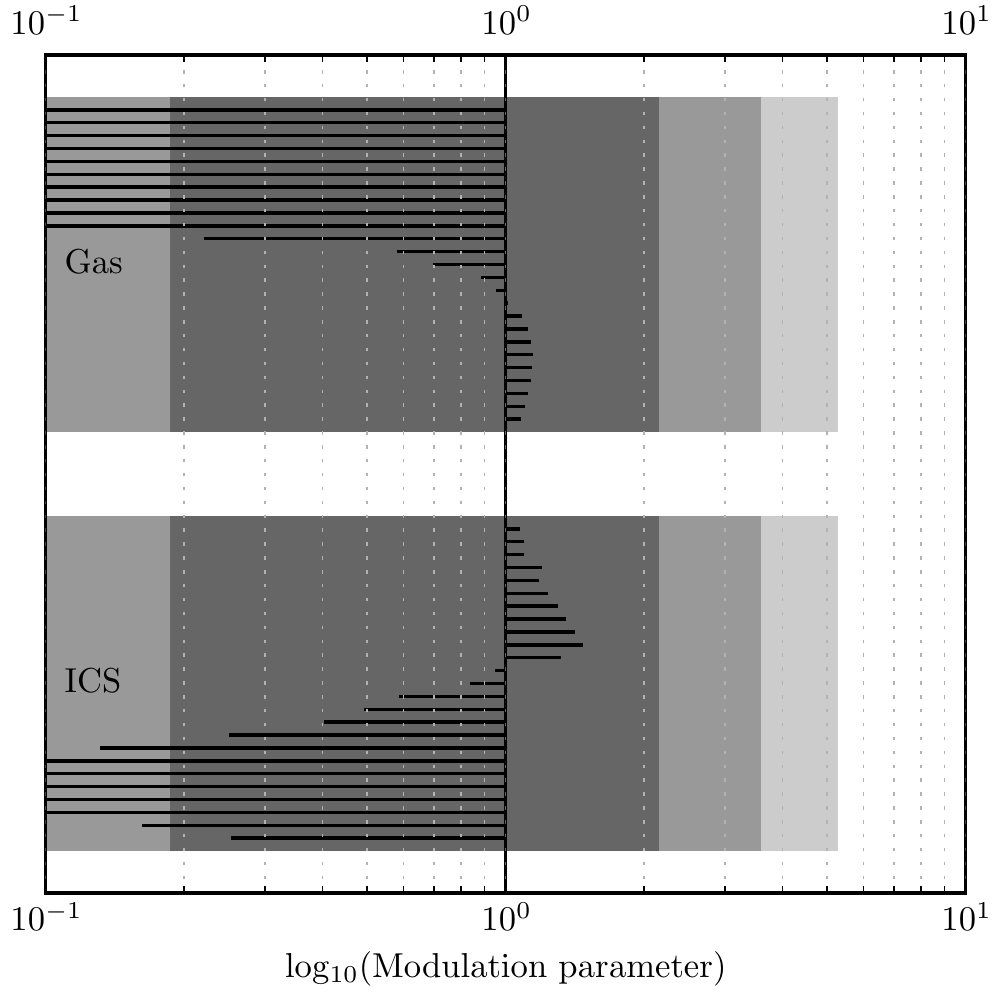}
  \includegraphics[width=0.32\linewidth]{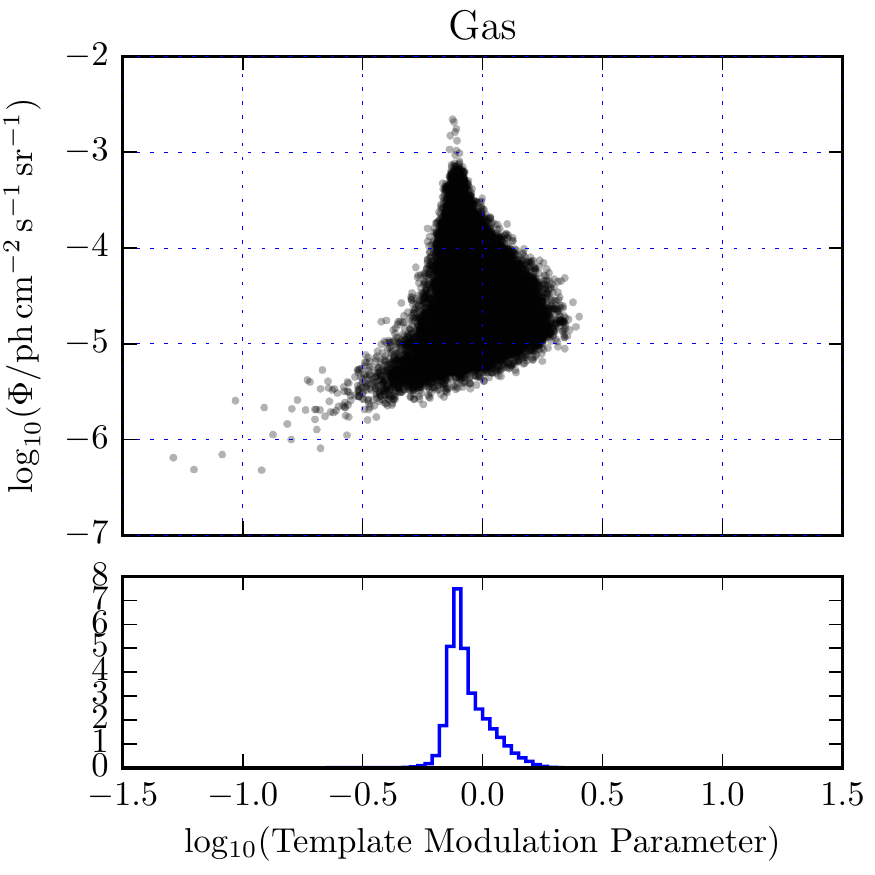}
  \includegraphics[width=0.32\linewidth]{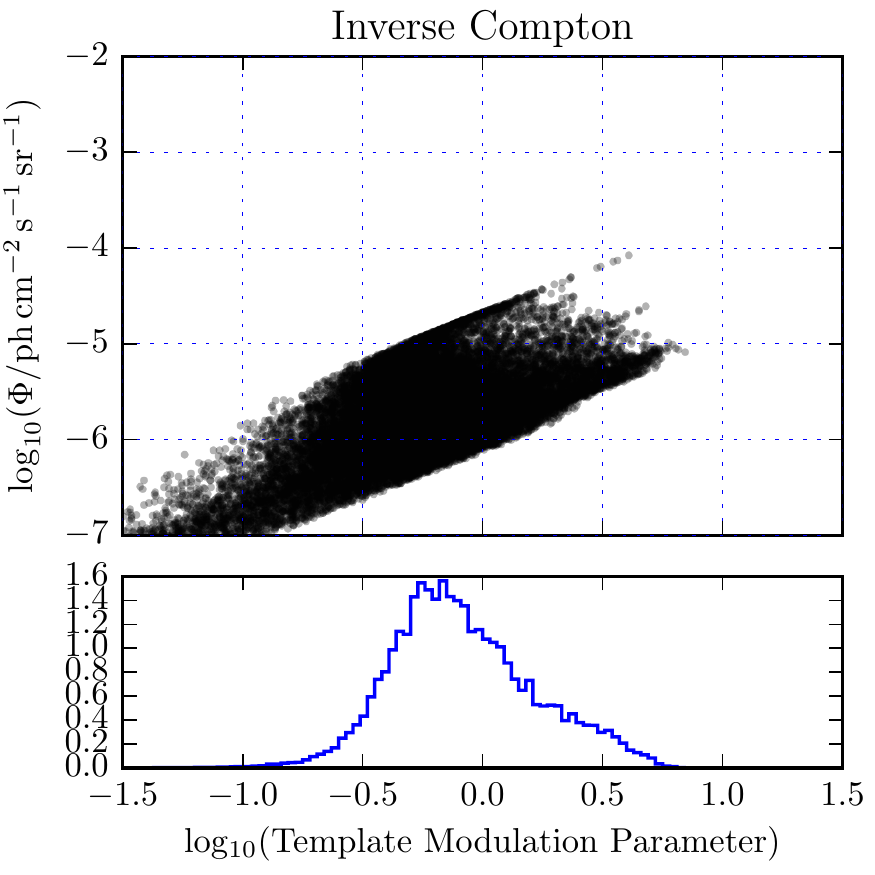}
  \caption{Rescaling parameters for spectra (left panel), and templates (center and right panels) for mock 2-component run (labeled ``free'' in table~\ref{tab:pi0_ICS_mock}) with essentially free templates and spectra, and strong smoothing ($\lambda,\lambda' = 1$, $\eta = 0$).}
  \label{fig:2comp_sm0p0_t1_s1_rsc}
\end{figure}

\begin{figure}
  \begin{minipage}{0.45\textwidth}  
    \includegraphics[width=0.8\linewidth]{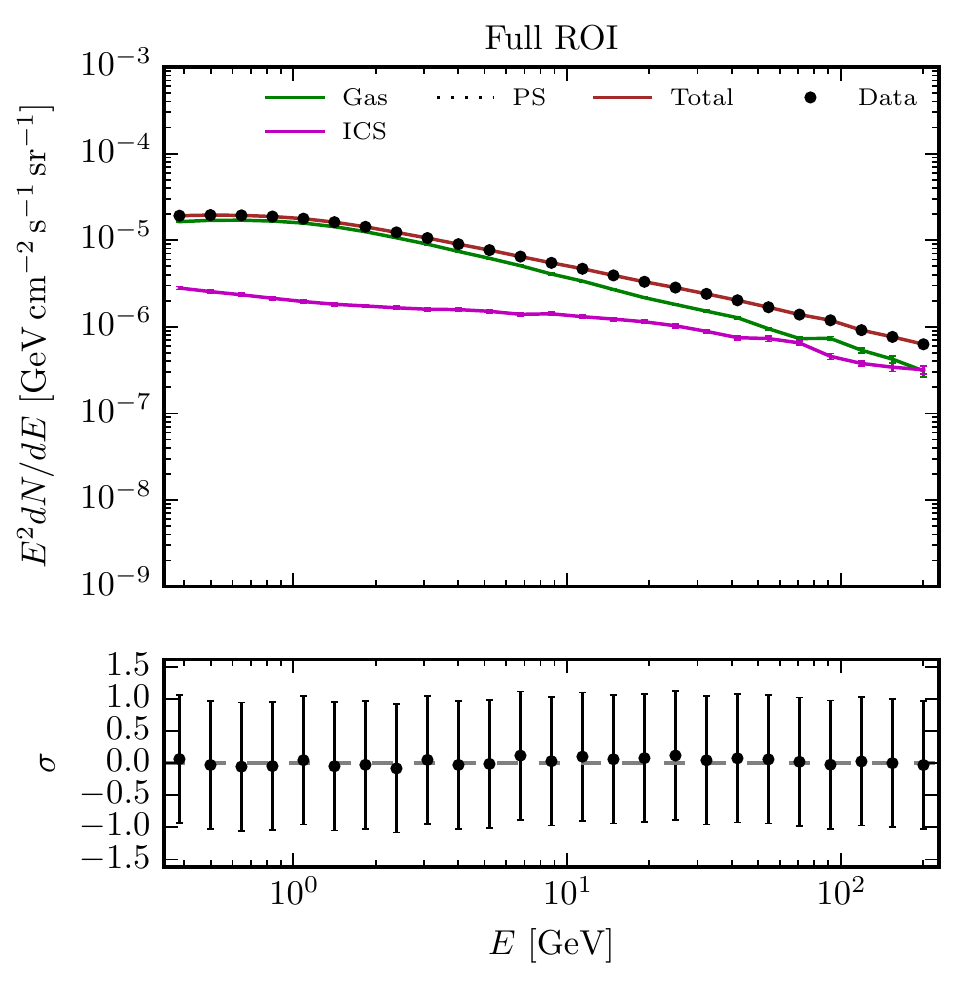}
  \end{minipage}
  \begin{minipage}{0.55\textwidth}  
    \includegraphics[width=0.99\linewidth]{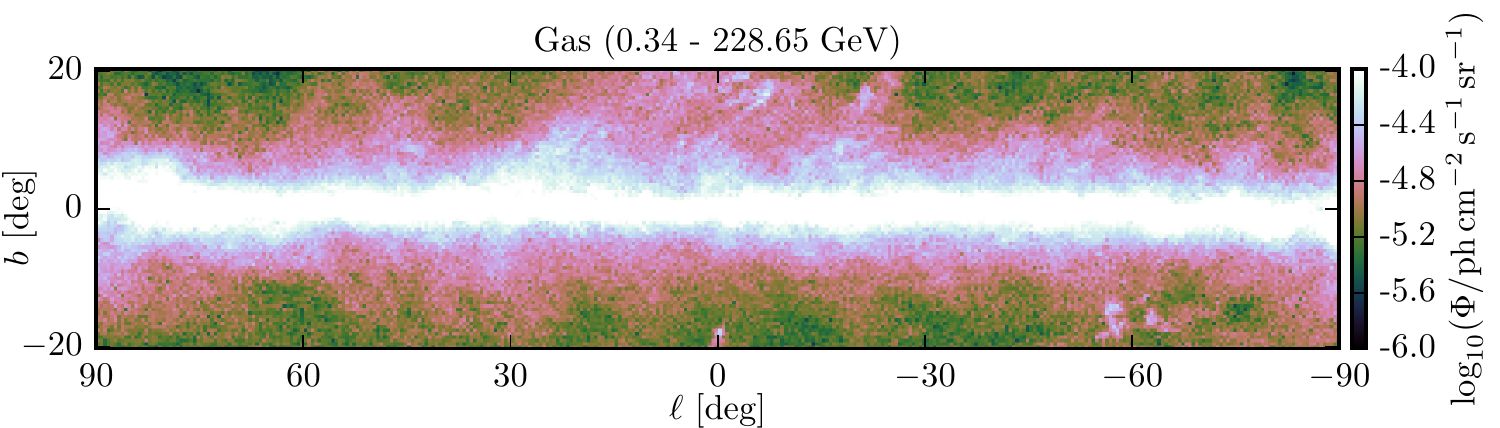}
    \includegraphics[width=0.99\linewidth]{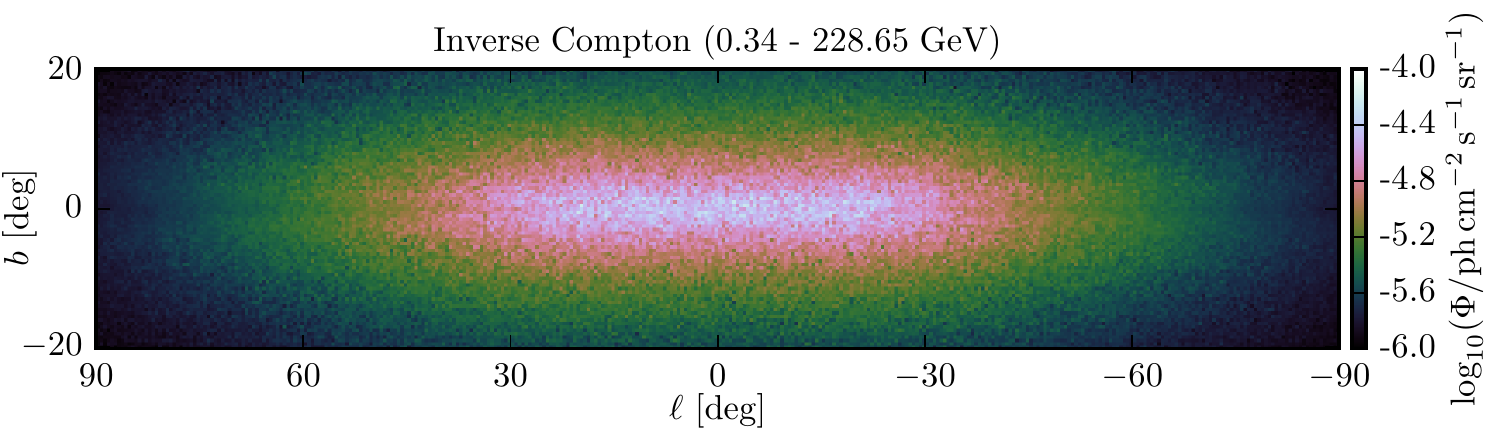}
  \end{minipage}
  \caption{Results for mock 2-component run (labeled ``constrained'' in table~\ref{tab:pi0_ICS_mock}) with stronger template and spectra regularization, and no smoothing ($\lambda,\lambda' = 4$, $\eta = 0$).}
  \label{fig:2comp_sm0p0_t0p5_s0p5}
\end{figure}

\begin{figure}
  \begin{minipage}{0.45\textwidth}
    \includegraphics[width=0.8\linewidth]{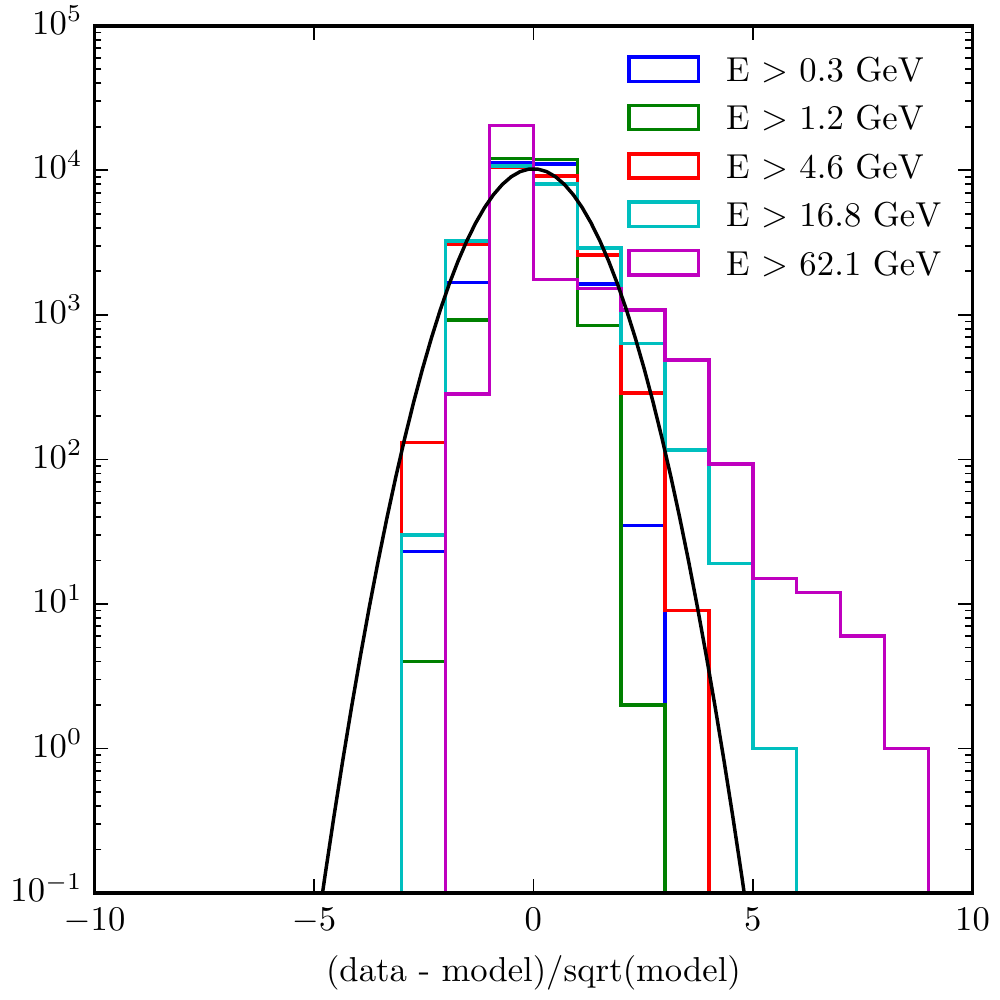}
  \end{minipage}
  \begin{minipage}{0.55\textwidth}
    \includegraphics[width=0.99\linewidth]{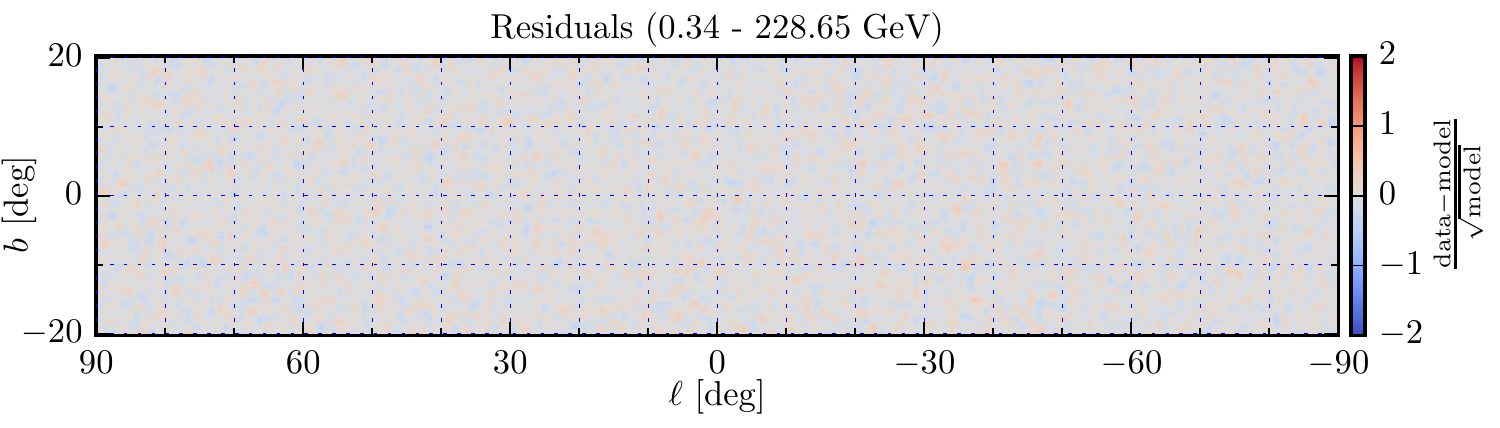}
    \includegraphics[width=0.99\linewidth]{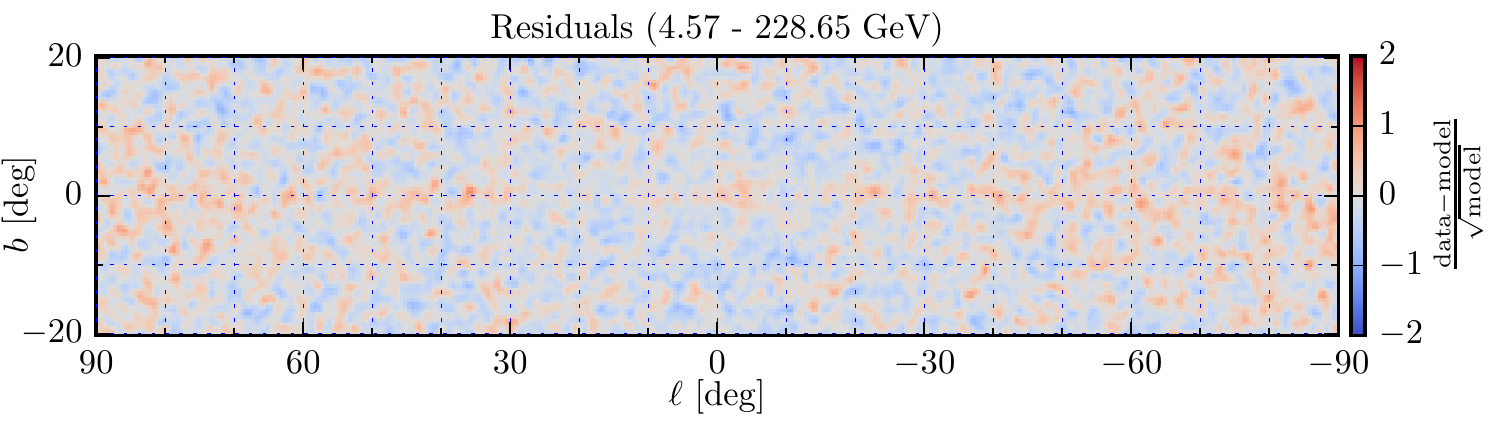}
  \end{minipage}
  \caption{Residuals for mock 2-component run (labeled ``constrained'' in table~\ref{tab:pi0_ICS_mock}) with stronger template and spectra regularization, and no smoothing ($\lambda,\lambda' = 4$, $\eta = 0$). Residual maps and histograms are in units of signficance.}
  \label{fig:2comp_sm0p0_t0p5_s0p5_res}
\end{figure}

\begin{figure}
  \centering
  \includegraphics[width=0.32\linewidth]{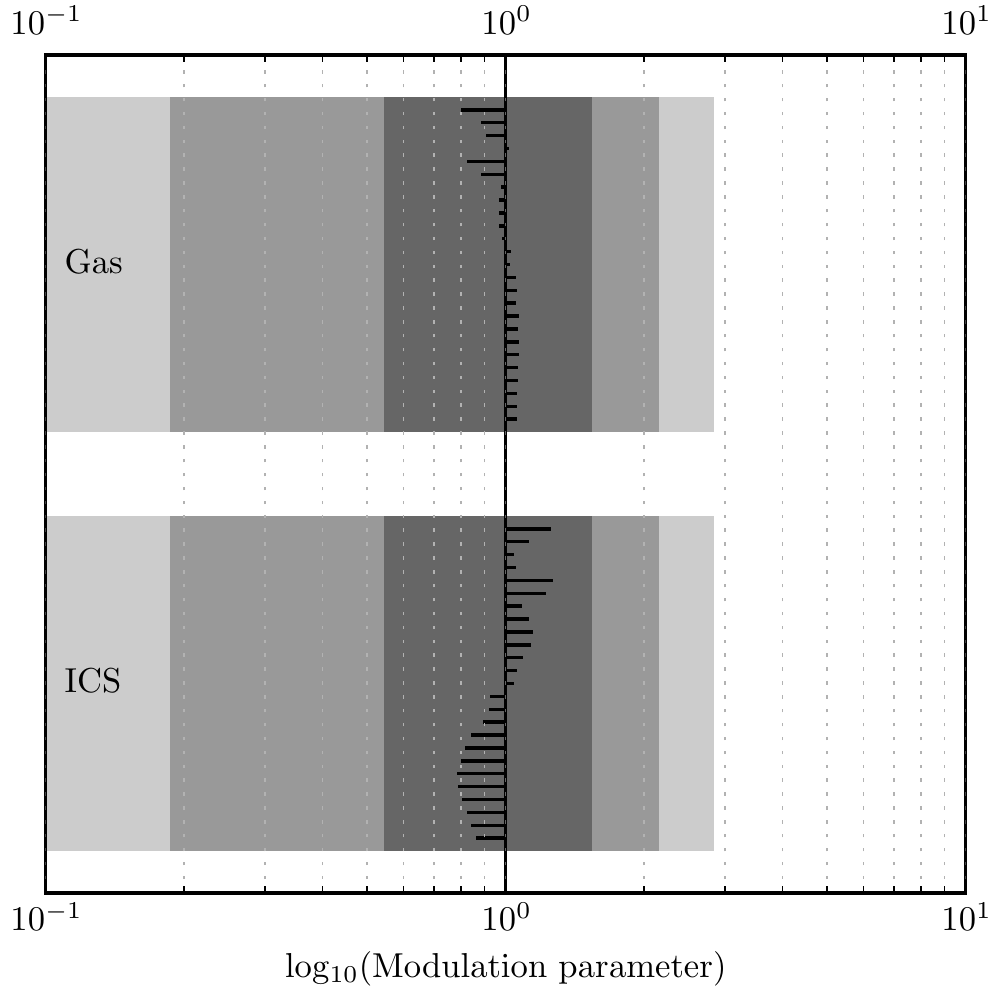}
  \includegraphics[width=0.32\linewidth]{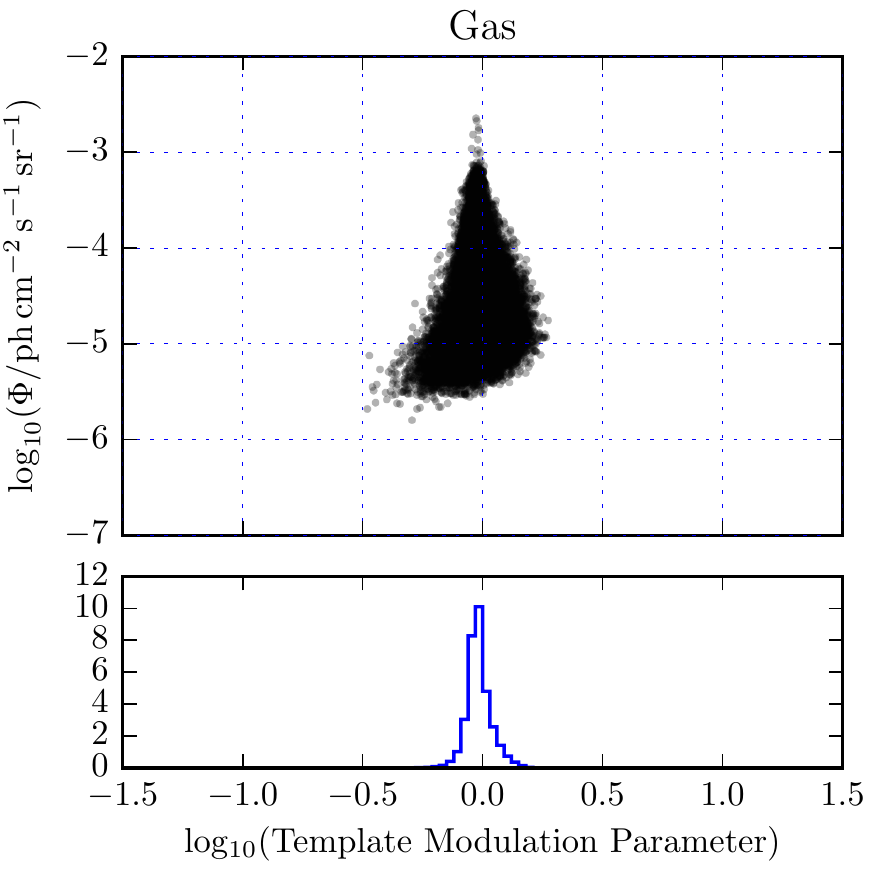}
  \includegraphics[width=0.32\linewidth]{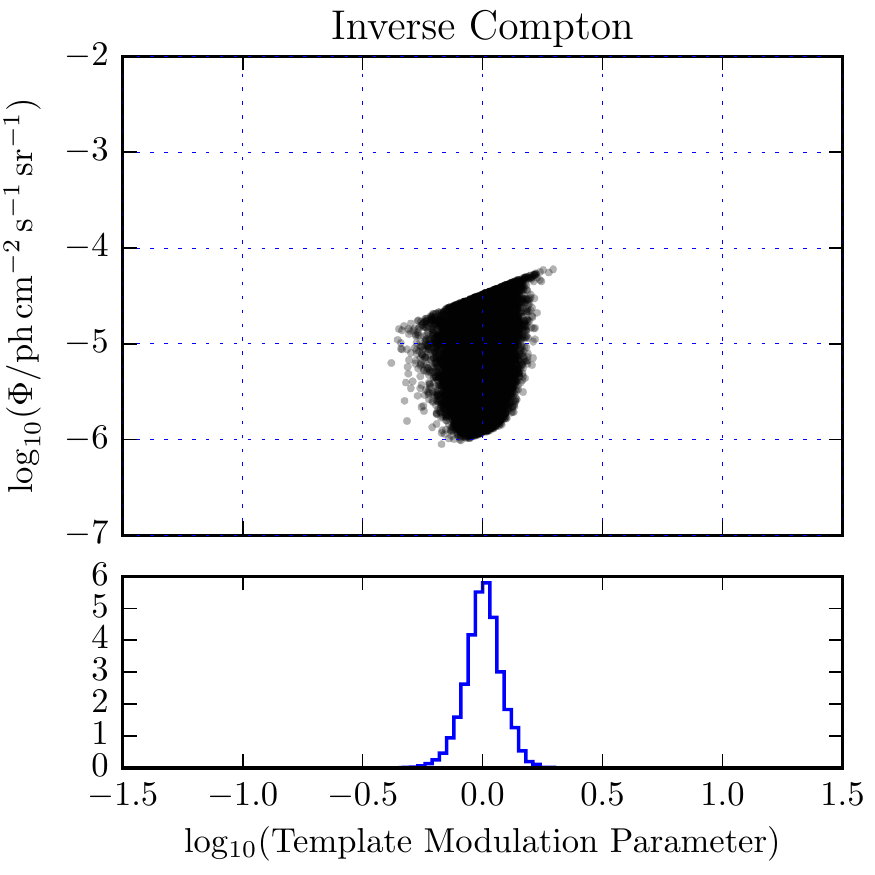}
  \caption{Rescaling parameters for spectra (left panel), and templates (center and right panels) for mock 2-component run (labeled ``constrained'' in table~\ref{tab:pi0_ICS_mock}) with essentially free templates and spectra, and strong smoothing ($\lambda,\lambda' = 4$, $\eta = 0$).}
  \label{fig:2comp_sm0p0_t0p5_s0p5_rsc}
\end{figure}

\begin{figure}
  \begin{minipage}{0.45\textwidth}
    \includegraphics[width=0.8\linewidth]{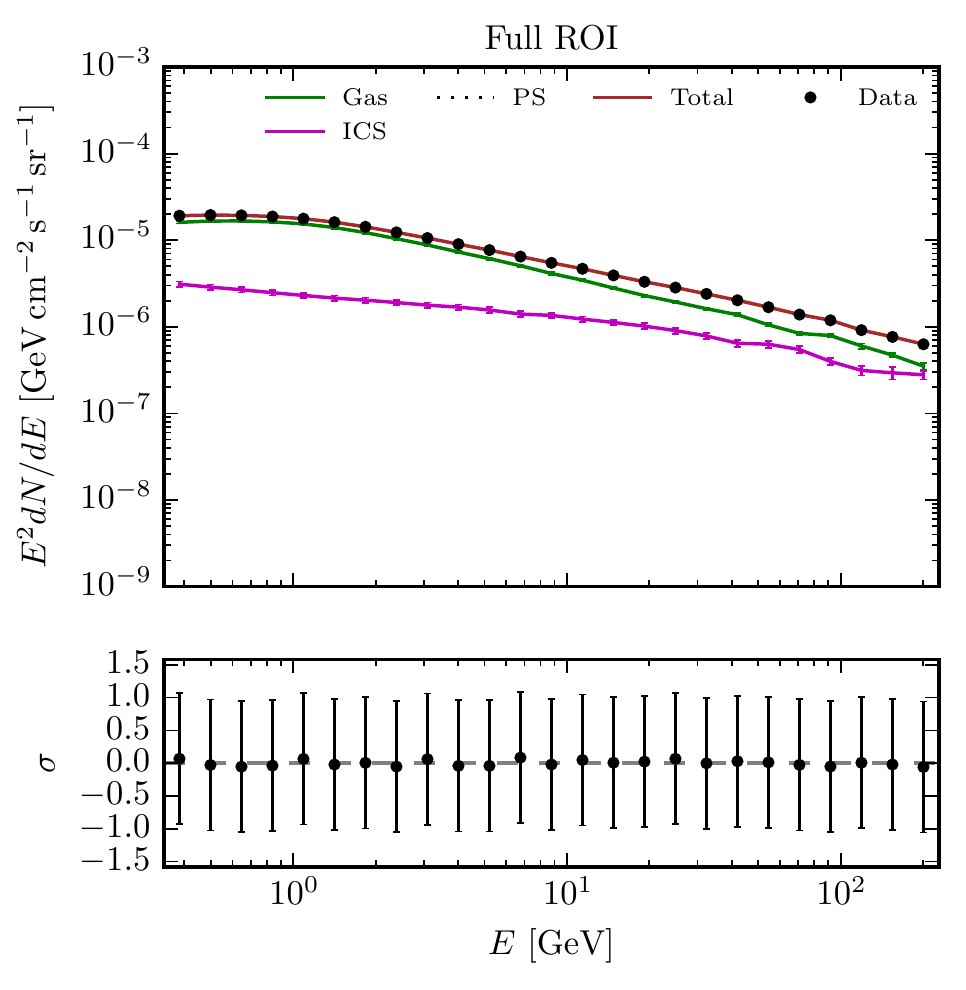}
  \end{minipage}
  \begin{minipage}{0.55\textwidth}
    \includegraphics[width=0.99\linewidth]{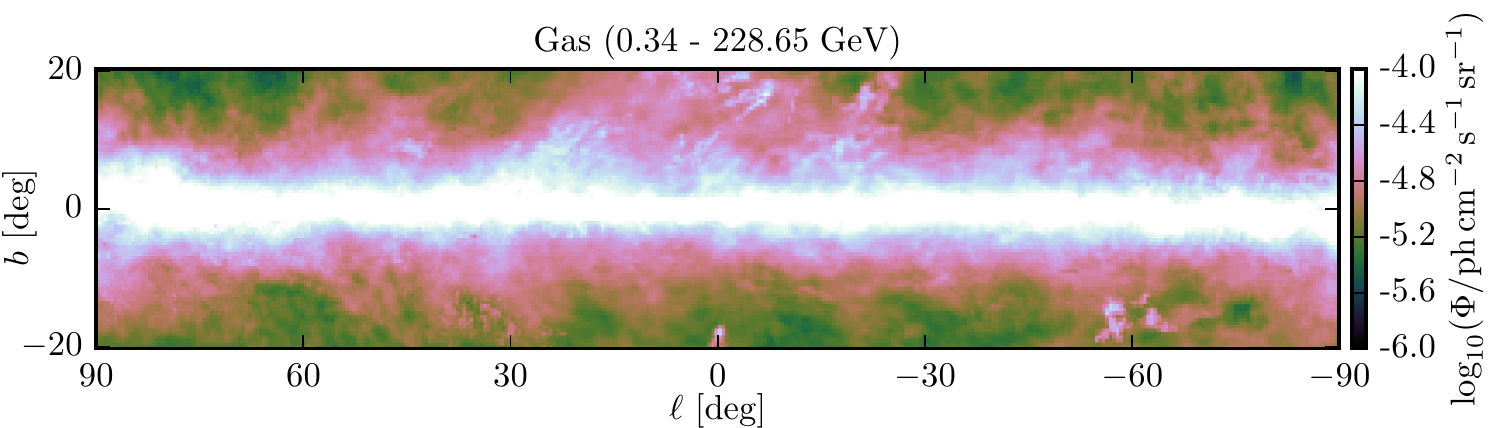}
    \includegraphics[width=0.99\linewidth]{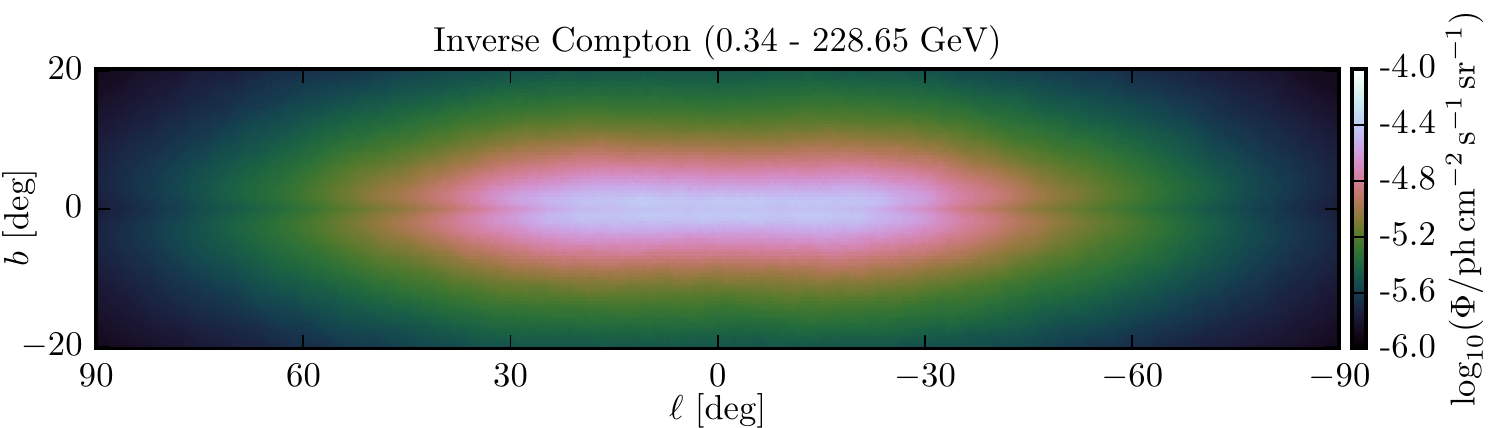}
  \end{minipage}
  \caption{Results for mock 2-component (labeled ``smoothed'' in table~\ref{tab:pi0_ICS_mock}) run with essentially free templates and spectra, and strong smoothing ($\lambda,\lambda' = 1$, $\eta = 100$).}
  \label{fig:2comp_sm0p1_t1_s1}
\end{figure}

\begin{figure}
  \begin{minipage}{0.45\textwidth}
    \includegraphics[width=0.8\linewidth]{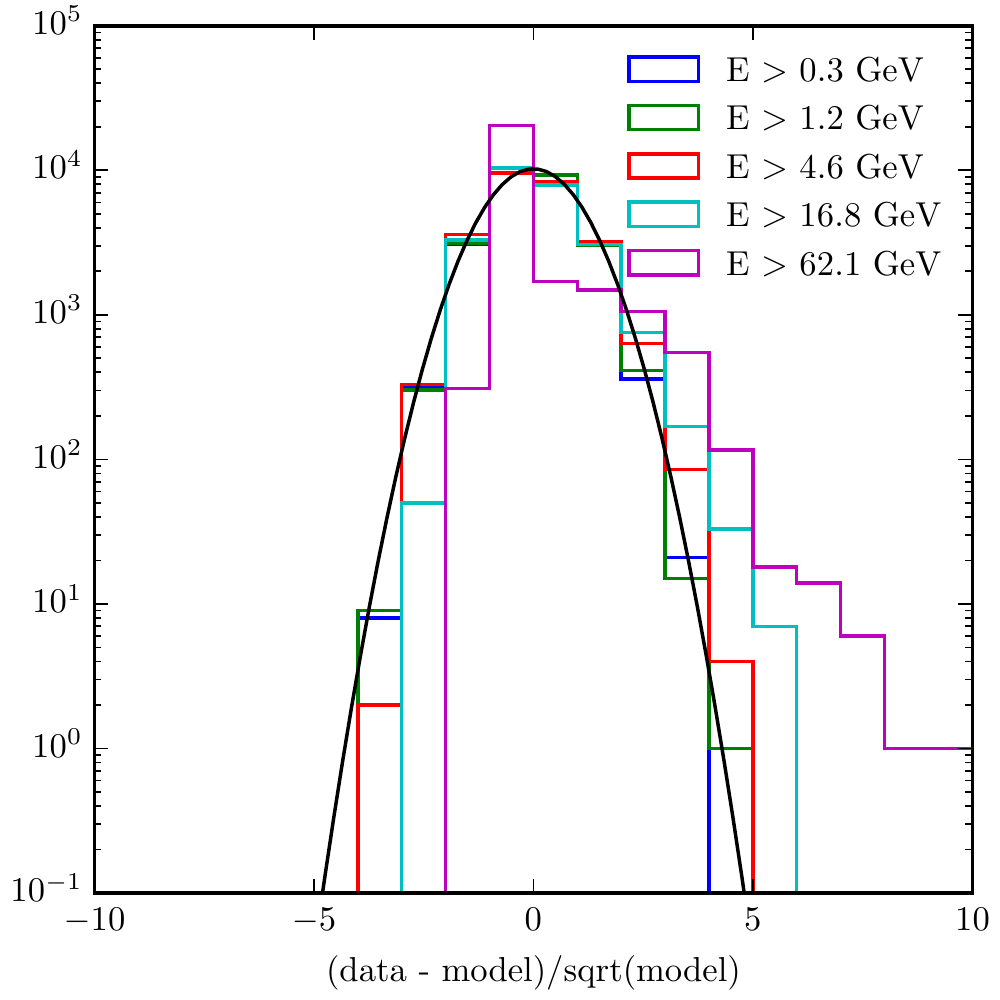}
  \end{minipage}
  \begin{minipage}{0.55\textwidth}
    \includegraphics[width=0.99\linewidth]{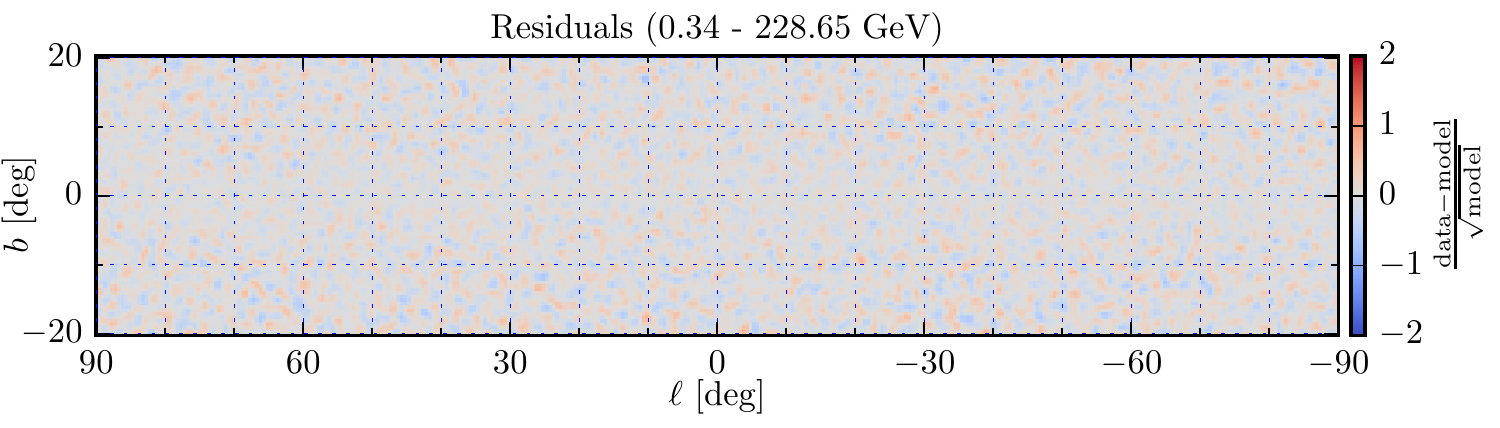}
    \includegraphics[width=0.99\linewidth]{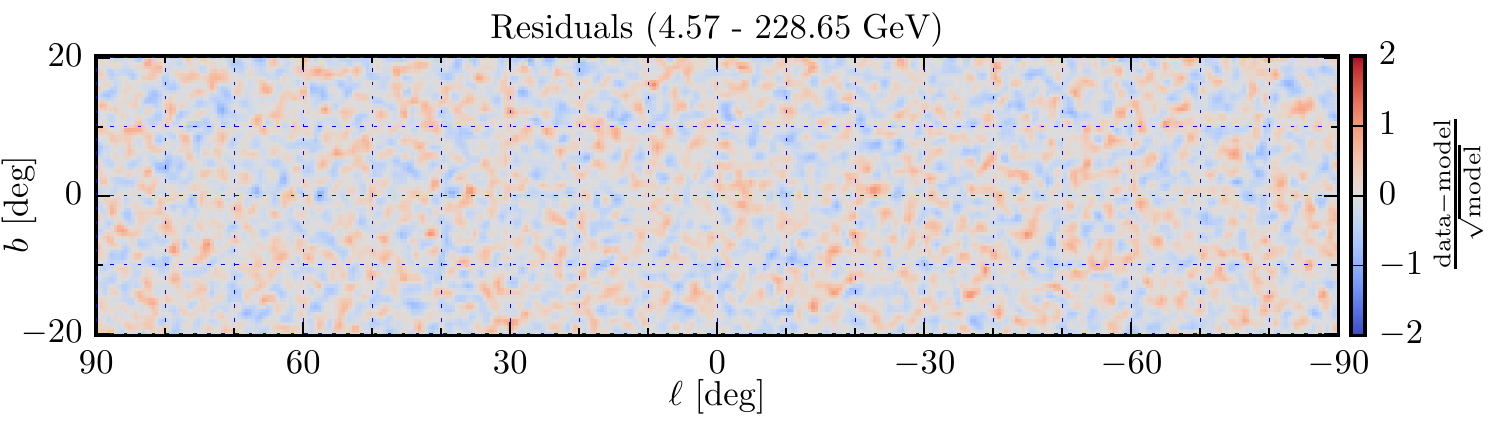}
  \end{minipage}
  \caption{Residuals for mock 2-component run (labeled ``smoothed'' in table~\ref{tab:pi0_ICS_mock}) with essentially free templates and spectra, and strong smoothing ($\lambda,\lambda' = 1$, $\eta = 100$). Residual maps and histograms are in units of significance.}
  \label{fig:2comp_sm0p1_t1_s1_res}
\end{figure}

\begin{figure}
  \centering
  \includegraphics[width=0.32\linewidth]{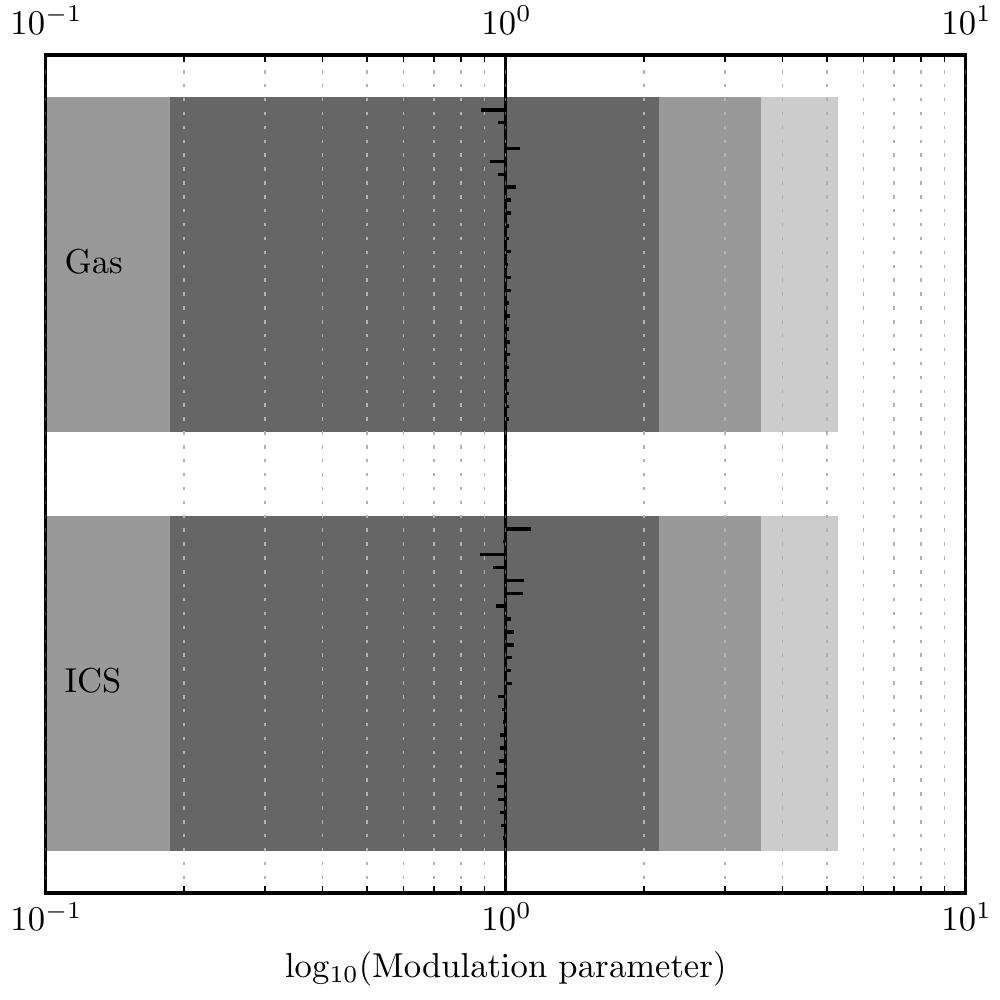}
  \includegraphics[width=0.32\linewidth]{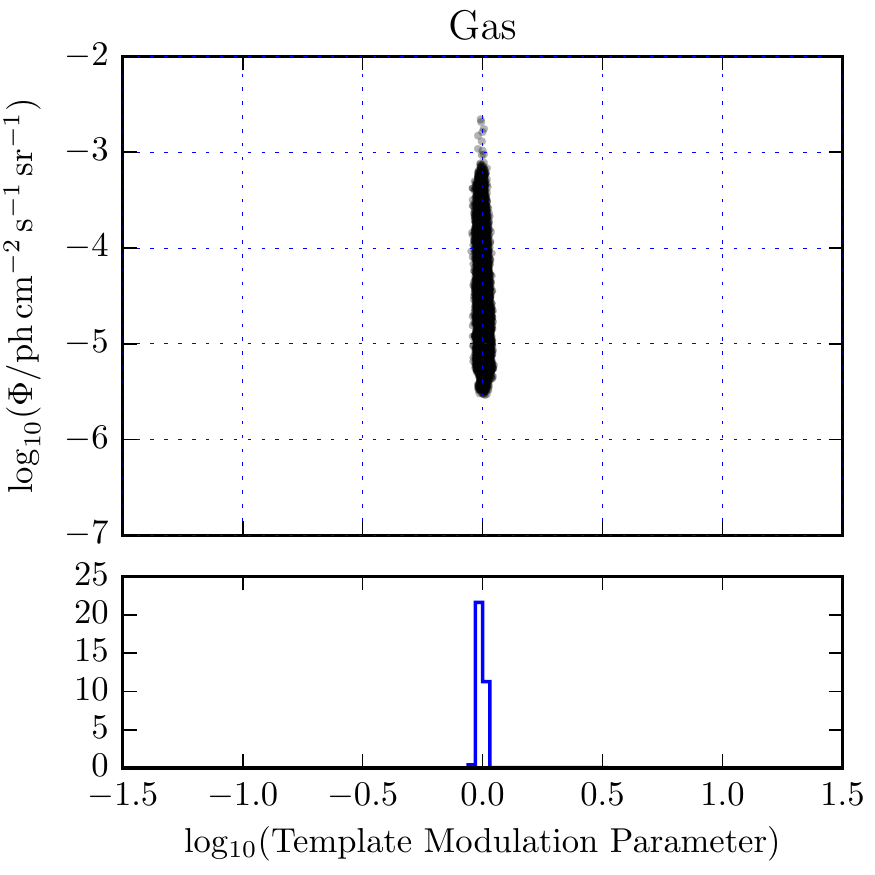}
  \includegraphics[width=0.32\linewidth]{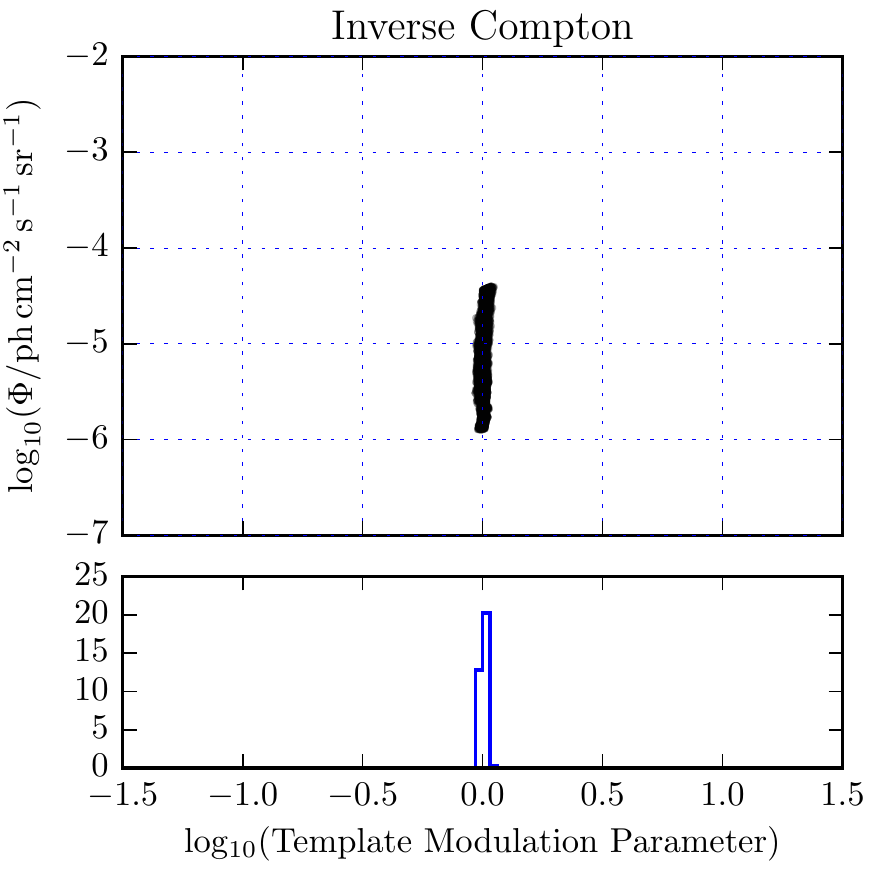}
  \caption{Rescaling parameters for spectra (left panel), and templates (center and right panels) for mock 2-component run (labeled ``smoothed'' in table~\ref{tab:pi0_ICS_mock}) with essentially free templates and spectra, and strong smoothing ($\lambda,\lambda' = 1$, $\eta = 100$).}
  \label{fig:2comp_sm0p1_t1_s1_rsc}
\end{figure}

For all of the following fits, we allow the overall normalization of each of the templates to freely vary ($\lambda'' = 0$). The templates and spectra are initialized to their original components described in section~\ref{sec:gde}. All runs were performed with MEM regularization. We show the hyper-parameter values for each fit in table~\ref{tab:pi0_ICS_mock}.

We first show the results of a run with moderate template and spectra modulation parameters, and no smoothing ($\lambda,\lambda' = 1$, $\eta,\eta' = 0$). As seen in figure~\ref{fig:2comp_sm0p0_t1_s1}, the fit fails to properly recover the two components, instead separating into a ``low-energy'' component and a ``high-energy'' component. If the modulation is set to be weaker, the same behavior is seen (the error bars on the spectra become larger). While a histogram of residuals and map of residuals summed over all energies indicates no substantial deviations, there is clear structure remaining in at least the higher energy residual map, all shown in figure~\ref{fig:2comp_sm0p0_t1_s1_res}. Additionally, as shown in figure~\ref{fig:2comp_sm0p0_t1_s1_rsc}, there are large deviations from the original spectral and spatial components. While this is formally not a bad fit (its likelihood is similar to those of the following runs), it is obvious from the figures that it is an unphysical result.

Next, we show the results of a run with stronger template and spectra modulation parameters, still with no smoothing ($\lambda,\lambda' = 4$, $\eta,\eta' = 0$); these regularization values are near those considered in fits to real data. In this case, the original model components are recovered well, as shown in figure~\ref{fig:2comp_sm0p0_t0p5_s0p5}. We show the residuals for this fit in figure~\ref{fig:2comp_sm0p0_t0p5_s0p5_res}. From the rescaling parameters shown in figure~\ref{fig:2comp_sm0p0_t0p5_s0p5_rsc}, we see very little deviation from the original spectra, and small deviations in the template rescaling

Strong smoothing can act to effectively constrain the allowed variation in the templates, even if the modulation parameters are relatively weak. In figure~\ref{fig:2comp_sm0p1_t1_s1}, we show the results from a fit with the same modulation parameters as in figure~\ref{fig:2comp_sm0p0_t1_s1}, but with strong smoothing applied to both templates (but still no smoothing on the spectra; $\lambda,\lambda' = 1$, $\eta=0.1$, $\eta'=0$). There are no strong residuals obvious in the maps or histograms in figure~\ref{fig:2comp_sm0p1_t1_s1_res}. As shown in figure~\ref{fig:2comp_sm0p1_t1_s1_rsc}, there are essentially no deviations from the original spectra or spatial templates.

\subsection{Tests with \run5 mock data}
\label{apx:run5}

\begin{figure}[h]
  \centering
  \includegraphics[width=0.6\linewidth]{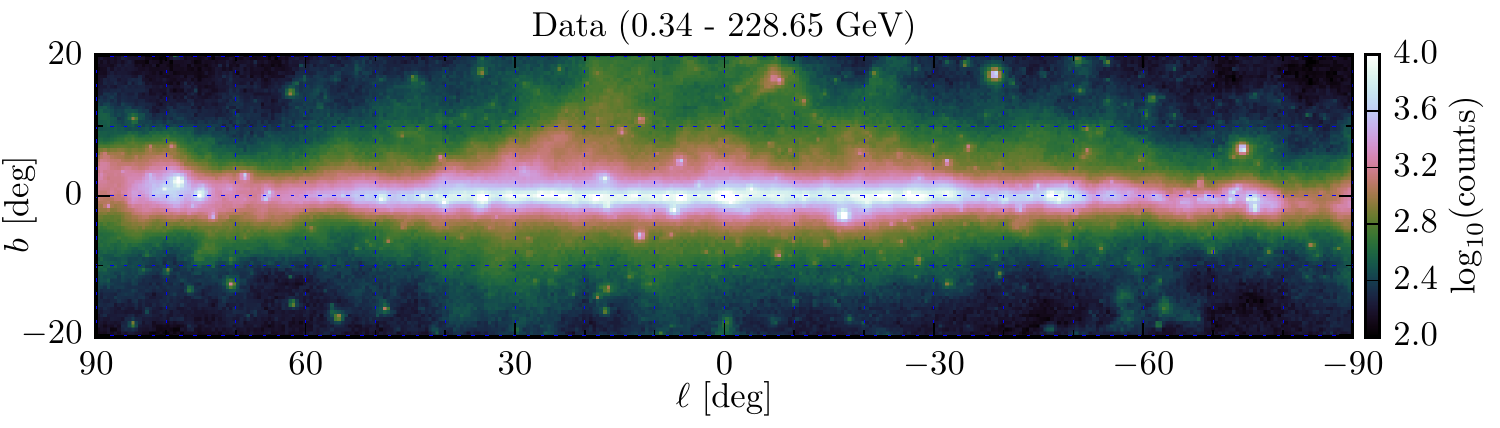}
  \includegraphics[width=0.4\linewidth]{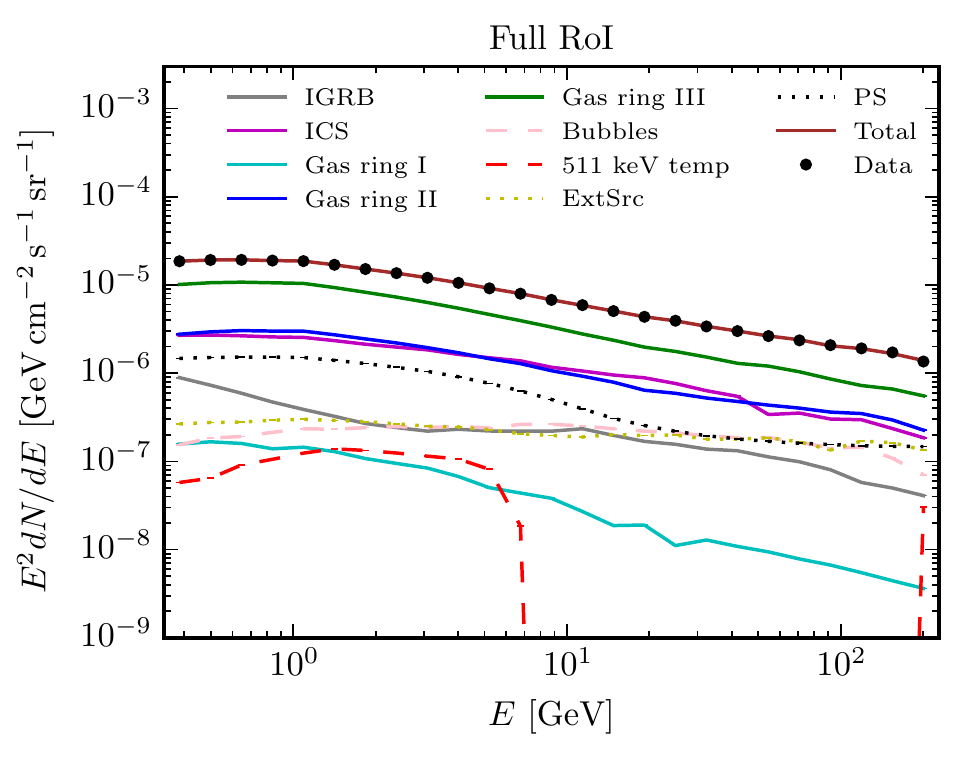}
    \caption{Top: counts map, summed over all energies. Bottom: spectral data components.}
    \label{fig:run5_mock_initial}
\end{figure}

For our final set of tests, we build a mock data set that is composed of the best-fit model components to run5, described in section~\ref{sec:min}. We show the counts map and spectra used for this set of tests in figure~\ref{fig:run5_mock_initial}.

\begin{figure}[h]
    \centering
    \includegraphics[width=0.4\linewidth]{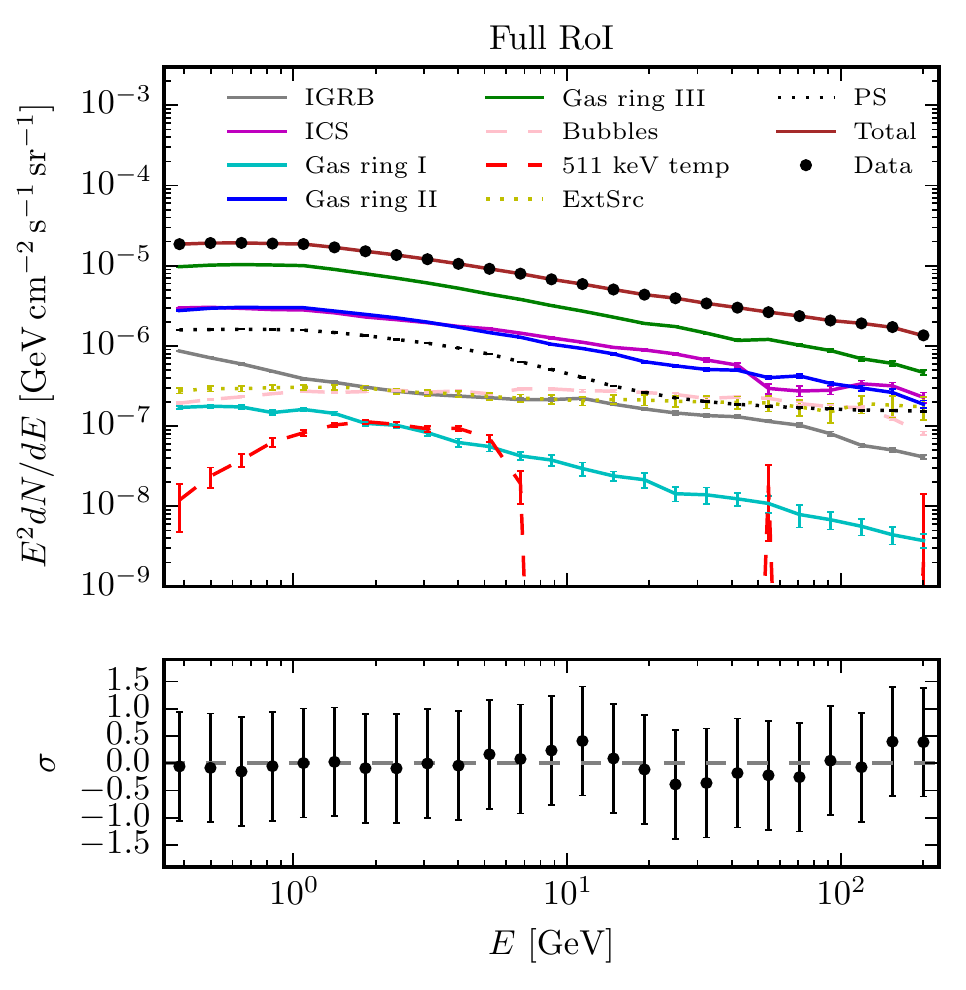}
    \includegraphics[width=0.4\linewidth]{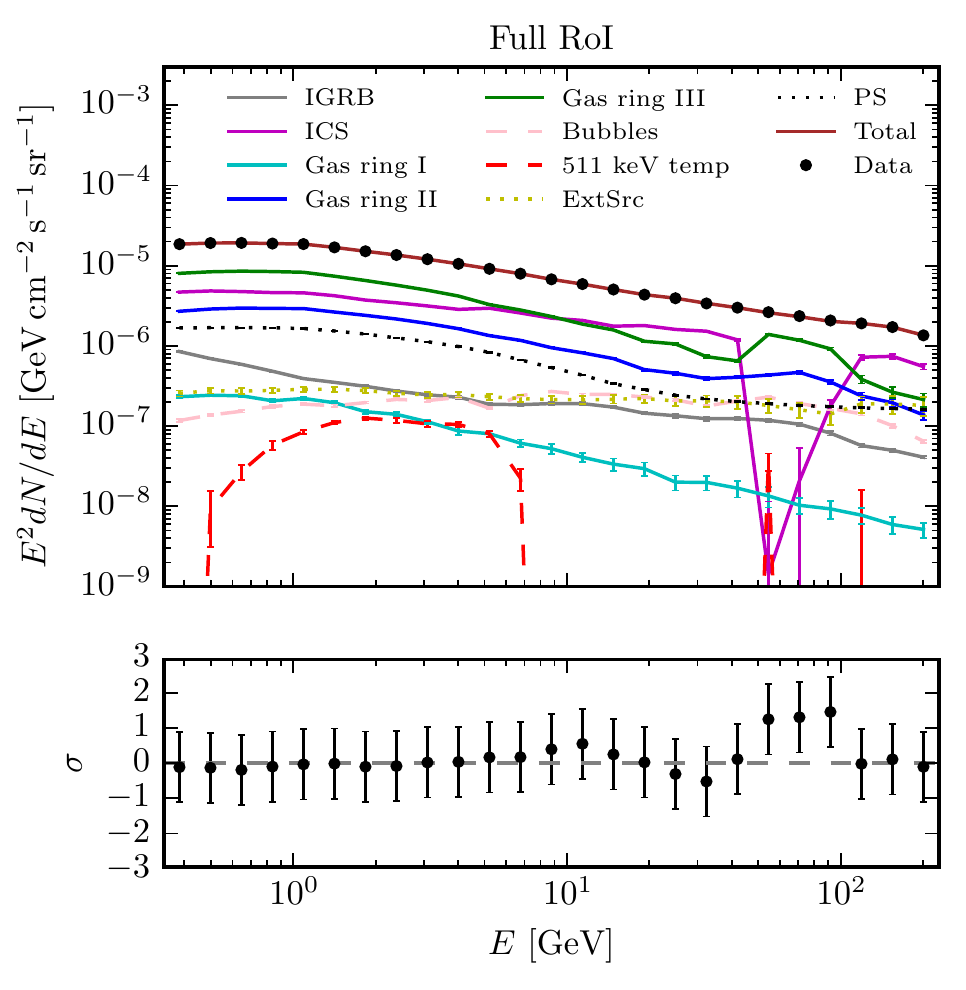}
    \includegraphics[width=0.4\linewidth]{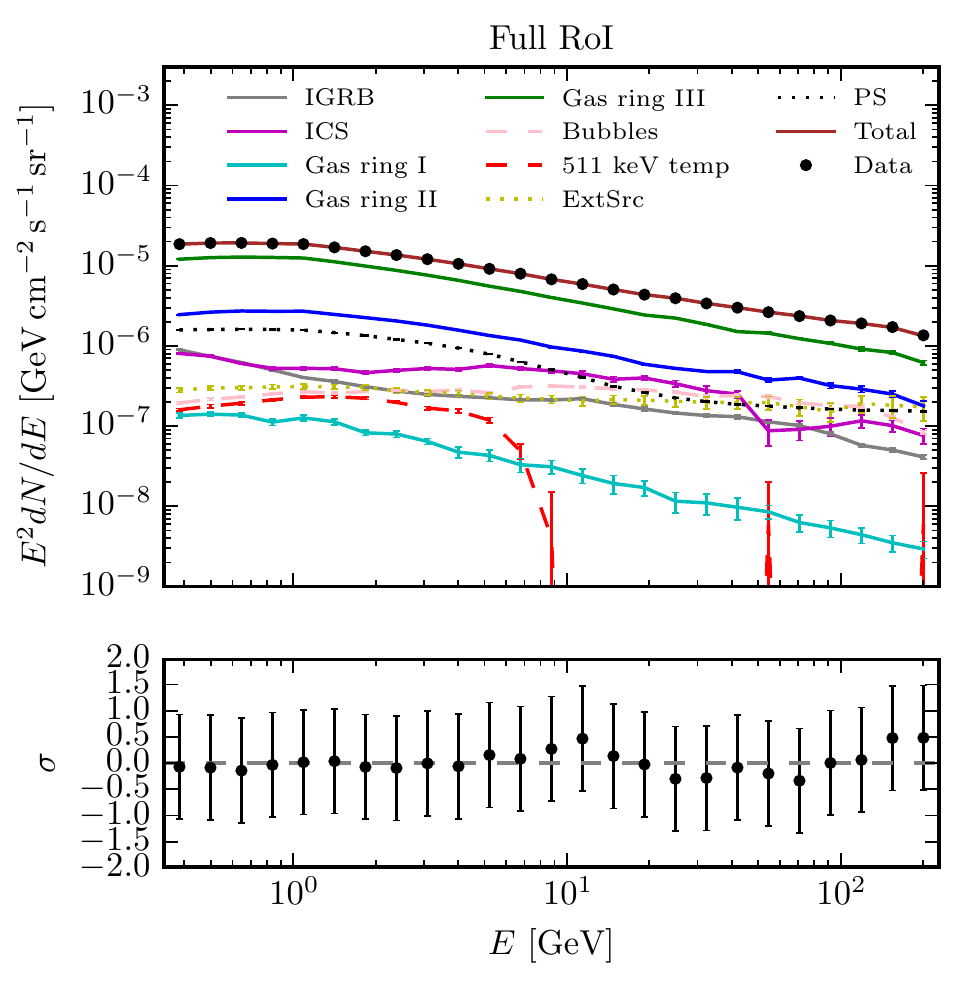}
    \includegraphics[width=0.4\linewidth]{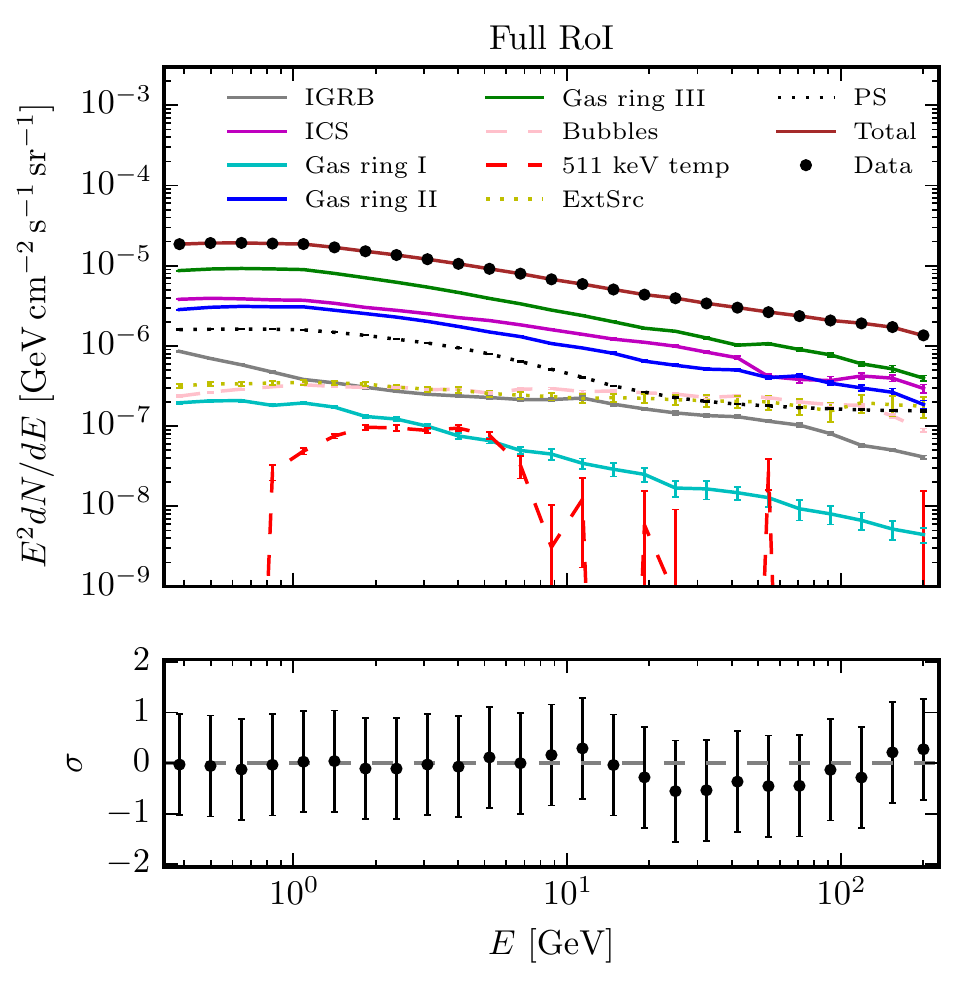}
    \caption{Comparison of spectra for the various tests performed with the run5 mock data set. Top-left: post-fit, original template modulation and smoothing parameters. Top-right: no smoothing. Bottom-left: weaker template modulation parameters. Bottom-right: stronger template modulation parameters.}
    \label{fig:run5_mock_spectra}
\end{figure}

\begin{figure}[h]
    \centering
    \includegraphics[width=0.4\linewidth]{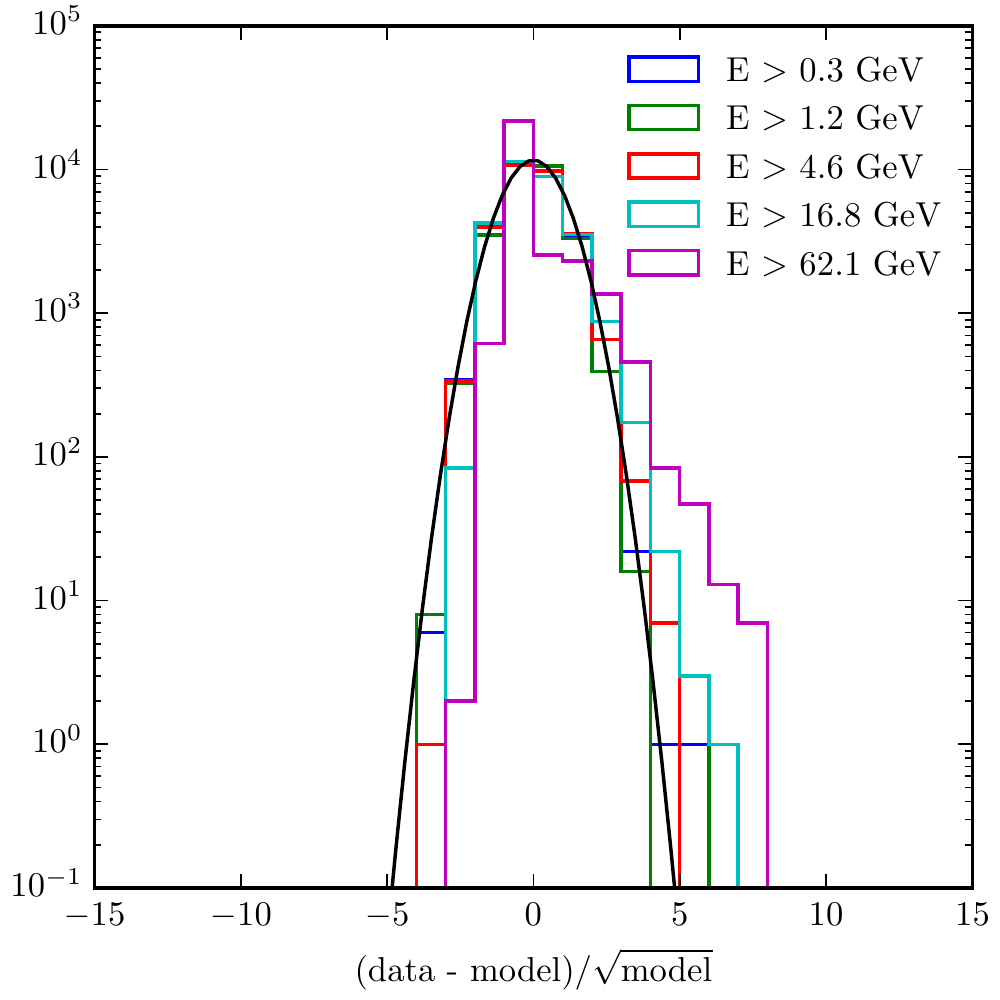}
    \includegraphics[width=0.4\linewidth]{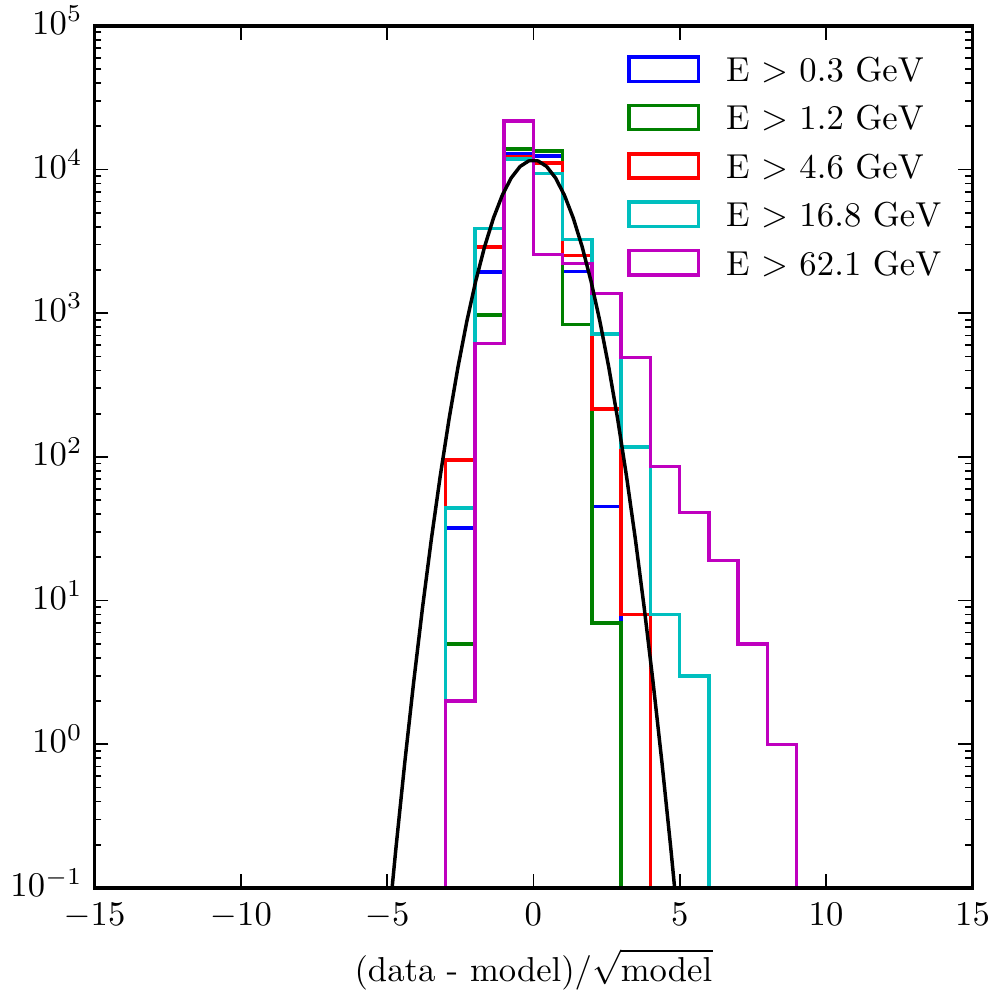}
    \includegraphics[width=0.4\linewidth]{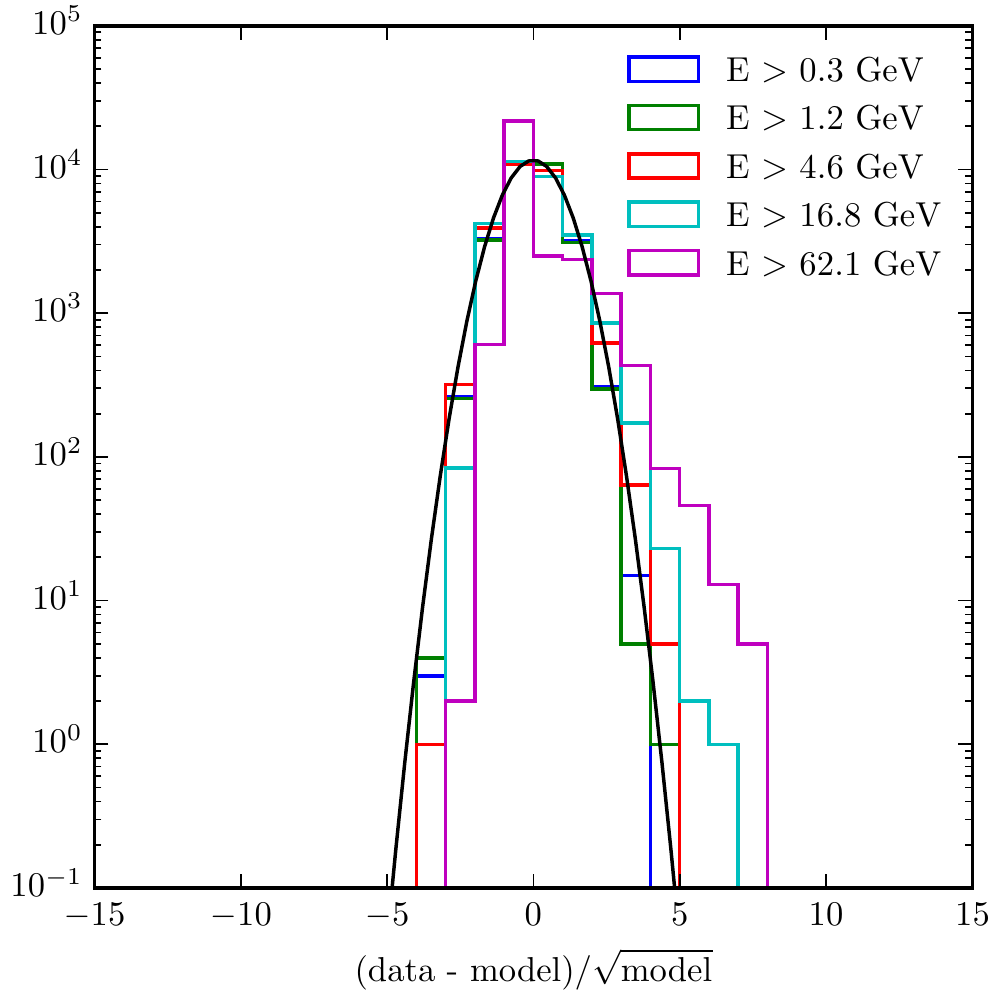}
    \includegraphics[width=0.4\linewidth]{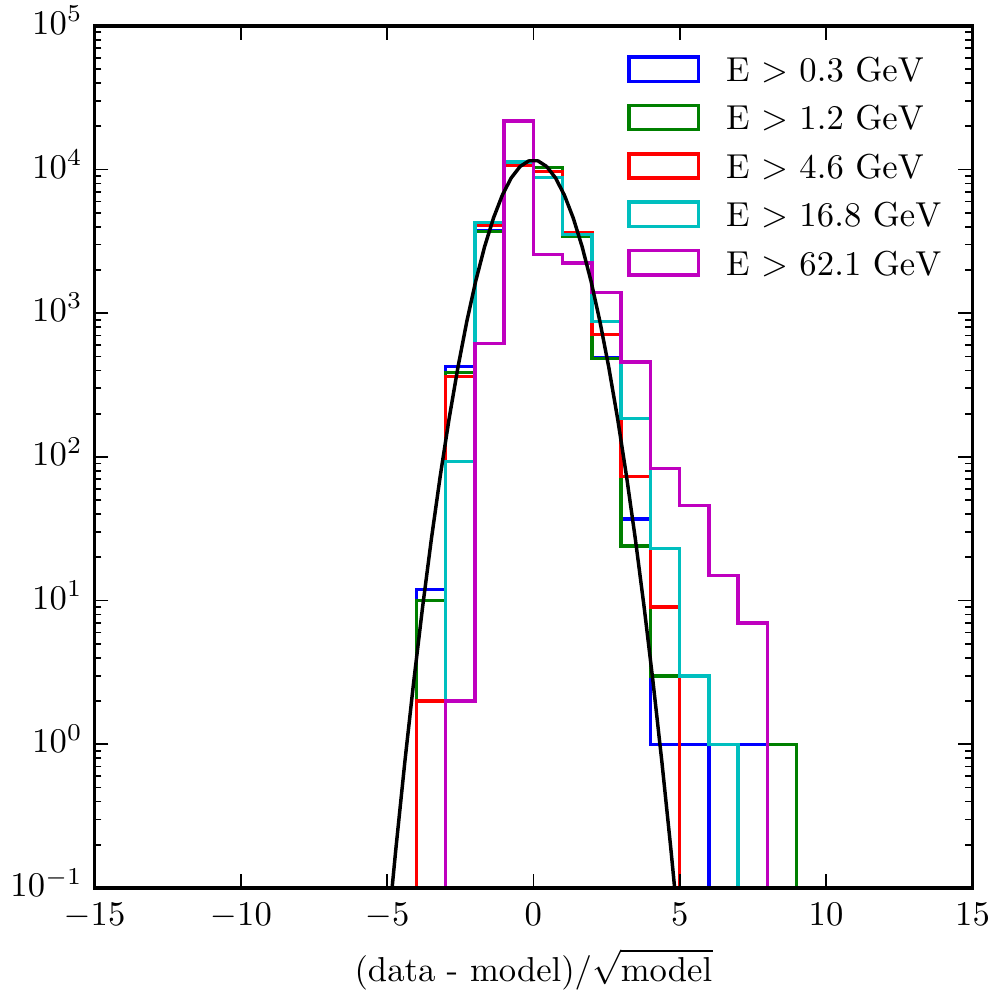}

    \caption{Comparison of histograms for the various tests performed with the run5 mock data set. Top-left: post-fit, original template modulation and smoothing parameters. Top-right: no smoothing. Bottom-left: weaker template modulation parameters. Bottom-right: stronger template modulation parameters.}
    \label{fig:run5_mock_hist}
\end{figure}

We first fit the mock data set, with the same model components as in \run5, but starting from the best-fit parameters for \run5 (see section ~\ref{sec:min}). As shown in in the top left panel of figure~\ref{fig:run5_mock_spectra}, the spectral components are almost perfectly recovered, to within $\sim30\%$ on average (except for a few obvious energy bins in the 511 keV spectrum, which is not smoothed). We show histograms of the residuals for this mock data set after fitting in the top left panel of figure~\ref{fig:run5_mock_hist}; this can be compared to the original fit for \run5 in figure~\ref{fig:residual_histogram}. The Poisson tail is clearly seen in the highest energy bin.

To test the robustness of our results for \run5 on real LAT data, we vary the template modulation and smoothing parameters for this mock data set. In our first test, we set the smoothing on all components to zero. The results of these tests are shown in the top-right panels of figures~\ref{fig:run5_mock_spectra} and \ref{fig:run5_mock_hist}. The most obvious feature in the spectra plot is the deviation in the ICS spectrum at one high energy bin. The mockdata set is Poisson-dominated at this energy, and there are few photons actually being fit. Additionally, the spatial and spectral modulation parameters are somewhat weaker for the ICS component than the other components, especially the gas, which, in the absence of of smoothing, means that the ICS component is permitted more variation. This deviation is clearly unphysical, and demonstrates the need for moderate regularization in low photon regimes, as was also seen in the previous two sections. Interestingly, the overall deviation from the data is also quite high, at the $\sim2\sigma$ level for three energy bins, indicating that these energies are not well fit. The overall likelihood is not significantly different, however, from any of the other mock data runs in this section.

We also perform fits where the spatial modulation parameters for the ICS and three gas components are set to be weaker and stronger by a factor of 4 relative to those in table~\ref{tab:fits}. The other modulation parameters and all smoothing parameters remain unchanged from the original \run5 set-up. The results of these tests are shown in the bottom panels of figures~\ref{fig:run5_mock_spectra} and \ref{fig:run5_mock_hist}. Again, in the case of the weaker modulation (bottom left), the ICS component is considerably suppressed while the outer gas ring III is enhanced relative to the original best-fit parameters; the 511 keV template is also somewhat enhanced. This illustrates how weakly constrained the ICS component is relative to the other components. The residual histograms in figure~\ref{fig:run5_mock_hist} all show reasonable overall fits.

\newpage
\bibliography{skyfact}
\bibliographystyle{JHEP}

\end{document}